\documentclass[11pt]{article}
\usepackage[utf8]{inputenc}
\usepackage{graphicx}
\usepackage{amsmath}
\usepackage{tabularx}
\usepackage{amssymb}
\usepackage{parskip}
\usepackage{textcomp}
\usepackage{bbm} 
\usepackage[dvipsnames]{xcolor}
\usepackage{hyperref}
\usepackage{cancel}
\usepackage{soul}
\usepackage{ragged2e}
\usepackage{bm}
\usepackage{romannum}
\usepackage{float}
\usepackage{tikz}
\usepackage[
linewidth=0.5pt,
middlelinecolor= black,
middlelinewidth=0.4pt,
roundcorner=1pt,
topline = false,
rightline = false,
bottomline = false,
rightmargin=0pt,
skipabove=0pt,
skipbelow=0pt,
leftmargin=-4mm,
innerleftmargin=3.5mm,
innerrightmargin=0pt,
innertopmargin=0pt,
innerbottommargin=0pt,
]{mdframed}
\usetikzlibrary{positioning}
\usetikzlibrary{decorations.text}
\usetikzlibrary{decorations.pathmorphing}
\usetikzlibrary{decorations.pathreplacing}
\usetikzlibrary{patterns}
\tikzset{snake it/.style={decorate, decoration={snake, segment length=1.5mm, amplitude=1pt}}}

\usepackage{geometry}
\AtBeginDocument{\pagenumbering{arabic}}

\title{\vspace{2mm}\LARGE Caustics in the spherically symmetric Einstein-dust system\vspace{-2mm}}
\author{David Bick\\ \\ \footnotesize
\textit{Department of Pure Mathematics and Mathematical Statistics,}\\ \footnotesize \textit{
University of Cambridge, Wilberforce Road, Cambridge CB3 0WB, United Kingdom}\vspace{5mm}}
\date{\today}

\begin{document}
\maketitle

\renewcommand{\labelenumi}{(\roman{enumi})}

\begin{center}
\begin{minipage}{0.85\textwidth}
\begin{center}
\footnotesize
\vspace{4mm}\textbf{Abstract}\vspace*{2mm}\\
\justifying
\hspace*{4mm}Caustics---envelopes formed by the trajectories of fluid particles---arise in proposed dynamical extensions for shell-crossing singularities occurring in the Einstein-dust system. In this study, a local existence result is established, describing the dynamics in a neighbourhood of such caustics. Specifically, we obtain spherically symmetric spacetimes $(M,g_{\mu\nu})$ containing a caustic $\mathcal{C}$, which, in the quotient $M/SO(3)$, is a timelike curve forming a singular boundary between a 2-dust region and a vacuum region. The spacetimes are constructed from solutions to a PDE problem posed with a spacelike direction of evolution. \\
\hspace*{4mm}Curvature invariants and energy densities diverge as the caustic is approached. Consequently the metric has limited regularity $g\in C^{1, 1/2}$ and is shown to satisfy Einstein's equation weakly. On the complement of the caustic, the metric is smooth and satisfies Einstein's equation classically. A (degenerate) coordinate system is identified in which the dynamical variables are smooth \textit{with extension to} the caustic. \\
\hspace*{4mm}Finally, a novel family of static, spherically symmetric spacetimes is identified, complementing the local construction above. Each spacetime contains an eternal annular 2-dust region bounded by a pair of caustics.
\end{center}
\end{minipage}
\end{center}
\vspace{1.2cm}
\Large\textbf{Contents}\small\vspace*{4mm}\\
\hyperlink{sec:1}{\textbf{1\hspace{3mm}Introduction}}\hfill \textbf{\pageref{1}} \\
\hyperlink{sec:1.1}{\hspace{5mm}\text{1.1\hspace{3mm}Shell-crossing singularities}}\dotfill \pageref{1.1}\\
\hyperlink{sec:1.2}{\hspace{5mm}\text{1.2\hspace{3mm}Dynamical extension via catastrophe folds}}\dotfill \pageref{1.2}\\
\hyperlink{sec:1.3}{\hspace{5mm}\text{1.3\hspace{3mm}Formulation of a restricted problem}}\dotfill \pageref{1.3}\\
\hyperlink{sec:1.4}{\hspace{5mm}\text{1.4\hspace{3mm}A new family of static examples}}\dotfill \pageref{1.4}\vspace*{2mm} \\
\hyperlink{sec:2}{\textbf{2\hspace{3mm}Preliminaries}}\hfill \textbf{\pageref{2}}\\
\hyperlink{sec:2.1}{\hspace{5mm}\text{2.1\hspace{3mm}Notation and gauge choices}}\dotfill \pageref{2.1}\\
\hyperlink{sec:2.2}{\hspace{5mm}\text{2.2\hspace{3mm}Constraint equations}}\dotfill \pageref{2.2}\\
\hyperlink{sec:2.3}{\hspace{5mm}\text{2.3\hspace{3mm}Specification of data and the domain of evolution}}\dotfill \pageref{2.3}\\
\hyperlink{sec:2.4}{\hspace{5mm}\text{2.4\hspace{3mm}Higher order boundary conditions}}\dotfill \pageref{2.5}\\
\hyperlink{sec:2.5}{\hspace{5mm}\text{2.5\hspace{3mm}Function spaces used in the iteration scheme}}\dotfill \pageref{2.5}\vspace*{2mm} \\
\hyperlink{sec:3}{\textbf{3\hspace{3mm}Analysis of the Einstein equations}}\hfill \textbf{\pageref{3}}\\
\hyperlink{sec:3.1}{\hspace{5mm}\text{3.1\hspace{3mm}Solving the linear wave equation with cusp-regular inhomogeneity}}\dotfill \pageref{3.1}\\
\hyperlink{sec:3.2}{\hspace{5mm}\text{3.2\hspace{3mm}Estimates for the Einstein equations}}\dotfill \pageref{3.2}\vspace*{2mm}\\
\hyperlink{sec:4}{\textbf{4\hspace{3mm}Analysis of the matter equations}}\hfill \textbf{\pageref{4}}\\
\hyperlink{sec:4.1}{\hspace{5mm}\text{4.1\hspace{3mm}Estimates for the geodesic equation}}\dotfill \pageref{4.1}\\
\hyperlink{sec:4.2}{\hspace{5mm}\text{4.2\hspace{3mm}Obtaining a smooth one-parameter family of geodesic tangents}}\dotfill \pageref{4.2}\\
\hyperlink{sec:4.3}{\hspace{5mm}\text{4.3\hspace{3mm}Forming a foliation of $\Sigma_\varepsilon$}}\dotfill \pageref{4.3}\\
\hyperlink{sec:4.4}{\hspace{5mm}\text{4.4\hspace{3mm}Existence and bounds for $\sigma$}}\dotfill \pageref{4.4}\\
\hyperlink{sec:4.5}{\hspace{5mm}\text{4.5\hspace{3mm}Proof of Proposition 4.1}}\dotfill \pageref{4.5}\vspace*{2mm} \\
\hyperlink{sec:5}{\textbf{5\hspace{3mm}Contraction bounds}}\hfill \textbf{\pageref{5}}\\
\hyperlink{sec:5.1}{\hspace{5mm}\text{5.1\hspace{3mm}Difference estimates for $\bm{u},\sigma$}}\dotfill \pageref{5.1}\\
\hyperlink{sec:5.2}{\hspace{5mm}\text{5.2\hspace{3mm}Contraction estimates for $\Omega^2,r$}}\dotfill \pageref{5.2}\vspace*{2mm}\\
\hyperlink{sec:6}{\textbf{6\hspace{3mm}Proof of Theorem 1.1}}\hfill \textbf{\pageref{6}}\\
\hyperlink{sec:6.1}{\hspace{5mm}\text{6.1\hspace{3mm}Preparatory lemmas}}\dotfill \pageref{6.1}\\
\hyperlink{sec:6.2}{\hspace{5mm}\text{6.2\hspace{3mm}Proof of Theorem 1.1}}\dotfill \pageref{6.2}\vspace*{2mm}\\
\hyperlink{sec:7}{\textbf{7\hspace{3mm}Static dust caustics}}\hfill \textbf{\pageref{7}}\\
\hyperlink{sec:7.1}{\hspace{5mm}\text{7.1\hspace{3mm}Reduction to ODE system}}\dotfill \pageref{7.1} \\
\hyperlink{sec:7.2}{\hspace{5mm}\text{7.2\hspace{3mm}Obtaining a 3-parameter family of solutions}}\dotfill \pageref{7.2} \\
\hyperlink{sec:7.3}{\hspace{5mm}\text{7.3\hspace{3mm}Proof of Theorem 1.2}}\dotfill \pageref{7.3}\vspace*{2mm} \\
\hyperlink{sec:A}{\textbf{A\hspace{2mm}Useful inequalities}}\hfill \textbf{\pageref{A}}\vspace*{2mm}\\
\hyperlink{sec:B}{\textbf{B\hspace{2mm}Weak formulation of Einstein's equation}}\hfill \textbf{\pageref{B}}\vspace*{5mm}\\

\Large\hypertarget{sec:1}{\textbf{1\hspace{4mm}Introduction}}\label{1}\normalsize\\ \\
A key measure of the strength of singularities occurring in the study of nonlinear partial differential equations is \textit{extendibility}, that is, whether a solution admits a suitable extension onto a larger domain, while remaining in a reasonable category of solutions to the original equation. For example, in analysis of the compressible Euler equations, the blowup of derivatives of the fluid variables marks the end of the $C^1$ development of initial data. However, solutions describing a propagating shock wave represent a dynamical extension within a class of weak solutions. In contrast, the curvature singularity in the Schwarzschild family of solutions to the Einstein vacuum equations of general relativity does not permit an extension even as a merely $C^0$ Lorentzian manifold \hyperlink{Sbi18}{$[$Sbi18$]$}.\\ \\ \\
\large\hypertarget{sec:1.1}{\textbf{1.1\hspace{4mm}Shell-crossing singularities}}\label{1.1}\normalsize\\ \\
In the gravitational collapse of a spherically symmetric inhomogeneous ball of dust, one may distinguish various types of `first singularities' that may develop\footnote[1]{This characterization can be made precise in the language of terminal indecomposable past sets.}. If the spherical fluid shells contract to zero area radius $r$, then in general the fluid energy density $\rho$ diverges. This may occur, firstly, on the closure of the centre of symmetry $\Gamma$, as was studied in Christodoulou's influential article \hyperlink{Chr84}{$[$Chr84$]$}. Secondly, if not on the closure of $\Gamma$, then the singularity is similar to the $r=0$ boundary in the Schwarzschild spacetime. This is the behaviour witnessed by non-central fluid shells in the seminal examples given by \textsc{Oppenheimer-Snyder} \hyperlink{OS39}{$[$OS39$]$}\footnote[2]{Note that, however, analysis of the interior region was not undertaken in that article.}. This article is concerned with a third type of first singularity, for which the area radius has a \textit{positive} limiting value. In this case---known as a \textit{shell-crossing singularity}---the congruence of timelike geodesics constituted by the fluid trajectories reaches a focal point\footnote[3]{It is reasonable to expect, moreover, that this type of singularity formation---on surfaces of codimension 2, that is---would be the generic kind for the Einstein-dust system \textit{outside of symmetry}.}.\\ \\
All three types of breakdown are pertinent to the cosmic censorship conjectures formulated by \textsc{Penrose} \hyperlink{Pen69}{$[$Pen69$]$}. Indeed, shell-crossing singularities were among the first examples to be identified \hyperlink{YSM73}{$[$YSM73$]$} of dynamical naked singularity formation from smooth initial data, representing a potential counterexample to weak cosmic censorship. However, even leaving aside reasonable criticism of the dust matter model, the validity of this interpretation is under question. Shell-crossing singularities---in contrast to the other cases---are thought to admit dynamical extensions beyond the singularity, allowing them to be presented as an interior point of a larger evolutionary spacetime, with only a mild loss in regularity.\\ \\ \\
\begin{minipage}{0.63\textwidth}
\large\hypertarget{sec:1.2}{\textbf{1.2\hspace{4mm}Dynamical extension via catastrophe folds}}\label{1.2}\normalsize\\ \\
One possible extension, studied in a degenerate case by \textsc{Papapetrou-Hamoui} \hyperlink{PH67}{$[$PH67$]$} and later in a heuristic study by \textsc{Clarke-O'Donnell} \hyperlink{COD92}{$[$COD92$]$}, we call the \textit{catastrophe fold continuation} and posits that, to the future of the first singularity, dust trajectories are continued into a new region, in which several dust species overlap. This \textit{multi-dust region} would be bounded by a pair of \textit{caustics} to which fluid trajectories are tangent. In the spherically symmetric quotient, these caustics descend to timelike curves, and we will consistently refer to them as such throughout this article, even though `upstairs' they correspond to hypersurfaces.\\ \\
As indicated in the Figure, each point of the multi-dust region is the intersection of \textit{three} fluid trajectories: one which is tangent to a caustic curve in the past, and two which are ta-\end{minipage}
\hspace{0.0\textwidth}
\begin{minipage}{0.35\textwidth}
\vspace{-6mm}
\begin{figure}[H]
\begin{center}
\begin{tikzpicture}[xscale=0.8,yscale=1.1]
\foreach \x in {0,0.5,...,5} {
\draw [gray,opacity=0.3] ({0.2*(\x)^1.7}, {(\x)}) -- ({0.2*(\x)^1.7-(1+(\x))*0.34*((\x)^0.7)}, {(\x)-(1+(\x))*1});
\draw [gray,opacity=0.3] ({-0.2*(\x)^1.7}, {(\x)}) -- ({-0.2*(\x)^1.7+(1+(\x))*0.34*((\x)^0.7)}, {(\x)-(1+(\x))*1});}
\foreach \x in {0,0.2,0.5,0.93,1.4,1.9,2.5,3,3.5,4,4.5,5} {
\draw [gray,opacity=0.3] ({0.2*(\x)^1.7}, {(\x)}) -- ({0.2*(\x)^1.7+(5-(\x))*0.34*((\x)^0.7)}, {(\x)+(5-(\x))*1});
\draw [gray,opacity=0.3] ({-0.2*(\x)^1.7}, {(\x)}) -- ({-0.2*(\x)^1.7-(5-(\x))*0.34*((\x)^0.7)}, {(\x)+(5-(\x))*1});}
\foreach \x in {0.5,1,...,6.5} {
\draw [gray,opacity=0.3] ({0.2*(5)^1.7-0.32*(\x)^1.4}, {5}) -- ({-1.04*5^0.7}, {(5)-(0.2*(5)^1.7+1.04*5^0.7)/(0.34*(5)^0.7)+0.65*(\x)});
\draw [gray,opacity=0.3] ({-0.2*(5)^1.7+0.32*(\x)^1.4}, {5}) -- ({1.04*5^0.7}, {(5)-(0.2*(5)^1.7+1.04*5^0.7)/(0.34*(5)^0.7)+0.65*(\x)});
}
\draw [gray,opacity=0.3] ({0.2*(5)^1.7-0.32*(0.5)^1.4}, {5}) -- ({-1.04*5^0.7}, {(5)-(0.2*(5)^1.7+1.04*5^0.7)/(0.34*(5)^0.7)+0.65*(0.5)});
\draw [black] ({0.2*(5)^1.7-0.32*(0.5)^1.4}, {5}) -- (-1,1.356);
\draw [darkgray,opacity=0.5] (-1.1,1.265) -- (-1,1.356);
\draw [darkgray,opacity=0.3] (-1.1,1.265) -- (-1.2,1.172);
\draw [black] ({-0.2*(0.2)^1.7}, {(0.2)}) -- ({-0.2*(0.2)^1.7-(5-(0.2))*0.34*((0.2)^0.7)}, {(0.2)+(5-(0.2)*1});
\draw [gray,opacity=0.3] ({-0.2*(4.5)^1.7}, {(4.5)}) -- ({-0.2*(4.5)^1.7+(1+(4.5))*0.34*((4.5)^0.7)}, {(4.5)-(1+(4.5))*1});
\draw [black] ({-0.2*(4.5)^1.7}, {(4.5)}) -- (0.7,1.135);
\draw [darkgray,opacity=0.5] (0.8,1.032) -- (0.7,1.135);
\draw [darkgray,opacity=0.3] (0.8,1.032) -- (0.9,0.930);
\draw [-stealth](0.21,2.47) -- (0.31,2.56);
\draw [-stealth](-0.6,2.47) -- (-0.7,2.57);
\draw [-stealth](-0.155,1.5) -- (-0.17,1.6);
\draw [-stealth](0,-0.7) -- (0,-0.4);
\draw [-stealth](-0.19,-0.7) -- (-0.13,-0.4);
\draw [-stealth](-0.38,-0.7) -- (-0.28,-0.4);
\draw [-stealth](0.19,-0.7) -- (0.13,-0.4);
\draw [-stealth](0.38,-0.7) -- (0.28,-0.4);
\draw [-stealth](-0.155,1.5) -- (-0.17,1.6);
\filldraw[color=black, fill=black](-0.22,2.08) circle (0.03);
\node[black, rotate=63] at (2.05, 3.62) [anchor = east] {$_\text{caustic curve}$};
\node[black, rotate=-63] at (-1, 2) [anchor = east] {$_\text{caustic curve}$};
\draw [fill=white,white,opacity=0.6] (0.6,-0.25) rectangle (2.9,0.5);
\draw [darkgray] (0.1,0) -- (0.7,0.1);
\node[black] at (1.8, 0.5) [anchor = north] {$_\text{shell-crossing}$};
\node[black] at (1.8, 0.2) [anchor = north] {$_\text{singularity}$};
\filldraw[color=black, fill=black](0,0) circle (0.03);
\draw [darkgray, thick, domain=0:5, samples=150] plot ({0.2*(\x)^(1.7)},{\x});
\draw [darkgray, thick, domain=0:5, samples=150] plot ({-0.2*(\x)^(1.7)},{\x});
\draw [fill=white,white,opacity=0.7] (-1.5,4.3) rectangle (1.5,4.6);
\node[darkgray] at (1.6, 4.45) [anchor = east] {$_\text{multi-dust region}$};
\end{tikzpicture}
\end{center}
\end{figure}
\end{minipage}\vspace*{4pt}\\
ngent to a caustic curve in the future. Hence, in this proposal, the interior of the multi-dust region would be described by a triple $(\bm{u}^\mu_i,\rho_i)$ ($i=1,2,3$), with each dust species possessing its own fluid velocity 4-vector field $\bm{u}^\mu_i$ and proper energy density $\rho_i$. Just as for a single dust, $\bm{u}^\mu_i$ is tangent to geodesic curves, and the particle current $\rho_i\bm{u}^\mu_i$ is covariantly conserved. Meanwhile, the energy-momentum tensor $T^{\mu\nu}=\sum_{i=1,2,3}\rho_i\bm{u}^\mu_i\bm{u}^\nu_i$ naturally receives contributions from all three dust species.\\ \\
As a consequence of the focusing of fluid trajectories needed to form caustic curves, and the conservation of particle current, the energy densities necessarily blowup as the caustic is approached. The correspondingly divergent contributions to the energy-momentum tensor entail, through the Einstein equations, that spacetime curvature is also unbounded near the caustic. \\ \\
Before going further, we define more precisely the notion of `caustic curve' and `dust caustic'. The \textit{caustic set} of a differentiable map is simply its critical value set, and a \textit{caustic curve} is a smooth curve arising as the caustic set of some differentiable map.\\ \\
\textbf{Example 1.1.}\textit{ The map $\mathbb{R}^2_{(\tau,\chi)}\to\mathbb{R}^2_{(t,x)}$ given by $$t(\tau,\chi)=\tau\qquad x(\tau,\chi)=(\tau-\chi)^2 $$
has caustic set $\{x=0\}$. This caustic curve is a prototype for the object obtained in \hyperlink{thm:1.1}{Theorem~1.1}.}\\ \\
\textbf{Example 1.2.}\textit{ The map $\mathbb{R}^2_{(\tau,\chi)}\to\mathbb{R}^2_{(t,x)}$ given by $$t(\tau,\chi)=\tau\qquad x(\tau,\chi)=\chi^3-3\chi\tau $$
has caustic set $\{x=\pm t^{3/2}\}$, a pair of intersecting caustic curves. This \ul{simple catastrophe fold} is the prototype for the situation depicted in the Figure above.}\\ \\
In the case of smooth maps $\mathbb{R}^2\to\mathbb{R}^2$, the caustic set is (in a suitable sense) generically a caustic \textit{curve}. In our case, we are more specifically interested in \textit{dust caustics}, which are caustic sets arising from parametrised families of timelike geodesics in a Lorentzian manifold $(M,g_{\mu\nu})$. Moreover, $(M,g_{\mu\nu})$ satisfies the equations of the Einstein-(multi-)dust system, with the timelike geodesics identified with dust trajectories in $(M,g_{\mu\nu})$. \\ \\
While the catastrophe fold continuation is an elegant proposal, a rigorous proof of existence remains outstanding, as emphasized in the heuristic study \hyperlink{COD92}{$[$COD92$]$} mentioned earlier. We may write down the following conjecture.\\ \\
\hypertarget{con:1.1}{\textbf{Conjecture 1.1.}} \textit{For \ul{generic}\footnote[4]{The degenerate caustics studied in \hyperlink{PH67}{$[$PH67$]$} are responsible for the word `generic'. These exceptional examples admit a well-defined continuation, but it does not involve a multi-dust region, only a single-dust region where the ordering of the fluid shells has reversed.} shell-crossing singularities in the spherically symmetric Einstein-dust system, there exists a unique local continuation as a catastrophe fold with the features described above.}
\\ \\ \\
\large\hypertarget{sec:1.3}{\textbf{1.3\hspace{4mm}Formulation of a restricted problem}}\label{1.3}\normalsize\\ \\
Three difficulties in particular may be identified as immediate challenges to the task of proving \hyperlink{con:1.1}{Conjecture 1.1}.\vspace*{3mm} \\
\textbf{Difficulties. }\textit{
\begin{enumerate}
\item[\textbf{1.}] Even after the first singularity, energy density and spacetime curvature remain (locally) unbounded. Since the geodesic spray is necessarily non-Lipschitz near the caustics, the very existence of `geodesic tangents' that generate each caustic is a priori unclear;
\item[\textbf{2.}] Dust caustics arise as free boundaries which must be solved for;
\item[\textbf{3.}] Analysis of the multi-dust region requires understanding of the interaction between dust species, the basic manifestation of which is the loss of a single comoving coordinate system in which the equations of motion are explicitly integrable.
\end{enumerate}}
\vspace{2mm}
The present article constitutes a \textit{preliminary} study of the catastrophe fold continuation, to precede a full resolution of the problem, and has three primary aims.\textit{
\begin{enumerate}
\item[\textbf{1.}] Understand the very tenability of dust caustics---that is, in view of problem 1 above, whether there is a non-empty set of examples of dust caustics;
\item[\textbf{2.}] Determine the salient geometrical and dynamical features of spacetime near dust caustics;
\item[\textbf{3.}] Develop techniques and estimates to be used in the full construction involving free boundaries.
\end{enumerate}}
\vspace{2mm}
To this end, we formulate and solve a \textit{restricted problem} which captures the fundamental element of  difficulty, namely the singular behaviour near the caustic curves. We eliminate the free-boundary aspect of the problem by taking a scattering approach: the location of the caustic curve is fixed in advance, and we solve for a spacetime slab representing a neighbourhood of the caustic. We also make the further simplification of removing the third dust species which passes transversally across the caustic\footnote[5]{This is an inessential reduction and primarily for reasons of brevity.}. Thus the caustic curve becomes also the boundary of a vacuum region, which via Birkhoff's theorem is isometric to a member of the Schwarzschild family.\\ \\
The task consists of solving a PDE problem involving wave-type equations for the metric components, posed with a spacelike direction of evolution. That this is possible relies heavily on the spherical symmetry of the spacetime manifold $(M,g_{\mu\nu})$ that we solve for. Indeed, the analysis takes place entirely on the $(1+1)$-dimensional quotient manifold $Q=M/SO(3)$, so that timelike data may be meaningfully posed. The wave-type Einstein equations are coupled to transport equations for the matter variables, which are degenerate in the sense that the transport operator is initially tangent to the surface on which data are prescribed.\\ \\
With the location of the caustic curve fixed in advance, it is natural to carry out the analysis in coordinates which are adapted to the caustic. A useful choice is one that \textit{straightens} the curve, making especially clear the features that arise on approach to the caustic. The Figure below shows which region of the catastrophe fold picture is coordinatized in this way. Note that the portion of the caustic curve under study is strictly to the future of the putative first singularity---behaviour near the intersection point of the two caustics is somewhat more singular: see remarks at the end of this Section. Our choice to study the curve on the left rather than on the right, however, is inessential. Since we only obtain a neighbourhood of the curve, the present theorem may be modified for that case with almost no changes. 
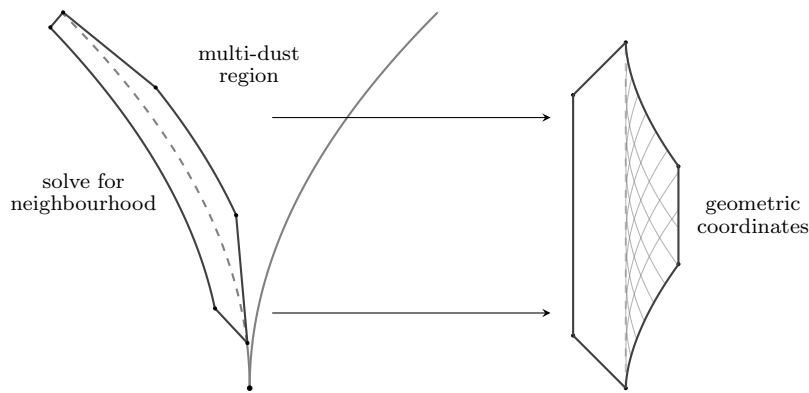
\vspace{3mm}
\begin{figure}[H]
\begin{center}
\begin{tikzpicture}
\draw [gray, thick, domain=0:5, samples=150] plot ({0.1*(\x)^(2)},{\x});
\draw [gray, thick, domain=0:0.6, samples=150] plot ({-0.1*(\x)^(2)},{\x});
\draw [gray, thick, dashed, domain=0.6:5, samples=150] plot ({-0.1*(\x)^(2)},{\x});
\draw [darkgray, thick, domain=1.06:4.8, samples=150] plot ({-0.1*(\x)^(2)-0.35},{\x});
\draw [darkgray, thick, domain=2.3:4, samples=150] plot ({-0.1*(\x)^(2)+0.35},{\x});
\draw [darkgray, thick] (-0.03, 0.6) -- (-0.46, 1.06);
\draw [darkgray, thick] (-2.48, 5) -- (-2.65, 4.8);
\draw [darkgray, thick] (-2.48, 5) -- (-1.25, 4);
\draw [darkgray, thick] (-0.03,0.6) -- (-0.18, 2.3);
\filldraw[color=black, fill=black](-0.03,0.6) circle (0.02);
\filldraw[color=black, fill=black](-0.18,2.3) circle (0.02);
\filldraw[color=black, fill=black](-1.25,4) circle (0.02);
\filldraw[color=black, fill=black](-2.48,5) circle (0.02);
\filldraw[color=black, fill=black](-2.65,4.8) circle (0.02);
\filldraw[color=black, fill=black](-0.46,1.06) circle (0.02);
\filldraw[color=black, fill=black](0,0) circle (0.03);
\draw [-stealth](0.3,1) -- (4,1);
\draw [-stealth](0.3,3.6) -- (4,3.6);
\node[black] at (-2.2, 3.0) [anchor = north] {$_\text{solve for}$};
\node[black] at (-2.2, 2.7) [anchor = north] {$_\text{neighbourhood}$};
\node[black] at (0, 4.7) [anchor = north] {$_\text{multi-dust}$};
\node[black] at (0,4.4) [anchor = north] {$_\text{region}$};
\node[black] at (6.7, 2.7) [anchor = north] {$_\text{geometric}$};
\node[black] at (6.7,2.4) [anchor = north] {$_\text{coordinates}$};
\filldraw[color=black, fill=black](5,0) circle (0.02);
\filldraw[color=black, fill=black](5,4.6) circle (0.02);
\filldraw[color=black, fill=black](4.3,0.7) circle (0.02);
\filldraw[color=black, fill=black](5.7,1.65) circle (0.02);
\filldraw[color=black, fill=black](5.7,2.95) circle (0.02);
\filldraw[color=black, fill=black](4.3,3.9) circle (0.02);
\draw [gray, domain=-0.2:1.87, samples=50, opacity=0.6] plot ({5+0.2*(\x)^2},{\x+0.5});
\draw [gray, domain=-0.6:1.87, samples=50, opacity=0.6] plot ({5+0.2*(\x)^2},{\x+1});
\draw [gray, domain=-0.85:1.66, samples=50, opacity=0.6] plot ({5+0.2*(\x)^2},{\x+1.5});
\draw [gray, domain=-1.37:1.18, samples=150, opacity=0.6] plot ({5+0.2*(\x)^2},{\x+2.5});
\draw [gray, domain=-1.61:0.9, samples=50, opacity=0.6] plot ({5+0.2*(\x)^2},{\x+3});
\draw [gray, domain=-1.86:0.65, samples=50, opacity=0.6] plot ({5+0.2*(\x)^2},{\x+3.5});
\draw [gray, domain=-1.86:0.4, samples=50, opacity=0.6] plot ({5+0.2*(\x)^2},{\x+4});
\draw [gray, domain=-1.1:1.42, samples=50, opacity=0.6] plot ({5+0.2*(\x)^2},{\x+2});
\draw [gray, thick, dashed, opacity=0.5] (5.0, 0) -- (5.0, 4.6);
\draw [darkgray, thick, domain=0:1.66, samples=150] plot ({0.3*(\x)^(1.7)+5},{\x});
\draw [darkgray, thick, domain=0:1.66, samples=150] plot ({0.3*(\x)^(1.7)+5},{4.6-\x});
\draw [darkgray, thick] (5.0, 0) -- (4.3, 0.7);
\draw [darkgray, thick] (5, 4.6) -- (4.3, 3.9);
\draw [darkgray, thick] (4.3,0.7) -- (4.3, 3.9);
\draw [darkgray, thick] (5.7, 1.65) -- (5.7, 2.95);
\end{tikzpicture}
\end{center}
\end{figure}
To fully determine the problem, only three independent pieces of data need to be prescribed.  One is the area radius $r_0$ as a function of proper time along the caustic. Together with the mass $m_0\in\mathbb{R}$ of the Schwarzschild geometry we intend to glue onto, this uniquely determines the caustic as a radial curve in Schwarzschild\footnote[6]{Of course, modulo translations along the flow generated by the usual Schwarzschild Killing vector field $T$.}. Finally, a third function $\sigma_0$ specifies precisely the blowup of energy density on approach to the caustic. This can be thought of as capturing the amount of dust becoming tangent to the caustic per unit proper time.
\vspace*{4mm} \\
\hypertarget{thm:1.1}{\textbf{Theorem 1.1. }}\textit{Given smooth positive functions $r_0,\sigma_0\in C^\infty[a,b]$ (subject to the conditions specified in \hyperlink{sec:2.3}{Section~2.3}), there exists a spherically symmetric spacetime $(M,g_{\mu\nu})$ with the following properties.
\begin{enumerate}
\item[1.] \begin{minipage}[t]{0.64\textwidth}The quotient manifold $Q$ is without boundary and covered by coordinates $(t,x)$ which take values in a connected open subset of $\mathbb{R}^2$, the interior of the region depicted in the opposite Figure.\vspace*{2mm} \\
$g_{\mu\nu}$ takes the form\vspace{-2mm}
$$g=\Omega^2(-dt^2+dx^2)+r^2g_{S^2} $$
where $\Omega^2,r$ are positive functions defined on $Q$ with $$\Omega^2|_{x=0}=1,\quad r|_{x=0}=r_0$$
In particular, the opposite Figure is a Penrose diagram. The constants $\varepsilon>0, \delta_1>0$ defining the coordinate chart depend only on norms on the functions $r_0,\sigma_0$.\end{minipage}\hspace{0.03\textwidth}
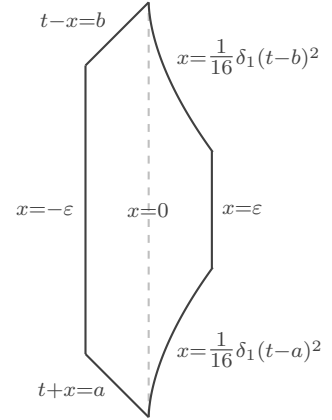
\begin{minipage}[t]{0.26\textwidth}
\vspace{-3mm}
\begin{figure}[H]
\begin{center}
\begin{tikzpicture}[scale=1.2]
\draw [gray, thick, dashed, opacity=0.5] (0.0, 0) -- (0.0, 4.6);
\draw [darkgray, thick, domain=0:1.66, samples=150] plot ({0.3*(\x)^(1.7)},{\x});
\draw [darkgray, thick, domain=0:1.66, samples=150] plot ({0.3*(\x)^(1.7)},{4.6-\x});
\draw [darkgray, thick] (0.0, 0) -- (-0.7, 0.7);
\draw [darkgray, thick] (0.0, 4.6) -- (-0.7, 3.9);
\draw [darkgray, thick] (-0.7,0.7) -- (-0.7, 3.9);
\draw [darkgray, thick] (0.7, 1.65) -- (0.7, 2.95);
\node[darkgray] at (0.7, 2.3) [anchor = west] {$_{x=\varepsilon}$};
\node[darkgray] at (-0.35, 2.3) [anchor = west] {$_{x=0}$};
\node[darkgray] at (-0.7, 2.3) [anchor = east] {$_{x=-\varepsilon}$};
\node[darkgray] at (-0.35, 0.3) [anchor = east] {$_{t+x=a}$};
\node[darkgray] at (-0.35, 4.4) [anchor = east] {$_{t-x=b}$};
\node[align=left, darkgray] at (0.2, 0.75) [anchor = west] {$_{x=\tfrac{1}{16}\delta_1(t-a)^2}$};
\node[align=left, darkgray] at (0.2, 4) [anchor = west] {$_{x=\tfrac{1}{16}\delta_1(t-b)^2}$};
\end{tikzpicture}
\end{center}
\end{figure}
\end{minipage}\vspace{1mm}
\item[2.] The subset $\{x>0\}$ is a smooth 2-dust spacetime: there exist smooth unit timelike vector fields $\bm{u}_\pm=\bm{u}^t_\pm\partial_t+\bm{u}^x_\pm\partial_x$ and smooth positive functions $\rho_\pm$, both spherically symmetric, for which the Einstein and matter equations are satisfied:\\
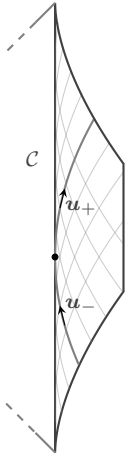
\begin{minipage}[t]{0.15\textwidth}\begin{figure}[H]
\begin{center}
\vspace{-4mm}
\begin{tikzpicture}[scale=1.3]
\draw [darkgray, thick] (0.0, 0) -- (0.0, 4.6);
\draw [darkgray, thick, domain=0:1.655, samples=150] plot ({0.3*(\x)^(1.7)},{\x});
\draw [gray, domain=-0.2:1.87, samples=150, opacity=0.4] plot ({0.2*(\x)^2},{\x+0.5});
\draw [gray, domain=-0.6:1.87, samples=150, opacity=0.4] plot ({0.2*(\x)^2},{\x+1});
\draw [gray, domain=-0.85:1.66, samples=150, opacity=0.4] plot ({0.2*(\x)^2},{\x+1.5});
\draw [gray, domain=-1.37:1.18, samples=150, opacity=0.4] plot ({0.2*(\x)^2},{\x+2.5});
\draw [gray, domain=-1.61:0.9, samples=150, opacity=0.4] plot ({0.2*(\x)^2},{\x+3});
\draw [gray, domain=-1.86:0.65, samples=150, opacity=0.4] plot ({0.2*(\x)^2},{\x+3.5});
\draw [gray, domain=-1.86:0.4, samples=150, opacity=0.4] plot ({0.2*(\x)^2},{\x+4});
\draw [gray, thick, domain=-1.1:1.42, samples=150] plot ({0.2*(\x)^2},{\x+2});
\draw [-stealth](0.05,2.5) -- (0.1,2.7);
\node[darkgray] at (0, 2.5) [anchor = west] {$_{\bm{u}_+}$};
\draw [-stealth] (0.1,1.3) -- (0.05,1.5);
\node[darkgray] at (0, 1.5) [anchor = west] {$_{\bm{u}_-}$};
\draw [darkgray, thick, domain=0:1.655, samples=150] plot ({0.3*(\x)^(1.7)},{4.6-\x});
\filldraw[color=black, fill=black](0,2) circle (0.03);
\draw [gray, thick] (0.0, 0) -- (-0.2, 0.2);
\draw [gray, thick, dashed] (-0.5, 0.5) -- (-0.2, 0.2);
\draw [gray, thick] (0.0, 4.6) -- (-0.2, 4.4);
\draw [gray, thick, dashed] (-0.2,4.4) -- (-0.5, 4.1);
\draw [darkgray, thick] (0.7, 1.65) -- (0.7, 2.95);
\node[darkgray] at (-0.4, 3) [anchor = west] {$_\mathcal{C}$};
\end{tikzpicture}
\end{center}
\end{figure}
\end{minipage}\hspace{0.03\textwidth}
\begin{minipage}[t]{0.75\textwidth}
$$\text{Ric}[g]^{\mu\nu}-\tfrac{1}{2}R[g]g^{\mu\nu}=2\rho_+\bm{u}^\mu_+\bm{u}^\nu_++2\rho_-\bm{u}^\mu_-\bm{u}^\nu_- $$
$$\bm{u}^\nu_\pm\nabla_\nu\bm{u}^\mu_\pm=0 \qquad\nabla_\mu\big(\rho_\pm \bm{u}^\mu_\pm\big)=0 $$
Moreover, $$\lim_{x\to0}\bm{u}^t_\pm=1,\quad\lim_{x\to0}\bm{u}^x_\pm=0,\quad\lim_{x\to0}(\rho_\pm x^{1/2})=\sigma_0$$ and $\bm{u}^x_+>0$, $\bm{u}^x_-<0$ hold on $\{x>0\}$. Each point of the timelike hypersurface $\mathcal{C}=\{x=0\}$ is the past endpoint of an integral curve of $\bm{u}_+$ and the future endpoint of an integral curve of $\bm{u}_-$, which glue together into a smooth tangent curve to $\mathcal{C}$.\end{minipage}\vspace{1mm}
\item[3.] The collection $(\Omega^2,r,\bm{u}_\pm,\rho_\pm x^{1/2})$ of variables on $\{x>0\}$ is smooth in coordinates $(t,\xi)$, where $\xi=x^{1/2}$, \ul{with extension to $\xi=0$}.\vspace{1mm}
\item[4.] The subset $\{x<0\}$ is isometric to a member of the Schwarzschild family. In particular, $\Omega^2,r$ are smooth on $\{x<0\}$.\vspace{1mm}
\item[5.] Globally, $\Omega^2,r$ have regularity $C^{1,1/2}$, but still satisfy Einstein's equation weakly.
\end{enumerate}}
\vspace{3mm}
\textbf{Remark 1.1.} \textit{The smooth tangent curves to $\mathcal{C}$ mentioned in 2$.$ are interpreted as the trajectories of fluid particles. Thus $\mathcal{C}$ is truly a caustic curve: an envelope of fluid trajectories.}
\\ \\
\textbf{Remark 1.2.} \textit{Having chosen the quantities $\varepsilon,\delta_1$, the spacetime $(M,g_{\mu\nu})$ with the above properties may be shown to be unique in a suitable sense. However, we do not claim that our domain is the largest one on which we have uniqueness. Indeed, $(M,g_{\mu\nu})$ may manifestly be extended near the quadratic parts of the boundary, which are not optimal, `cutting off' fluid trajectories near the endpoints of the caustic curve.}\\ \\
\hypertarget{rem:1.3}{\textbf{Remark 1.3.}} \textit{The satisfaction of Einstein's equation weakly (discussed in Appendix B) is sensitive to the choice of smooth structure on $M$, which we take to be the one induced by $(t,x)$ coordinates. If one instead uses area radius $r$ to provide a coordinate / smooth structure (as in numerous studies in spherical symmetry), committing in particular to the smoothness of $r$ on $M$, then the Christoffel symbols fail to be locally square integrable and Einstein's equation is not satisfied weakly.}\\ \\
\hypertarget{rem:1.4}{\textbf{Remark 1.4.}} \textit{The caustic curve $\mathcal{C}$ is not a} thin shell: \textit{Hawking mass $m$ is continuous across $\mathcal{C}$ and the weak solution of Einstein's equation does not involve $\delta$ functions sourced along $\mathcal{C}$.}
\\ \\
To elaborate on \hyperlink{rem:1.4}{Remark~1.4}, we mention that, in addition to the catastrophe fold continuation, another viable proposal exists (see $[$\hyperlink{Nol03}{Nol03}; \hyperlink{Teg11}{Teg11}$]$ and the upcoming \hyperlink{Smi25}{$[$Smi25$]$}). In the \textit{thin shell continuation}, fluid shells accumulate on a timelike curve on which positive mass is concentrated, with junction conditions \hyperlink{Isr65}{$[$Isr65$]$} determining the subsequent motion of the thin shell. Einstein's equation can once again be solved weakly, with the metric belonging to the lower regularity class $C^{0,1}$.\\ \\
Where does this place us on the path to settling \hyperlink{con:1.1}{Conjecture~1.1}? With \hyperlink{thm:1.1}{Theorem~1.1} in hand, a plausible next step would be to provide examples of prescribed caustic curves which extend all the way to the putative intersection point of the two caustic curves defining the catastrophe fold. To achieve this, new kinds of blowup need to be managed. From calculations in \hyperlink{COD92}{$[$COD92$]$} (see equation (5)), we expect, for example, the behaviour
$$|\sigma_0(t)|\gtrsim t^{-1/4}\qquad |r''_0(t)|\gtrsim t^{-1/2}$$
to hold, if $t=0$ represents the point of intersection. We speculate that a suitably $t$-weighted version of our norms $\|\cdot\|_{C^k_\text{cusp}}$ (see \hyperlink{sec:2.3}{Section~2.3}) may provide the correct control to close a local existence argument similar to the one used in \hyperlink{thm:1.1}{Theorem~1.1}.\\ \\ \\
\large\hypertarget{sec:1.4}{\textbf{1.4\hspace{4mm}A new family of static examples}}\label{1.4}\normalsize\\ \\
To supplement the general results of \hyperlink{thm:1.1}{Theorem~1.1}, we also present a novel family of `global' spacetimes which contain caustic curves, and which are particularly easy to construct and analyze. These spacetimes contain an eternal annular 2-dust region bounded by a pair of caustic curves. Outside the annulus, we again have a vacuum spacetime, locally isometric to a Schwarzschild solution. However, the interior vacuum region has a negative mass parameter. One may think of this as providing a `repulsive gravitational force', needed to maintain a static configuration of matter which has no charge or angular momentum. See already \hyperlink{rem:1.5}{Remark~1.5} below.
\newpage
\hypertarget{thm:1.2}{\textbf{Theorem 1.2.}}\textit{ There exists a 3-parameter family of static, spherically symmetric spacetimes $(M,g_{\mu\nu})$ with the following properties.
\begin{enumerate}
\item[1.] \begin{minipage}[t]{0.5\textwidth}$(M,g_{\mu\nu})$ consists of three regions $\{0\leq r<r_\text{in}\}$, $\{r_\text{in}\leq r\leq r_\text{out}\}$, $\{r_\text{out}<r\}$. The first and third are locally isometric to Schwarzschild spacetimes with masses $m_\text{in}<0$ and $m_\text{out}>0$ respectively. The second is an annular 2-dust region bounded by caustic curves at $\{r=r_\text{in}\}$ and $\{r=r_\text{out}\}$. \vspace*{2mm} \\
The dust trajectories oscillate between the caustic curves, changing from outgoing to ingoing and vice versa.\end{minipage}\hspace{0.00\textwidth}
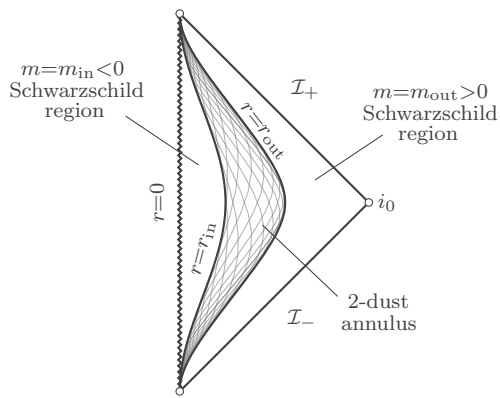
\begin{minipage}[t]{0.50\textwidth}
\vspace{-10mm}
\begin{figure}[H]
\begin{center}
\begin{tikzpicture}[scale=1]
\foreach \y in {0,24,...,360}
{
\draw [gray, domain=-4*pi:4*pi, samples=100, opacity=0.6] plot ({0.8*(rad(atan((\x)+0.8+0.4*cos(deg(\x)+\y)))-rad(atan((\x)-0.8-0.4*cos(deg(\x)+\y)))},({0.8*(rad(atan((\x)+0.8+0.4*cos(deg(\x)+\y)))+rad(atan((\x)-0.8-0.4*cos(deg(\x)+\y)))});
}
\draw [darkgray, thick, domain=-8*pi:8*pi, samples=500] plot ({0.8*(rad(atan((\x)*abs(\x)+0.4))-rad(atan((\x)*abs(\x)-0.4))},({0.8*(rad(atan((\x)*abs(\x)+0.4))+rad(atan((\x)*abs(\x)-0.4))});
\draw [darkgray, thick, domain=-8*pi:8*pi, samples=500] plot ({0.8*(rad(atan((\x)*abs(\x)+1.2))-rad(atan((\x)*abs(\x)-1.2))},({0.8*(rad(atan((\x)*abs(\x)+1.2))+rad(atan((\x)*abs(\x)-1.2))});
\draw [darkgray, thick] (0, -0.8*pi) -- (0.8*pi,0);
\draw [darkgray, thick] (0.8*pi,0) -- (0,0.8*pi);
\node[darkgray] at (1.35, 1.5) [anchor = west] {$_{\mathcal{I}_+}$};
\node[darkgray] at (1.3, -1.55) [anchor = west] {$_{\mathcal{I}_-}$};
\node[darkgray] at (2.5, 0) [anchor = west] {$_{i_0}$};
\node[darkgray,rotate=90] at (-0.3, -0.4) [anchor = west] {$_{r=0}$};
\node[darkgray,rotate=-52] at (0.75, 1.4) [anchor = west] {$_{r=r_\text{out}}$};
\node[darkgray,rotate=76] at (0.2, -1.1) [anchor = west] {$_{r=r_\text{in}}$};
\node[darkgray] at (2.43, 1.47) [anchor = west] {$_{m=m_\text{out}>0}$};
\node[darkgray] at (2.3, 1.2) [anchor = west] {$_\text{Schwarzschild}$};
\node[darkgray] at (2.77, 0.9) [anchor = west] {$_\text{region}$};
\draw [darkgray] (2.7, 0.9) -- (1.8,0.3);
\node[darkgray] at (2.45-4.7, 1.47+0.3) [anchor = west] {$_{m=m_\text{in}<0}$};
\node[darkgray] at (2.3-4.7, 1.2+0.3) [anchor = west] {$_\text{Schwarzschild}$};
\node[darkgray] at (2.77-4.7, 0.9+0.3) [anchor = west] {$_\text{region}$};
\draw [darkgray] (-0.72, 1.15) -- (0.25,0.5);
\node[darkgray] at (2.1, -1.3) [anchor = west] {$_\text{2-dust}$};
\node[darkgray] at (2.01, -1.6) [anchor = west] {$_\text{annulus}$};
\draw [darkgray] (2.1, -1.25) -- (1.1,-0.3);
\draw[darkgray, decorate,decoration={zigzag,segment length=0.8mm, amplitude=.2mm},thick] (0, -0.8*pi) -- (0, 0.8*pi);
\filldraw[color=darkgray, fill=white](0,0.8*pi) circle (0.05);
\filldraw[color=darkgray, fill=white](0,-0.8*pi) circle (0.05);
\filldraw[color=darkgray, fill=white](0.8*pi,0) circle (0.05);
\end{tikzpicture}
\end{center}
\end{figure}\end{minipage}
\item[2.] The local properties near the caustic curves agree with those concluded in \hyperlink{thm:1.1}{Theorem~1.1}.
\item[3.] $(M,g_{\mu\nu})$ is free of trapped and anti-trapped surfaces, and fails to be globally hyperbolic.
\end{enumerate}
Finally, the 3 parameters are $(r_\text{in},m_\text{in},\varsigma)$, belonging to the range $r_\text{in},\varsigma>0$, $m_\text{in}<0$, where $\varsigma$ is a constant capturing the density of the fluid annulus.}\\ \\
\begin{minipage}[t]{0.49\textwidth}
\hypertarget{rem:1.5}{\textbf{Remark 1.5.}} \textit{It is necessary to have $m_\text{in}<0$ in the above examples. However, one may construct globally hyperbolic examples, either by using a positive cosmological constant $\Lambda>0$ or by making the dust carry charge. In the latter, the charged dust annulus would surround a Reissner-Nordstrom black hole, which can be subextremal, provided the dust particles have sufficiently large fundamental charge $\mathfrak{e}$. The region exterior to the fluid annulus has strictly greater Reissner-Nordstr\"om parameters. See the opposite Penrose diagram for comparison with \hyperlink{thm:1.2}{Theorem~1.2}. These examples are omitted from the present article for brevity.}\end{minipage}
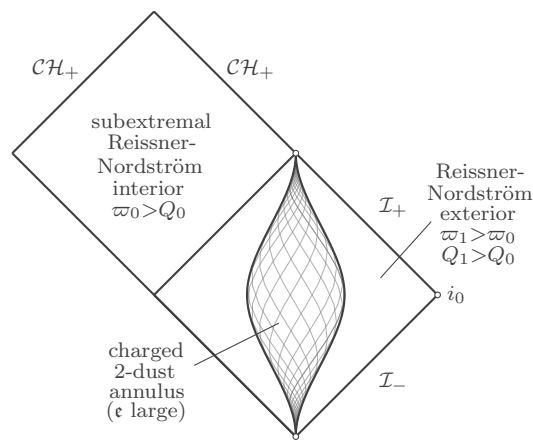
\begin{minipage}[t]{0.50\textwidth}
\vspace{-8mm}
\begin{figure}[H]
\begin{center}
\begin{tikzpicture}[scale=0.75]
\foreach \y in {0,24,...,360} {
\draw [gray, domain=-4*pi:4*pi, samples=100, opacity=0.6] plot ({0.8*(rad(atan((\x)+0.6*cos(deg(\x)+\y)))-rad(atan((\x)-0.6*cos(deg(\x)+\y)))},({0.8*(rad(atan((\x)+0.6*cos(deg(\x)+\y)))+rad(atan((\x)-0.6*cos(deg(\x)+\y)))});
}
\draw [darkgray, thick, domain=-8*pi:8*pi, samples=500] plot ({0.8*(rad(atan((\x)*abs(\x)+0.6))-rad(atan((\x)*abs(\x)-0.6))},({0.8*(rad(atan((\x)*abs(\x)+0.6))+rad(atan((\x)*abs(\x)-0.6))});
\draw [darkgray, thick, domain=-8*pi:8*pi, samples=500] plot ({0.8*(rad(atan((\x)*abs(\x)-0.6))-rad(atan((\x)*abs(\x)+0.6))},({0.8*(rad(atan((\x)*abs(\x)-0.6))+rad(atan((\x)*abs(\x)+0.6))});
\draw [darkgray, thick] (0, -0.8*pi) -- (0.8*pi,0);
\draw [darkgray, thick] (0.8*pi,0) -- (0,0.8*pi);
\draw [darkgray, thick] (-0.8*pi,0) -- (0,0.8*pi);
\draw [darkgray, thick] (-0.8*pi,0) -- (0,-0.8*pi);
\draw [darkgray, thick] (-0.8*pi,0) -- (-1.6*pi,0.8*pi);
\draw [darkgray, thick] (-0.8*pi,1.6*pi) -- (-1.6*pi,0.8*pi);
\draw [darkgray, thick] (-0.8*pi,1.6*pi) -- (0,0.8*pi);
\draw [darkgray, thick] (-0.8*pi,0) -- (0,-0.8*pi);
\node[darkgray] at (1.3, 1.55) [anchor = west] {$_{\mathcal{I}_+}$};
\node[darkgray] at (1.3, -1.55) [anchor = west] {$_{\mathcal{I}_-}$};
\node[darkgray] at (2.5, 0) [anchor = west] {$_{i_0}$};
\node[darkgray] at (-1.4, 4) [anchor = west] {$_\mathcal{CH_+}$};
\node[darkgray] at (-4.85, 4) [anchor = west] {$_\mathcal{CH_+}$};
\node[darkgray] at (2.3, 2.2) [anchor = west] {$_\text{Reissner-}$};
\node[darkgray] at (2.15, 1.82) [anchor = west] {$_\text{Nordstr\"{o}m}$};
\node[darkgray] at (2.38, 1.45) [anchor = west] {$_\text{exterior}$};
\node[darkgray] at (2.37, 1.05) [anchor = west] {$_{\varpi_1>\varpi_0}$};
\node[darkgray] at (2.4, 0.7) [anchor = west] {$_{Q_1>Q_0}$};
\draw [darkgray] (2.4, 1.3) -- (1.5,0.3);
\node[darkgray] at (-3.8, 3.1) [anchor = west] {$_\text{subextremal}$};
\node[darkgray] at (-3.52, 2.72) [anchor = west] {$_\text{Reissner-}$};
\node[darkgray] at (-3.7, 2.30) [anchor = west] {$_\text{Nordstr\"{o}m}$};
\node[darkgray] at (-3.43, 1.9) [anchor = west] {$_\text{interior}$};
\node[darkgray] at (-3.45, 1.5) [anchor = west] {$_{\varpi_0>Q_0}$};
\node[darkgray] at (-3.5, -1.0) [anchor = west] {$_\text{charged}$};
\node[darkgray] at (-3.4, -1.35) [anchor = west] {$_\text{2-dust}$};
\node[darkgray] at (-3.49, -1.7) [anchor = west] {$_\text{annulus}$};
\node[darkgray] at (-3.55, -2.08) [anchor = west] {$_{(\mathfrak{e}\text{ large})}$};
\draw [darkgray] (-1.8, -1.25) -- (-0.3,-0.5);
\filldraw[color=darkgray, fill=white](0,0.8*pi) circle (0.05);
\filldraw[color=darkgray, fill=white](0,-0.8*pi) circle (0.05);
\filldraw[color=darkgray, fill=white](0.8*pi,0) circle (0.05);
\end{tikzpicture}
\end{center}
\end{figure}\end{minipage}
\vspace{3mm} \\
\textbf{Remark 1.6.} \textit{While the proof uses a radial parametrization of the governing ODE system (i$.$e$.$ choosing area radius $r$ as the independent variable), the examples agree with \hyperlink{rem:1.3}{Remark~1.3} in that Einstein's equation is not satisfied weakly if a smooth structure is chosen with respect to which $r$ is smooth.} \\ \\
We give finally a structural overview of the paper and the techniques used. The majority of the article addresses \hyperlink{thm:1.1}{Theorem~1.1}. As in many proofs of local-wellposedness, an iterative scheme is designed which converges to a solution of the coupled system of equations. A key ingredient is the used of weighted norms to handle the expected blowup of derivatives near the caustic. Here, it is essential that the blowup rate $\rho\propto x^{-1/2}$ is \textit{integrable} in $x$, making the whole problem tractable. In Section~2, the preliminaries of the problem are reviewed, covering gauge choices, prescription of data, function spaces and the weighted norms used in the argument. Bounds are then obtained in these norms for the Einstein and matter equations, in Sections 3 and 4 respectively, which lead to a well-defined iteration map. In Section~5, contraction estimates are derived for the iteration map, and in Section~6, we prove \hyperlink{thm:1.1}{Theorem~1.1} by tying together the details and gluing to the Schwarzschild region. Section~7 is a self-contained, straightforward, chapter which establishes \hyperlink{thm:1.2}{Theorem~1.2}. After reducing the dynamics to a first-order ODE system in two variables, elementary theory is used to demonstrate the existence of a 3-parameter family of solutions with the desired properties. \\ \\
\textbf{Acknowledgements. }The author would like to express his gratitude to his advisor Mihalis Dafermos for his support, encouragement and guiding insight; and to Mahir Had\^{z}i\'{c}, Claude Warnick, Matt Smith and Lili Feh\'{e}rt\'{o}i-Nagy for several clarifying discussions. This work was funded by EPSRC. The author also thanks Princeton University for hosting him as a VSRC in April 2024.\\ \\ \\
\Large\hypertarget{sec:2}{\textbf{2\hspace{4mm}Preliminaries}}\label{2}\normalsize\\ \\
\large\hypertarget{sec:2.1}{\textbf{2.1\hspace{4mm}Notation and gauge choices}}\label{2.1}\normalsize\\ \\
We choose the metric signature $(-++\hspace{2pt}+)$, and choose units such that the speed of light is $c=1$ and Newton's constant is $G=1/4\pi$. Thus the Einstein equation reads
\begin{equation}
\text{Ric}[g]_{\mu\nu}-\tfrac{1}{2}R[g]g_{\mu\nu}=2 T_{\mu\nu} \tag{\hypertarget{eqn:2.1}{2.1}}
\end{equation}
We use a boldface $\bm{u}$ for the fluid velocity 4-vector, and its components, to distinguish it from the null coordinate $u$. The notation $\bm{u}(\cdot)$ denotes the directional derivative $\bm{u}^\mu\nabla_\mu(\cdot)$.\\ \\
Two species of dust---one ingoing (tangent to the caustic in the future) and one outgoing (tangent to the caustic in the past)---are notated respectively with a $+$ and $-$ sign, along with their corresponding energy densities $\rho_\pm$.  (Note that the terms ingoing/outgoing do not refer to the sign of $\bm{u}(r)$.) A subscript $\pm$ is used to compactly refer to both/either. The energy momentum tensor associated to these two species takes the following form, written with indices raised. 
$$T^{\mu\nu}=\rho_+\bm{u}^\mu_+\bm{u}^\nu_++\rho_-\bm{u}^\mu_-\bm{u}^\nu_- $$
We now state with more precision the notion of spherical symmetry that we will need. Since we are solely concerned with behaviour away from the centre of symmetry, we are able to omit discussion of the centre entirely.\\ \\
\hypertarget{def:2.1}{\textbf{Definition 2.1}} (Spherical symmetry)\textbf{.} \textit{We say that a connected, time-oriented (3+1)-dimensional Lorentzian manifold $(M,g_{\mu\nu})$ is \ul{spherically symmetric} if it admits an $SO(3)$-action by isometries, the orbits of which are spacelike 2-spheres, and moreover if $M$ splits diffeomorphically as $M=Q\times S^2$, where $(Q,g_Q)$ is a connected, time-oriented (1+1)-dimensional Lorentzian manifold, with the metric taking the form 
$$g=g_Q+r^2g_{S^2} $$
Here, $r:Q\to\mathbb{R}_{>0}$ is the \ul{area radius} of the 2-spheres.}\\ \\
We are interested in spherically symmetric spacetimes for which the quotient manifold $Q$ admits a global coordinate system $(t,x)$ with respect to which $g_Q$ takes the form
$$g_Q=\Omega^2(-dt^2+dx^2) $$
This can always be achieved locally on any (1+1)-dimensional Lorentzian manifold $Q$. Since we are performing a local construction, this assumption is without loss of generality. Of course, this is nothing but a 45$^{\circ}$ degree rotation of a \textit{double null} coordinate system $(u,v)$, with the two related by 
$$ u=t-x,\qquad v=t+x$$
We then have 
$$\partial_u\equiv\tfrac{1}{2}(\partial_t-\partial_x),\qquad \partial_v=\tfrac{1}{2}(\partial_t+\partial_x) $$
and we use the $\partial_t,\partial_x,\partial_u,\partial_v$ variously throughout the article with these relations understood. Though we will occasionally carry out calculations in the $(u,v)$ coordinates (especially when we study the Einstein equations), we primarily work with $(t,x)$. The reason is that we are particularly interested in behaviour near $x=0$, where data are posed. For example, the energy-momentum tensor component $T_{tt}$ diverges as $x\to0$ at the characteristic rate $x^{-1/2}$, but it is a crucial fact (see \hyperlink{sec:2.2}{Section~2.2}) that the components $T_{xx},T_{tx}$ are bounded, and in fact vanish at $x=0$. Similarly, in view of the expected blowup of $\rho_\pm$, conveyed in \hyperlink{thm:1.1}{Theorem~1.1} by the condition that $\lim_{x\to0}(\rho_\pm x^{1/2})=\sigma_0$, we consider the renormalised variable
$$\sigma_\pm:=\rho_\pm x^{1/2} $$
The gauge freedom in this coordinate choice is completely used up by asserting that the caustic curve $\mathcal{C}$ occurs at $x=0$, and that $\Omega^2|_{x=0}=1$, making $t$ a proper time parameter along $\mathcal{C}$.\\ \\
Recall the \textit{Hawking mass} function $m:Q\to\mathbb{R}$ which is defined by $$1-\frac{2m}{r}=g(\nabla r,\nabla r)\vspace{-2mm} $$
In the $(t,x)$ coordinates, we have
\begin{equation}
m=\frac{r}{2}\big(1+\Omega^{-2}((\partial_tr)^2-(\partial_xr)^2)\big)\tag{\hypertarget{eqn:2.2}{2.2}}
\end{equation}
For convenience, we also fix notation for two Christoffel symbols 
$$\Gamma_x:=\tfrac{1}{2}\partial_x\log\Omega^2\qquad \Gamma_t:=\tfrac{1}{2}\partial_t\log\Omega^2 $$
We lastly present the full coupled system of Einstein and matter equations, written in terms of $(t,x)$ coordinates and matter variables $(\bm{u}^t_\pm,\bm{u}^x_\pm,\sigma_\pm)$. We omit angular components $T_{AB}$ of the energy-momentum tensor in all expressions, since they vanish identically in our model. Taking respectively the `$tx$' and `$xx$' components of equation \hyperlink{eqn:2.1}{(2.1)} yields
\begin{align*}
\partial_t\partial_x r-\Gamma_t\partial_xr-\Gamma_x\partial_tr&=-rT_{tx}\equiv -r\Omega^4x^{-1/2}\big(\sigma_+\bm{u}^t_+\bm{u}^x_++\sigma_-\bm{u}^t_-\bm{u}^x_-\big)\tag{\hypertarget{eqn:2.3}{2.3}}\\
\partial^2_tr-\Gamma_t\partial_tr-\Gamma_x\partial_xr+\frac{\Omega^2m}{r^2}&=-rT_{xx}\equiv r\Omega^4x^{-1/2}\big(\sigma_+(\bm{u}^x_+)^2+\sigma_-(\bm{u}^x_-)^2\big)\tag{\hypertarget{eqn:2.4}{2.4}}
\end{align*}
which are constraint equations associated to the timelike $\{x=\text{const.}\}$ hypersurfaces. These involve only the bounded components $T_{tx},T_{xx}$, and in particular may be evaluated on $x=0$ (see \hyperlink{sec:2.2}{Section~2.2}). We then have semilinear wave equations for the geometry
\begin{align*}
(-\partial^2_x+\partial^2_t)\log\Omega^2-\frac{4\Omega^2m}{r^3}&=-2(T_{tt}-T_{xx})\equiv -4\Omega^2 x^{-1/2}\big(\sigma_++\sigma_-\big)\tag{\hypertarget{eqn:2.5}{2.5}}\\
(-\partial^2_x+\partial^2_t)r+\frac{2\Omega^2m}{r^2}&=r(T_{tt}-T_{xx})\equiv 2r\Omega^2 x^{-1/2}\big(\sigma_++\sigma_-\big)\tag{\hypertarget{eqn:2.6}{2.6}}
\end{align*}
and transport equations for the matter variables
\begin{equation*}
(\bm{u}^t_\pm\partial_t+\bm{u}^x_\pm\partial_x)\bm{u}^t_\pm+\Gamma_t((\bm{u}^t_\pm)^2+(\bm{u}^x_\pm)^2)+2\Gamma_x\bm{u}^t_\pm\bm{u}^x_\pm =0\tag{\hypertarget{eqn:2.7}{2.7}}
\end{equation*}
\begin{equation*}
(\bm{u}^t_\pm\partial_t+\bm{u}^x_\pm\partial_x)\bm{u}^x_\pm+\Gamma_x((\bm{u}^t_\pm)^2+(\bm{u}^x_\pm)^2)+2\Gamma_t\bm{u}^t_\pm\bm{u}^x_\pm =0\tag{\hypertarget{eqn:2.8}{2.8}}
\end{equation*}
\begin{equation*}
(\bm{u}^t_\pm\partial_t+\bm{u}^x_\pm\partial_x)\sigma_\pm+\sigma_\pm\bigg(\partial_t\bm{u}^t_\pm+\partial_x\bm{u}^x_\pm-\frac{\bm{u}^x_\pm}{2x}+2\bm{u}^t_\pm\Gamma_t+2\bm{u}^x_\pm\Gamma_x+\frac{2}{r}\big(\bm{u}^t_\pm\partial_tr+\bm{u}^x_\pm\partial_xr\big)\bigg)=0\tag{\hypertarget{eqn:2.9}{2.9}} 
\end{equation*}
The constraints \hyperlink{eqn:2.3}{(2.3)}-\hyperlink{eqn:2.4}{(2.4)} determine data for $(\Omega^2,r)$ at $x=0$, while $(\bm{u}^t_\pm,\bm{u}^x_\pm,\sigma_\pm)$ are to obey
$$\bm{u}^t_\pm|_{x=0}=1\qquad \bm{u}^x_\pm|_{x=0}=0\qquad \sigma_\pm|_{x=0}=\sigma_0$$
$$\qquad\bm{u}^x_-<0,\quad \bm{u}^x_+>0 \qquad \text{on }x>0$$
It is the wave equations \hyperlink{eqn:2.5}{(2.5)}-\hyperlink{eqn:2.6}{(2.6)} that we \textit{solve} (coupled to the matter equations \hyperlink{eqn:2.7}{(2.7)}-\hyperlink{eqn:2.9}{(2.9)}) and we achieve propagation of the constraints \hyperlink{eqn:2.3}{(2.3)}-\hyperlink{eqn:2.4}{(2.4)} once a solution to the coupled system has been obtained. We do not separately need to impose that $\bm{u}_\pm$ are unit vectors, since this necessarily holds at $x=0$, and is then preserved by the geodesic equations.
\vspace*{2mm} \\ \\
\large\hypertarget{sec:2.2}{\textbf{2.2\hspace{4mm}Constraint equations}}\label{2.2}\normalsize\\ \\
In order to ensure that equations \hyperlink{eqn:2.3}{(2.3)}-\hyperlink{eqn:2.4}{(2.4)} hold already at $x=0$, we must understand the behaviour of the components $T_{tx},T_{xx}$ as $x\to0$, which appear on the right-hand side. In fact, they vanish at $x=0$, which follows from a cancellation between the energy-momentum contributions of the two dusts, and which we now demonstrate. This fact is essential for consistency: if we took limits as $x\to 0$ from below (i$.$e$.$ from the vacuum side), then \hyperlink{eqn:2.3}{(2.3)}-\hyperlink{eqn:2.4}{(2.4)} would also have a vanishing right-hand side. 
\\ \\
The components take the form
\begin{align*}
T_{tx}&=-\Omega^4x^{-1/2}\big(\sigma_+\bm{u}^x_+\bm{u}^t_++\sigma_-\bm{u}^x_-\bm{u}^t_-\big)\\
T_{xx}&=\Omega^4x^{-1/2}\big(\sigma_+(\bm{u}^x_+)^2+\sigma_-(\bm{u}^x_-)^2\big)
\end{align*}
As well as the boundary values for $\Omega^2$, $\bm{u}^t_\pm$ and $\sigma_\pm$ listed above, we also have the more precise limit (which will be clarified in \hyperlink{sec:4.5}{Section~4.5}---see `\textsc{Base Case}')
$$\lim_{x\to0}\bm{u}^x_\pm x^{-1/2}=\pm2\big(-(\Gamma_x)_0\big)^{1/2} $$
Given these facts, it is now easily seen that $\lim_{x\to0}T_{xx}=\lim_{x\to0}T_{tx}=0 $. For the first one, we use that $\bm{u}^x_\pm=O(x^{1/2})$, while $\sigma_\pm,\Omega^2$ are bounded. For the second one, the two summands converge to the same finite limit, but with the opposite sign, so the sum vanishes in the limit.\\ \\
Evaluating equations \hyperlink{eqn:2.3}{(2.3)}-\hyperlink{eqn:2.4}{(2.4)} now at $x=0$, we have seen that the $T$ terms disappear. Since, in our gauge choice, $\Omega^2|_{x=0}=1$, the terms in $\Gamma_t=\tfrac{1}{2}\partial_t\log\Omega^2$ also vanish, leaving us with
\begin{align*}
\partial_t\partial_xr=\Gamma_x\partial_tr\tag{\hypertarget{eqn:2.10}{2.10}}\\
\partial^2_tr=\Gamma_x\partial_xr-\frac{m_0}{r^2}\tag{\hypertarget{eqn:2.11}{2.11}}
\end{align*}
which will now be used to obtain initial data for the wave equations.\\ \\ \\
\large\hypertarget{sec:2.3}{\textbf{2.3\hspace{4mm}Specification of data and the domain of evolution}}\label{2.3}\normalsize\\ \\
Let $[a,b]\subset\mathbb{R}$ be a closed, bounded interval. Let $m_0\in\mathbb{R}$ be the mass of the Schwarzschild geometry we intend to glue onto. We supply smooth positive functions $r_0,\sigma_0\in C^\infty[a,b]$, which we can choose arbitrarily except for two requirements:
\begin{equation*}
r''_0(t)<-\frac{m_0}{r_0(t)^2}\tag{\hypertarget{eqn:2.12}{2.12}}
\end{equation*}
\begin{equation*}
1-\frac{2m_0}{r_0(t)}+r'_0(t)^2>0\tag{\hypertarget{eqn:2.13}{2.13}}
\end{equation*}
The first condition \hyperlink{eqn:2.12}{(2.12)} conveys the \textit{convexity} necessary to be a caustic curve, and should be compared with the behaviour of radial timelike geodesics in Schwarzschild, which obey the ODE
$$\frac{d^2r}{d\tau^2}=-\frac{m_0}{r(\tau)^2}$$
where $\tau$ is proper time. (Note that, by our gauge choice $\Omega^2|_{x=0}=1$, the differentiation in \hyperlink{eqn:2.12}{(2.12)} is also with respect to proper time.) The second condition \hyperlink{eqn:2.13}{(2.13)} is satisfied by all functions $r_0(t)$ arising from unit speed timelike curves in Schwarzschild, even in the interior region (if $m_0>0$), as a calculation in Bondi coordinates $(v,r)$ easily verifies.\\ \\
We define $(\partial_xr)_0$ through \hyperlink{eqn:2.2}{(2.2)} as the positive square root of the expression in \hyperlink{eqn:2.13}{(2.13)}
$$(\partial_xr)_0(t):=\sqrt{1-\frac{2m_0}{r_0(t)}+r'_0(t)^2} $$
and $(\Gamma_x)_0\in C^\infty[a,b]$ through the constraint equation \hyperlink{eqn:2.11}{(2.11)}
$$r''_0(t)=(\Gamma_x)_0(t)(\partial_xr)_0(t)-\frac{m_0}{r_0(t)^2} $$
which, as a consequence of \hyperlink{eqn:2.12}{(2.12)}, is a strictly negative, smooth function on $[a,b]$. We define the positive constants 
$$\delta_1:=\inf_{[a,b]}|(\Gamma_x)_0|>0\qquad \delta_2=\inf_{[a,b]} r_0>0$$
The collection $(r_0,\sigma_0,(\partial_xr)_0,(\Gamma_x)_0)$ is then an \textit{admissible set of data} for our problem, and we are ready to define the domains on which our analysis will be undertaken.\\
\begin{minipage}{0.8\textwidth}
\hypertarget{def:2.2}{\textbf{Definition 2.2. }}(Domains of evolution) \textit{For each positive $\varepsilon>0$, denote by $\Sigma_\varepsilon$ the open subset of the $(t,x)$-plane bounded by the lines $x=0$, $x=\varepsilon$ and the quadratic curves}$$x=\tfrac{1}{16}\delta_1(t-a)^2\qquad x=\tfrac{1}{16}\delta_1(t-b)^2 $$
\textit{(The $x=\varepsilon$ portion of the boundary may be empty.)}\\ \\
Each $\Sigma_\varepsilon$ of course depends on $a,b$ and $\delta_1$ as well. Whereas we regard $[a,b]$ and $\delta_1$ as fixed throughout the analysis, we will consider different values of $\varepsilon$, hence we include only $\varepsilon$ in the notation. We also restrict to values of $\varepsilon$ small enough so that the quadratic curves remain timelike, and we always take $\varepsilon\leq1$, which is convenient for several estimates.
\end{minipage}
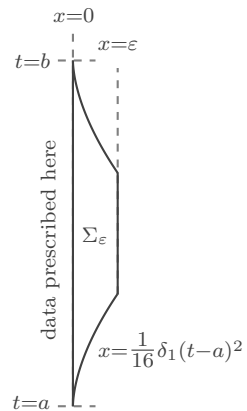
\begin{minipage}{0.2\textwidth}
\vspace{-2mm}
\begin{figure}[H]
\begin{center}
\begin{tikzpicture}{scale=0.9}
\draw [darkgray, thick, domain=0:1.5, samples=150] plot ({0.3*(\x)^(1.7)},{\x});
\draw [darkgray, thick, domain=0:1.5, samples=150] plot ({0.3*(\x)^(1.7)},{4.6-\x});
\draw [gray, thick, dashed] (0, 0) -- (0, 5);
\draw [gray, thick, dashed] (-0.2, 4.6) -- (0.3, 4.6);
\draw [gray, thick, dashed] (-0.2, 0.0) -- (0.3, 0.0);
\draw [gray, thick, dashed] (0.595, 1.49) -- (0.595, 4.5);
\draw [darkgray, thick] (0.0, 0) -- (0.0, 4.6);
\draw [darkgray, thick] (0.595, 1.49) -- (0.595, 3.11);
\node[darkgray] at (-0.4, 5.2) [anchor = west] {$_{x=0}$};
\node[darkgray] at (0.195, 4.8) [anchor = west] {$_{x=\varepsilon}$};
\node[darkgray] at (-0.15, 0.05) [anchor = east] {$_{t=a}$};
\node[darkgray] at (-0.15, 4.6) [anchor = east] {$_{t=b}$};
\node[darkgray, opacity=0] at (-0.8, 4.6) [anchor = east] {$_{t=b}$};
\node[darkgray, rotate=90] at (-0.3, 3.7) [anchor = east] {$_\text{data prescribed here}$};
\node[darkgray, align=left] at (0.63, 2.27) [anchor = east] {$_{\Sigma_\varepsilon}$};
\node[align=left, darkgray] at (0.2, 0.75) [anchor = west] {$_{x=\tfrac{1}{16}\delta_1(t-a)^2}$};
\end{tikzpicture}
\end{center}
\end{figure}
\end{minipage}
\vspace{0.2cm} \\
\hypertarget{def:2.3}{\textbf{Definition 2.3. }}(Cusp-regularity) \textit{Let $U$ be an open subset of the right half-plane $\{(t,x):x> 0\}$. An alternative coordinate chart for $U$ is $(t,\xi)$ where $\xi=x^{1/2}$, which is smoothly equivalent to $(t,x)$. Let $f:U\to\mathbb{R}$ be a smooth function. Coordinate derivatives of $f$ in this alternative chart, say $\partial^i_\xi\partial^j_tf$, are also smooth functions on $U$. We say that $f$ is \ul{cusp-regular} if such derivatives, to arbitrary order, admit limits on the boundary---that is, each $\partial^i_\xi\partial^j_tf$ extends to a continuous function on $$U\cup\big(\partial U\cap\{\xi=0\}\big) $$}
\hspace{-1.5mm}\textit{We denote by $C^\infty_\text{cusp}(U)$ the vector space of cusp-regular functions on $U$. Similarly, for each $k\in\mathbb{N}$, denote by $C^k_\text{cusp}(U)$ the vector space of functions $f\in C^k(U)$ whose $(t,\xi)$ coordinate derivatives (up to order $k$) extend continuously to $\{\xi=0\}$. We associate to each $C^k_\text{cusp}(U)$ the (complete) norm}
$$\|f\|_{C^k_\text{cusp}}:=\sum_{i+j\leq k}\sup_U|\partial^i_\xi\partial^j_tf| \vspace{0mm}$$
For example, $\xi=x^{1/2}$ itself, defined on any such $U$, is cusp-regular. Alternatively, any power series in powers of $t,\xi$ is cusp-regular---but there is of course no requirement of analyticity in the definition above. Functions which smoothly extend (in the usual sense) to $\{x=0\}$ are cusp-regular, but not vice versa, as $\xi$ itself shows. In view of extendibility, we freely refer to values of cusp-regular functions, and their derivatives, on the boundary $\partial U\cap\{\xi=0\}$ (if nonempty), for example when discussing boundary data.\\ \\
This is nothing but the usual notion of smoothness-with-extension, understood with respect to another chart. All of the functions describing spacetime near a caustic curve have the sort of regularity captured by $C^\infty_\text{cusp}$, and our estimates will all be in terms of the above norms. On the other hand, it has not proved useful to dispense with $x$ entirely, and recoordinatize $\xi=x^{1/2}$ once and for all. If the equations \hyperlink{eqn:2.5}{(2.5)}-\hyperlink{eqn:2.9}{(2.9)} are written entirely in coordinates $(t,\xi)$, then one obtains a \textit{Fuchsian} system. Fuchsian equations are characterized by having coefficients that diverge at an inverse polynomial rate as one approaches the surface on which data are posed: in the present case, this would be inverse powers of $\xi$. There is a well-developed existence theory for Fuchsian systems $[$\hyperlink{BG73}{BG73}; \hyperlink{Man00}{Man00}$]$ and a history of applying these results to problems in general relativity $[$\hyperlink{Mon81}{Mon81}; \hyperlink{Ren04}{Ren04}$]$. However, the most general existence results available are in the analytic class. While limited results in the smooth class do exist in the literature $[$\hyperlink{Kic96}{Kic96}; \hyperlink{CN98}{CN98}; \hyperlink{Ren00}{Ren00}$]$, we believe that they are too specific to be easily adapted to the present system, or not without at least as much analysis as is required to execute our direct proof.\\ \\
Using the identities $$\tfrac{1}{2\xi}\partial_\xi\equiv\partial_x,\qquad 2\xi\bm{u}^\xi\equiv\bm{u}^x,\qquad\partial_\xi\bm{u}^\xi\equiv\partial_x\bm{u}^x-\tfrac{1}{2x}\bm{u}^x$$
we see that the unbounded terms in \hyperlink{eqn:2.9}{(2.9)} cancel in favour of the bounded $\partial_\xi\bm{u}^\xi$. 
\\ \\ \\
\large\hypertarget{sec:2.4}{\textbf{2.4\hspace{4mm}Higher order boundary conditions}}\label{2.4}\normalsize\\ \\
Solutions to the joint system \hyperlink{eqn:2.5}{(2.5)}-\hyperlink{eqn:2.9}{(2.9)} are required to satisfy the stipulated boundary conditions at $x=0$. However, the limiting values of their higher derivatives, for example $\lim_{\xi\to0}\partial^k_\xi\sigma_\pm$, are also uniquely determined, and are expressible algebraically as suitable functions of the data $(r_0,\sigma_0,(\partial_xr)_0,(\Gamma_x)_0)$ and derivatives. During the iteration scheme, it is necessary to keep track of whether the iterates already satisfy suitable higher order limits.\\ \\
\hypertarget{def:2.4}{\textbf{Definition 2.4.}} \textit{Let $k\in\mathbb{N}_0$ be a non-negative integer. The collection $(\Omega^2,r,\bm{u}^t_\pm,\bm{u}^x_\pm,\sigma_\pm)$ of functions defined on $\Sigma_\varepsilon$ \ul{satisfies $k^\text{th}$ order boundary conditions} if, for each $i=0,\dots,k$, the limits $$\partial^i_\xi\Gamma_t\big|_{\xi=0},\quad\partial^i_\xi\Gamma_x\big|_{\xi=0},\quad\partial^i_\xi(\partial_tr)\big|_{\xi=0},\quad\partial^i_\xi(\partial_xr)\big|_{\xi=0},\quad\partial^i_\xi\bm{u}^t_\pm\big|_{\xi=0},\quad\partial^i_\xi\bm{u}^x_\pm\big|_{\xi=0},\quad\partial^i_\xi\sigma_\pm\big|_{\xi=0}$$
coincide with the unique boundary values attained by bona fide solutions to the joint system \hyperlink{eqn:2.5}{(2.5)}-\hyperlink{eqn:2.9}{(2.9)}.}\\ \\
(For why we see the awkward combination of $\partial_x$ and $\partial_\xi$ derivatives, see already comments at the end of \hyperlink{sec:2.5}{Section~2.5}.) We use the above condition to prove bounds on the iterates in terms of ($k^\text{th}$ order norms) on initial data. It is therefore essential that satisfying $k^\text{th}$ order boundary conditions is propagated by the iteration scheme. Note that we do not need to separately demand anything about limits of $\partial_t$ derivatives, because all iterates are cusp-regular (see \hyperlink{def:2.3}{Definition~2.3}) and so we can always commute $\lim_{\xi\to0}$ with $\partial_t$.\\ \\ \\
\large\hypertarget{sec:2.5}{\textbf{2.5\hspace{4mm}Function spaces used in the iteration scheme}}\label{2.5}\normalsize\\ \\
We are now ready to define the function space to which iterates of the metric components and matter variables are to belong. Fix an admissible set of data $(r_0,\sigma_0,(\partial_xr)_0,(\Gamma_x)_0)$ as described in \hyperlink{sec:2.3}{Section~2.3}, with their associated constants $\delta_1,\delta_2>0$. Let $k\in\mathbb{N}$ be a positive integer, denoting the $k^\text{\hspace{1pt}th}$ level of differentiability, and let $\varepsilon,B,C>0$ be positive constants.\\ \\
\hypertarget{def:2.5}{\textbf{Definition 2.5. }}\textit{The collection $(\Omega^2,r,\bm{u}^t_\pm,\bm{u}^x_\pm,\sigma_\pm)$ of functions defined on $\Sigma_\varepsilon$ belongs to $\mathcal{A}(k,\varepsilon,B,C)$ if the following conditions hold.
\begin{itemize}
\item[$\circ$] \ul{cusp-regularity}\vspace*{2mm} \\
\hspace*{6mm}$\Omega^2,r,\bm{u}^t_\pm,\bm{u}^x_\pm,\sigma_\pm$ belong to the regularity class $C^\infty_\text{cusp}(\Sigma_\varepsilon)$.
\item[$\circ$] \ul{matter equations}\vspace*{2mm} \\
\hspace*{6mm}On the background given by $(\Omega^2,r)$, $\bm{u}_\pm:=\bm{u}^t_\pm\partial_t+\bm{u}^x_\pm\partial_x$ are geodesic vector fields, and\\ \hspace*{6mm}$\sigma_\pm x^{-1/2}\bm{u}_\pm$ are divergence-free:
$$\bm{u}^\nu_\pm\nabla_\nu\bm{u}^\mu_\pm=0 \qquad\nabla_\mu\big(\sigma_\pm x^{-1/2}\bm{u}^\mu_\pm\big)=0 $$
\hspace{6mm}Moreover, $\bm{u}^x_+>0$, $\bm{u}^x_-<0$ hold throughout $\Sigma_\varepsilon$.
\item[$\circ$] \ul{foliation by integral curves of $\bm{u}_\pm$}\vspace*{2mm} \\
\hspace*{6mm}$\Sigma_\varepsilon$ is foliated by integral curves of $\bm{u}_+$ (resp. $\bm{u}_-$), each having past (resp. future) endpoint \\
\hspace*{6mm}on $\{x=0\}\cap\overline{\Sigma}_\varepsilon$, where $\bm{u}^x_\pm\to 0$. We have the lower and upper bounds
$$\tfrac{1}{4}\delta_1\leq|\bm{u}_\pm(\bm{u}^x_\pm)|\leq 4C\vspace{0mm} $$
\item[$\circ$] \ul{$k^{\hspace{1pt}th}$-order boundary conditions}\vspace*{2mm} \\
\hspace*{6mm}As in \hyperlink{def:2.4}{Definition~2.4}.
\item[$\circ$] \ul{bounds}\vspace*{2mm} \\
\hspace*{6mm}The following bounds hold for the norm of the matter variables at the $k^{\hspace{1pt}th}$ level: 
$$\|\sigma_\pm\|_{C^k_\text{cusp}},\hspace{1mm}\|\bm{u}_\pm(\sigma_\pm)\|_{C^k_\text{cusp}},\hspace{1mm}\|\bm{u}^t_\pm\|_{C^{k+1}_\text{cusp}},\hspace{1mm}\|\bm{u}^\xi_\pm\|_{C^{k+1}_\text{cusp}}\leq B $$
\hspace{6mm}The constants $C,\delta_i>0$ control the metric components above and below at the $k^{\hspace{1pt}th}$ level:
$$\|\Gamma_x\|_{C^{k+1}_\text{cusp}},\hspace{4pt}\|\Gamma_t\|_{C^{k+1}_\text{cusp}},\hspace{4pt}\|\partial_xr\|_{C^{k+1}_\text{cusp}},\hspace{4pt}\|\partial_tr\|_{C^{k+1}_\text{cusp}}\leq C$$ $$ \inf_{\Sigma_\varepsilon}|\Gamma_x|\geq\tfrac{1}{2}\delta_1,\hspace{4pt}\sup_{\Sigma_\varepsilon}|\Gamma_t|\leq\tfrac{1}{2}\delta_1,\hspace{4pt}\inf_{\Sigma_\varepsilon} r\geq\tfrac{1}{2}\delta_2$$
\end{itemize}}
Note that $\partial_x$ is usually a `bad' derivative: for $\psi\in C^\infty_\text{cusp}$, we have generically that $\partial_x\psi=\tfrac{1}{2\xi}\partial_\xi\psi$ is \vspace*{-1.5pt}\\
unbounded. However, this is not true for the metric components $\log\Omega^2, r$. These variables are to satisfy a wave equation and have prescribed $\partial_x$-derivatives at $x=0$. In other words, $\partial_\xi\log\Omega^2$, $\partial_\xi r$ linearly vanish at $\xi=0$. It turns out to be cleaner to estimate the norms of $\partial_x\log\Omega^2$, $\partial_x r$ in this case.
\\ \\ \\
\Large\hypertarget{sec:3}{\textbf{3\hspace{4mm}Analysis of the Einstein equations}}\label{3}\normalsize\\ \\
In this section, we show that, even in the presence of unbounded energy-momentum terms, the Einstein equations \hyperlink{eqn:2.5}{(2.5)}-\hyperlink{eqn:2.6}{(2.6)} can be solved anyway, with bounds in the norms $\|\cdot\|_{C^k_\text{cusp}}$. These will be sufficient to show later that the iteration scheme for the joint Einstein-matter system is well-defined.\\ \\ \\
\large\hypertarget{sec:3.1}{\textbf{3.1\hspace{4mm}Solving the linear wave equation with cusp-regular inhomogeneity}}\label{3.1}\normalsize\\ \\
We begin with a straightforward lemma which applies to a class of inhomogeneities for the standard linear wave equation, including in particular the energy-momentum contributions of the type expected above. For now, let $C^\infty_\text{cusp}$ denote cusp-regular functions defined on the whole half-plane $\{(t,x):x>0\}$.\\ \\
\hypertarget{lem:3.1}{\textbf{Lemma 3.1}}\textbf{. }\textit{Let $\phi_0,(\partial_x\phi)_0\in C^\infty(\mathbb{R})$ be smooth data and let $\psi\in C^\infty_\text{cusp}$. Then there exists a cusp-regular solution $\phi\in C^\infty_\text{cusp}$, unique in the class $C^1(x\geq 0)\cap C^2(x>0)$, to the linear inhomogeneous wave equation}
$$\square\phi:=(-\partial^2_x+\partial^2_t)\phi=\psi x^{-1/2};\qquad \phi|_{x=0}=\phi_0,\quad \partial_x\phi|_{x=0}=(\partial_x\phi)_0  \vspace{4mm}$$
\textbf{Remark 3.1. }\textit{This equation is of course explicitly solvable (via d'Alembert's formula) whenever the inhomogeneity is locally integrable. The content of the lemma is that $\phi$ still inherits the $C^\infty_\text{cusp}$ regularity of $\psi$, despite the fact that the bicharacteristic cone of $\square$ degenerates at $x=0$ in the coordinates $(t,x^{1/2})$, and despite the presence of the divergent factor $x^{-1/2}$. In particular, applying the lemma to $$\square\phi=\psi\equiv (\psi x^{1/2})x^{-1/2} $$
yields that $\phi\in C^\infty_\text{cusp}$ whenever $\square\phi\in C^\infty_\text{cusp}$---one could say that $\square^{-1}$ preserves $C^\infty_\text{cusp}$.}\\ \\
\begin{minipage}{0.8\textwidth}
\textit{Proof. }We write down d'Alembert's formula and claim it has the desired properties. At each point $(t,x)$, let $\triangleright(t,x)=\{(\tilde{t},\tilde{x}):\tilde{x}\geq 0;\hspace{2pt}|t-\tilde{t}|\leq x-\tilde{x}\}$, and define $\phi(t,x)$ by\vspace{-1mm}
$$\phi(t,x):=\tfrac{1}{2}\big(\phi_0(t+x)+\phi_0(t-x)\big)+\tfrac{1}{2}\int^{t+x}_{t-x}(\partial_x\phi)_0(s)ds+\tilde{\phi}(t,x) \vspace{-3mm}$$
$$\text{where }\quad \tilde{\phi}(t,x):=-\tfrac{1}{2}\int_{\triangleright(t,x)}\psi(\tilde{t},\tilde{x})\tilde{x}^{-1/2}d\tilde{x}d\tilde{t}\vspace{2mm}$$
In the definition of $\tilde{\phi}$, the integrand is dominated by a multiple of $x^{-1/2}$ which is integrable on compact subsets of the $(t,x)$ plane, and along null segments. It follows straightforwardly that $\tilde{\phi}(t,x)$ exists, and is moreover $C^1$, with $\partial_x\tilde{\phi}$, $\partial_t\tilde{\phi}$ converging to zero as $x\to 0$. It follows that $\phi(t,x)$ exists and takes the given data.\vspace{6mm}
\end{minipage}
\hfill
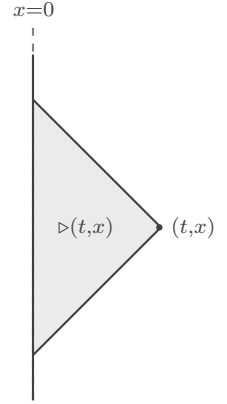
\begin{minipage}{0.17\textwidth}
\vspace{-16mm}
\begin{figure}[H]
\begin{center}
\begin{tikzpicture}
\filldraw [white, fill=lightgray, domain=0:3, variable=\x, thick, opacity=0.3]
      (0, 4.0)
      -- (1.7,2.3)
      -- (0, 0.6)
      -- cycle;
\draw [gray, thick, dashed] (0, 0) -- (0, 5);
\draw [darkgray, thick] (0.0, 0) -- (0.0, 4.6);
\draw [darkgray, thick] (0, 0.6) -- (1.7, 2.3);
\draw [darkgray, thick] (1.7, 2.3) -- (0, 4.0);
\node[darkgray] at (-0.4, 5.2) [anchor = west] {$_{x=0}$};
\node[darkgray, align=left] at (0.2,2.3) [anchor = west] {$_{\triangleright(t,x)}$};
\node[align=left, darkgray] at (1.7,2.3) [anchor = west] {$_{(t,x)}$};
\draw[darkgray, fill=darkgray] (1.68,2.3) circle (1pt) {};
\end{tikzpicture}
\end{center}
\end{figure}
\end{minipage}
Applying $\partial_t$ preserves the class $C^\infty_\text{cusp}$, and commutes through the integral $\int_{\triangleright(t,x)}$, yielding $$\partial_t\tilde{\phi}=-\tfrac{1}{2}\int_{\triangleright(t,x)}\partial_t\psi(\tilde{t},\tilde{x})\tilde{x}^{-1/2}d\tilde{x}d\tilde{t} $$ 
Repeating the logic of the paragraph above, we get that $\partial_t\phi$ is $C^1$ too. Differentiating successively in $\partial_t$ and applying this argument, we find that all the derivatives $\partial^k_t\phi$, $\partial^k_t\partial_x\phi$ exist on $x>0$.\\ \\
We can now use the wave equation itself, which holds classically, to upgrade the regularity in $x$. Indeed,  we can argue by induction as follows. The wave equation allows us to convert $\partial^2_x$ into $\partial^2_t$ repeatedly. The (already proven) existence of the $\partial_t$ derivatives then allows us to conclude that $\phi$ is smooth on the domain $x>0$.
\\ \\
It remains to show not just that $\phi\in C^\infty(\{x>0\})$, but also $\phi\in C^\infty_\text{cusp}$, i$.$e$.$ all derivatives $\partial^k_\xi\partial^l_t$ converge as $\xi\to0$. For convenience, we record the relations
$$\partial_x\equiv\tfrac{1}{2\xi}\partial_\xi\qquad\partial^2_x\equiv-\tfrac{1}{4\xi^3}\partial_\xi+\tfrac{1}{4\xi^2}\partial^2_\xi $$
Writing $\square\phi=\psi x^{-1/2}$ in coordinates $(t,\xi)$, multiplying by $4\xi^2$ and rearranging, we have
\begin{equation}\partial^2_\xi\phi=4\xi^2\partial^2_t\phi-4\xi\psi+2\partial_x\phi \tag{\hypertarget{eqn:3.1}{3.1}}
\end{equation}
Every term on the right-hand side converges as $\xi\to0$, and so $\partial^2_\xi\phi$ also converges. For higher derivatives, we would like to differentiate successively in $\xi$ and appeal to the assumed cusp-regularity of $\psi$ to deduce a limit. However, there is an inverse factor $\xi^{-1}$ hidden in the last term of \hyperlink{eqn:3.1}{(3.1)}, so we cannot quite differentiate freely in $\xi$. Multiplying through by $\xi$, differentiating again with $\partial_\xi$, and finally dividing through by $\xi$, we obtain
\begin{equation}
\partial^3_\xi\phi=12\xi\partial^2_t\phi+4\xi^2\partial^2_t\partial_\xi\phi-8\psi-4\xi\partial_\xi\psi
\vspace{2mm}\tag{\hypertarget{eqn:3.2}{3.2}}
\end{equation}
With these two equations in hand, we have everything needed to show that all derivatives $\partial^k_\xi\partial^l_t\phi$ converge as $\xi\to0$. We can differentiate \hyperlink{eqn:3.2}{(3.2)} freely in $\xi$ without generating any $\xi^{-1}$ factors. It then follows by another straightforward induction, using the assumed cusp-regularity of $\psi$ and commuting with $t$ whenever needed, to establish that all derivatives converge. \hfill$\square$\\ \\
Though this is stated for the whole half-plane $x>0$, we can equally well choose any domain $\Sigma_s$ with $s$ sufficiently small---we only need each $\triangleright(t,x)$ to be contained in $\Sigma_s$  (see remarks following \hyperlink{def:2.2}{Definition~2.2}).\\ \\ \\
\large\hypertarget{sec:3.2}{\textbf{3.2\hspace{4mm}Estimates for the Einstein equations}}\label{3.2}\normalsize\\ \\
In the iteration scheme for the joint Einstein-matter system, we generate new iterates for the metric components $\Omega^2,r$ by solving the wave-type Einstein equations \hyperlink{eqn:2.5}{(2.5)}-\hyperlink{eqn:2.6}{(2.6)} with all terms on the right-hand side replaced by previous iterates (see \hyperlink{eqn:3.3}{(3.3)}-\hyperlink{eqn:3.4}{(3.4)} below). Thus \hyperlink{lem:3.1}{Lemma~3.1} applies to these equations, guaranteeing existence and (cusp-)regularity of solutions. The next result obtains precise bounds for these solutions at an arbitrary level of differentiability.\\ \\
\hypertarget{pro:3.2}{\textbf{Proposition 3.2}} (Estimates for the Einstein equations)\textbf{. }\textit{Fix $k\in\mathbb{N}$ and an admissible set of data $(r_0,\sigma_0,(\partial_xr)_0,(\Gamma_x)_0)$. Define the constants $\delta_i,N_i$ controlling the data at the k$^{\hspace{1pt}th}$ level:}
$$\delta_1=\inf_{[a,b]}|(\Gamma_x)_0|>0\qquad\delta_2=\inf_{[a,b]}r_0>0 $$
$$N_1=\|(\Gamma_x)_0\|_{C^{k+1}[a,b]}\hspace{0.5cm} N_2=\max\big\{\|r_0\|_{C^{k+2}[a,b]},\|(\partial_xr)_0\|_{C^{k+1}[a,b]}\big\}\hspace{0.5cm} N_3=\|\sigma_0\|_{C^k[a,b]}\vspace{2mm} $$
\textit{Then there exists $C=C(k,\delta_i,N_i)>0$ such that, for all positive constants $B>0$, there exists $\varepsilon>0$ with the following property. For all $(\Omega^2,r,\bm{u}^t_\pm,\bm{u}^x_\pm,\sigma_\pm)\in \mathcal{A}(k,\varepsilon,B,C)$ (see \hyperlink{def:2.5}{Definition~2.5}), if we take the unique cusp-regular solution $\overline{\Omega}^2,\overline{r}\in C^\infty_\text{cusp}(\Sigma_\varepsilon)$ to}
\begin{equation}
(-\partial^2_x+\partial^2_t)(\log\overline{\Omega}^2)=\frac{4\Omega^2m}{r^3}-4\Omega^2(\sigma_++\sigma_-) x^{-1/2}\hspace{9.5mm} \tag{\hypertarget{eqn:3.3}{3.3}}
\end{equation}
\begin{equation}
(-\partial^2_x+\partial^2_t)\hspace{1pt}\overline{r}=-\frac{2\Omega^2m}{r^2}+2r\Omega^2(\sigma_++\sigma_-) x^{-1/2}\tag{\hypertarget{eqn:3.4}{3.4}}
\end{equation}
\vspace{1mm}
$$\log\overline{\Omega}^2|_{x=0}=0,\quad \partial_x\log\overline{\Omega}^2|_{x=0}=2(\Gamma_x)_0, \qquad \overline{r}|_{x=0}=r_0,\quad \partial_x\overline{r}|_{x=0}=(\partial_xr)_0\vspace{1mm}$$
\textit{given by \hyperlink{lem:3.1}{Lemma~3.1}, then the same bounds propagate:}
$$\|\overline{\Gamma}_x\|_{C^{k+1}_\text{cusp}},\hspace{4pt}\|\overline{\Gamma}_t\|_{C^{k+1}_\text{cusp}},\hspace{4pt}\|\partial_x\overline{r}\|_{C^{k+1}_\text{cusp}},\hspace{4pt}\|\partial_t\overline{r}\|_{C^{k+1}_\text{cusp}}\leq C$$ $$ \inf_{\Sigma_\varepsilon}|\overline{\Gamma}_x|\geq\tfrac{1}{2}\delta_1,\hspace{4pt}\sup_{\Sigma_\varepsilon}|\overline{\Gamma}_t|\leq\tfrac{1}{2}\delta_1,\hspace{4pt}\inf_{\Sigma_\varepsilon} \overline{r}\geq\tfrac{1}{2}\delta_2$$
\textit{and $\overline{\Omega}^2,\overline{r}$ satisfy $k^\text{th}$ order limits. $(\hspace{1pt}\overline{\Gamma}_t,\overline{\Gamma}_x$ are the Christoffel symbols corresponding to the new $\overline{\Omega}^2.)$ Finally, we can increase $C$ if needed, and/or decrease $\varepsilon$, and the above continues to hold.}\\ \\
Before beginning the proof, we explain briefly the role played by this proposition in \hyperlink{thm:1.1}{Theorem 1.1}---saving the full discussion for \hyperlink{sec:6}{Section~6}. We are concerned to ensure that when we iterate the metric components $(\Omega^2,r)\mapsto (\overline{\Omega}^2,\overline{r})$, the new iterates $(\overline{\Omega}^2,\overline{r})$ obey the same bounds satisfied by $(\Omega^2,r)$. The outcome of this proposition is that we \textit{can} commit to bounds $C$ for the metric components that depend only on (norms on) data and the level of differentiability $k$. Whatever bounds $B$ one has for the matter variables, we can still recover the intended bounds $C$ for the new iterate $(\overline{\Omega}^2,\overline{r})$, if we reduce $\varepsilon$ enough. (We are happy also for the width $\varepsilon$ of the slab to depend on $B$.) Hence we obtain the final constants in the following order:
\begin{enumerate}
\item[\textbf{1.}] Obtain $C=C(k,\delta_i,N_i)>0$ from \hyperlink{pro:3.2}{Proposition~3.2}.
\item[\textbf{2.}] In \hyperlink{pro:4.1}{Proposition~4.1}, obtain bounds $B=B(k,\delta_i,N_i,C)>0$ for the matter variables.
\item[\textbf{3.}] Re-insert into \hyperlink{pro:3.2}{Proposition~3.2}, yielding $\varepsilon=\varepsilon(k,\delta_i,N_i,C,B)>0$.
\end{enumerate}
Putting it together, we get an iteration scheme that closes at the $k^\text{\hspace{1pt}th}$ level, with $B$, $C$ and $\varepsilon$ all depending only $k$, $\delta_i$, $N_i$ at the end of the day.\\ \\
We now record some calculations involved in the proof of \hyperlink{pro:3.2}{Proposition 3.2}. The proof involves estimating terms which are integrated along null segments, with one end on the boundary $\{x=0\}$. The following lemma allows us to control the $\|\cdot\|_{C^k_\text{cusp}}$ norm of such terms, in terms of lower derivatives (order $\leq k-1$) evaluated at the boundary. \\ \\
\hypertarget{lem:3.3}{\textbf{Lemma 3.3 }}(Control of integrated terms)\textbf{.} \textit{For each $k\in\mathbb{N}_0$, there exists a constant $C_k>0$ with the following property. Let $\psi\in C^\infty_\text{cusp}$ and suppose $\|\psi\|_{C^k_\text{cusp}}<\infty$. We define, in double null coordinates (see again \hyperlink{sec:2.1}{Section~2.1}) 
$$F_\psi(u,v)= \int^v_u\psi(u,\tilde{v}) x^{-1/2}d\tilde{v}$$
where $2x=\tilde{v}-u$ in the integrand. Then for all $\gamma>0$, there exists $\varepsilon=\varepsilon(k,\gamma,\|\psi\|_{C^k_\text{cusp}})>0$ such that the following bound holds.}
\begin{equation*}\hspace{3cm}\|F_\psi\|_{C^k_\text{cusp}}\leq\begin{cases}\hspace{1.5mm}\gamma\hspace{4.85cm}k=0\\ \hspace{1.5mm}\gamma+C_k\hspace{-1.5mm}\sum\limits_{i+j\leq k-1}\hspace{0.5mm}\sup\limits_{x=0}\big|\partial^{i}_\xi\partial^j_t\psi\big| \qquad k>0
\end{cases}\tag{\hypertarget{eqn:3.5}{3.5}}
\end{equation*}
\textit{Finally, the above estimate holds} mutatis mutandis \textit{when integrating instead with respect to $u$ along lines $\{v=\text{const.}\}$}.
\\ \\
\textit{Proof. }We proceed by induction, taking the base case $k=0$ to be clear. First, we decompose $\partial_\xi$ into the sum (see again remarks in \hyperlink{sec:2.1}{Section 2.1} on notation for coordinate derivatives)
$$\partial_\xi\equiv2\xi \partial_x\equiv2\xi\big(2\partial_v-\partial_t\big) $$
in order to write 
$$\partial_\xi\int^v_u\psi x^{-1/2}d\tilde v=4\psi-2\xi\int^v_u\partial_t\psi\hspace{2pt} x^{-1/2}d\tilde v $$
Differentiating $k$ further times and applying the Leibniz rule yields the identity
\begin{equation*}
\partial^{k+1}_\xi\hspace{-4pt}\int^v_u\psi x^{-1/2}d\tilde v=4\partial^k_\xi\psi-2\xi\partial^k_\xi\int^v_u\psi x^{-1/2}d\tilde v-2k\partial^{k-1}_\xi\hspace{-4pt}\int^v_u\partial_t\psi\hspace{2pt} x^{-1/2}d\tilde v\tag{\hypertarget{eqn:3.6}{3.6}}
\end{equation*}
which will be the basis for our inductive step.\vspace*{2mm} \\
Suppose now that for some $k\in\mathbb{N}_0$, such a $C_k$ exists (if $k=0$, we can think of $C_0=0$), and fix $\psi\in C^\infty_\text{cusp}$ and $\gamma>0$. By the inductive hypothesis, choose $\varepsilon\in(0,1)$ so that inequality \hyperlink{eqn:3.5}{(3.5)} holds, for this $k$ and for \textit{both} $\psi$ and $\partial_t\psi$, and moreover $$4\varepsilon^{1/2}\|\partial^{k+1}_\xi\psi\|_{C^0(\Sigma_\varepsilon)}\leq\gamma $$
We estimate $$\bigg\|\int^v_u\psi x^{-1/2}d\tilde v\bigg\|_{C^{k+1}_\text{cusp}(\Sigma_\varepsilon)}\leq \bigg\|\int^v_u\psi x^{-1/2}d\tilde v\bigg\|_{C^k_\text{cusp}(\Sigma_\varepsilon)}+\bigg\|\int^v_u\partial_t\psi\hspace{2pt} x^{-1/2}d\tilde v\bigg\|_{C^k_\text{cusp}(\Sigma_\varepsilon)}$$
$$\hspace{3cm}+\hspace{2mm}\bigg\|\partial^{k+1}_\xi\hspace{-4pt}\int^v_u\psi x^{-1/2}d\tilde v\bigg\|_{C^0(\Sigma_\varepsilon)}  $$
To handle the last term, we apply the identity \hyperlink{eqn:3.6}{(3.6)} above, writing also
$$\partial^k_\xi\psi(t,\xi)=\partial^k_\xi\psi(t,0)+\int^\xi_0\partial^{k+1}_\xi\psi \hspace{2pt}d\xi $$
so that 
$$\bigg\|\partial^{k+1}_\xi\hspace{-4pt}\int^v_u\psi x^{-1/2}d\tilde v\bigg\|_{C^0(\Sigma_\varepsilon)}\leq 4\sup_{x=0}|\partial^k_\xi\psi|+4\varepsilon^{1/2}\|\partial^{k+1}_\xi\psi\|_{C^0(\Sigma_\varepsilon)}+(2k+2)\bigg\|\int^v_u\psi x^{-1/2}d\tilde v\bigg\|_{C^k_\text{cusp}(\Sigma_\varepsilon)}$$
Putting it together, we can achieve the bound
$$\bigg\|\int^v_u\psi x^{-1/2}d\tilde v\bigg\|_{C^{k+1}_\text{cusp}(\Sigma_\varepsilon)}\leq(5+2k)\gamma+\big(4+(2k+4)C_k\big)\sum\limits_{i+j\leq k}\hspace{0.5mm}\sup\limits_{x=0}\big|\partial^{i}_\xi\partial^j_t\psi\big| $$
so that, after rescaling $\gamma$, we get the desired inequality \hyperlink{eqn:3.5}{(3.5)} for $k+1$ by setting
$$C_{k+1}=4+(2k+4)C_k\vspace{-8mm} $$
\hfill$\square$\\ \\
Next, recall that our aim in \hyperlink{pro:3.2}{Proposition~3.2} is to control $k+1$ derivatives of $\overline{\Gamma}_t,\overline{\Gamma}_x$ and $\partial_t\overline{r}, \partial_x\overline{r}$---that is, $k+2$ derivatives of $\log\overline{\Omega}^2$ and $\overline{r}$. For solutions of the standard (inhomogenous) wave equation, this would require control of $k+1$ derivatives of the inhomogeneity---yet we only have access to $k$ derivatives of $\sigma_\pm$ here. Fortunately, there is a standard technique to gain a derivative in this case. Namely, we exploit the fact that the troubling terms are components of an (exactly conserved) energy-momentum tensor: recall in \hyperlink{def:2.5}{Definition~2.5} that the matter equations are satisfied exactly on the background $(\Omega^2,r)$. The technique involves commuting null derivatives and integrating along null segments, and so we rotate coordinates back to $(u,v)$ , adopt the notation
$$\Gamma_u=\partial_u\log\Omega^2\qquad \Gamma_v=\partial_v\log\Omega^2 \vspace{1mm}$$
Rewriting equation \hyperlink{eqn:3.3}{(3.3)} in null coordinates $(u,v)$, we have $$\partial_u\overline{\Gamma}_v=\frac{\Omega^2m}{r^3}-\tfrac{1}{2}\Omega^4 T^{uv} $$
We also record, in terms of coordinate derivatives, the conservation of $T^{\mu\nu}$:
\begin{equation}
0=\partial_u T^{uu}+\partial_vT^{uv}+\big(2\Gamma_u+\partial_u\log r^2\big)T^{uu}+\big(\Gamma_v+\partial_v\log r^2\big)T^{uv}\tag{\hypertarget{eqn:3.7}{3.7}}
\end{equation}
\begin{equation}
0=\partial_u T^{uv}+\partial_vT^{vv}+\big(2\Gamma_v+\partial_v\log r^2\big)T^{vv}+\big(\Gamma_u+\partial_u\log r^2\big)T^{uv}\vspace{2mm}\tag{\hypertarget{eqn:3.8}{3.8}}
\end{equation}
This allows us to exchange $\partial_u$, $\partial_v$ derivatives and ultimately eliminate differentiated components of $T^{\mu\nu}$:
\begin{align*}
\partial_u\big(\partial_v\overline{\Gamma}_v-\tfrac{1}{2}\Omega^4 T^{uu}\big)&=\partial_v\bigg(\frac{\Omega^2m}{r^3}-\tfrac{1}{2}\Omega^4T^{uv}\bigg)-\tfrac{1}{2}\partial_u(\Omega^4 T^{uu})\\
&=\partial_v\bigg(\frac{\Omega^2m}{r^3}\bigg)-\tfrac{1}{2}(\partial_u\Omega^4)T^{uu}-\tfrac{1}{2}(\partial_v\Omega^4)T^{uv}-\tfrac{1}{2}\Omega^4(\partial_vT^{uv}+\partial_uT^{uu})\\
&=\partial_v\bigg(\frac{\Omega^2m}{r^3}\bigg)+\tfrac{1}{2}\Omega^4T^{uu}\partial_u\log r^2+\tfrac{1}{2}\Omega^4T^{uv}(\partial_v\log r^2-\Gamma_v)
\end{align*}
where, in the last equality, we applied \hyperlink{eqn:3.7}{(3.7)} and simplified the expression.\vspace*{2mm} \\
We now study the limit, on the caustic $\{u=v\}$, of the expression $\partial_v\overline{\Gamma}_v-\Omega^4 T^{uu}$ (differentiated on the left-hand side above), observing that 
\begin{align*}
\partial_v\overline{\Gamma}_v-\tfrac{1}{2}\Omega^4 T^{uu}&=\tfrac{1}{2}(\partial^2_t\log\overline{\Omega}^2+\partial_t\partial_x\log\overline{\Omega}^2)-\tfrac{1}{4}(-\partial^2_x+\partial^2_t)\log\overline{\Omega}^2-\tfrac{1}{2}\Omega^4T^{uu}\\
&=\tfrac{1}{2}(\partial^2_t\log\overline{\Omega}^2+\partial_t\partial_x\log\overline{\Omega}^2)-\frac{\Omega^2m}{r^3}+\tfrac{1}{2}\Omega^4(T^{uv}-T^{uu})
\end{align*}
so that
$$\lim_{u\to v}(\partial_v\overline{\Gamma}_v-\tfrac{1}{2}\Omega^4 T^{uu})=(\Gamma_x)'_0(t)-\frac{m_0}{r_0(t)^3}+\tfrac{1}{2}\lim_{u\to v}(T^{uv}-T^{uu}) $$
using that $\log\overline{\Omega}^2|_{x=0}=0$. Integrating now along the segment $\{v=\text{const.}\}$, we have, at each point $(u,v)$:
$$\partial_v\overline{\Gamma}_v(u,v)=\tfrac{1}{2}\Omega^4(u,v)T^{uu}(u,v)+(\Gamma_x)'_0(v)-\frac{m_0}{r_0(v)^3}+\tfrac{1}{2}\lim_{u\to v}(T^{uv}-T^{uu})(v)\hspace{3cm} $$
\begin{equation}\hspace{2cm}-\int^v_u\bigg[\partial_v\bigg(\frac{\Omega^2m}{r^3}\bigg)+\tfrac{1}{2}\Omega^4T^{uu}\partial_u\log r^2+\tfrac{1}{2}\Omega^4T^{uv}(\partial_v\log r^2-\Gamma_v)\bigg]du \tag{\hypertarget{eqn:3.9}{3.9}}
\end{equation}
This is routine so far, and applies to a completely general conserved energy-momentum tensor. We are ready to begin our proof.\\ \\
\textit{Proof of Proposition 3.2.}  We will use \hyperlink{eqn:3.9}{(3.9)}, and similar identities for $\partial_u\overline{\Gamma}_u,\partial^2_u\overline{r},\partial^2_v\overline{r}$ in the sequel. In our specific case, the components $T^{uu},T^{uv},T^{vv}$ are each unbounded. This reflects the fact that $\partial_u$,$\partial_v$ are `bad' derivatives of the cusp-regular variable $\overline{\Gamma}_v$. When we combine $\partial_u\overline{\Gamma}_v$, $\partial_v\overline{\Gamma}_v$ so as to achieve a `good' derivative, the unbounded contributions cancel. Thus, for example,
\begin{align*}
\partial_t\overline{\Gamma}_v&=\partial_u\overline{\Gamma}_v+\partial_v\overline{\Gamma}_v\\
&=-\tfrac{1}{2}\Omega^4(T^{uv}-T^{uu})+\text{bounded terms}\\
&=-2\Omega^4\big(\sigma_+\bm{u}^u_+\bm{u}^\xi_++\sigma_-\bm{u}^u_-\bm{u}^\xi_-\big)+\text{bounded terms}
\end{align*}
We now directly estimate $\|\cdot\|_{C^k_\text{cusp}}$ for $\partial_t\overline{\Gamma}_v$, $\partial_\xi\overline{\Gamma}_v$, $\partial_t\overline{\Gamma}_u$, $\partial_\xi\overline{\Gamma}_u$, and similarly for derivatives of $\overline{r}$. We give details only for $\partial_t\overline{\Gamma}_v$, since all have the same form, and in particular all involve dealing with three types of terms, as now indicated:\vspace{-2mm}
\begin{align*}
\partial_t\overline{\Gamma}_v= &\overbrace{4\Omega^4\xi\big(\sigma_+(\bm{u}^\xi_+)^2+\sigma_-(\bm{u}^\xi_-)^2\big)-2\Omega^4\big(\sigma_+\bm{u}^t_+\bm{u}^\xi_++\sigma_-\bm{u}^t_-\bm{u}^\xi_-\big)+\frac{\Omega^2 m}{r^3}}^{\textcircled{\footnotesize1}}\\
&+\underbrace{(\Gamma_x)'_0(v)-\frac{m_0}{r_0(v)^3}}_{\textcircled{\footnotesize2}}-\underbrace{\int^v_u\bigg[\partial_v\bigg(\frac{\Omega^2m}{r^3}\bigg)+\tfrac{1}{2}\Omega^4T^{uu}\partial_u\log r^2+\tfrac{1}{2}\Omega^4T^{uv}(\partial_v\log r^2-\Gamma_v)\bigg]du}_{\textcircled{\footnotesize3}}
\end{align*}
Note that $\lim_{u\to v}(T^{uv}-T^{uu})=0$ in our case (see \hyperlink{sec:2.2}{Section~2.2}), explaining the disappearance of that term from the earlier expression.\\ \\
We will show that each type of term \textcircled{\footnotesize1}-\textcircled{\footnotesize3} is bounded above by a constant depending only on $\delta_i,N_i$ (the constants controlling our data) \textit{and} possibly an additional quantity that can be made arbitrarily small by reducing $\varepsilon$.\\ \\
\underline{Step 1: Estimates for \textcircled{\footnotesize1}:} The collection labelled \textcircled{\footnotesize1} is a polynomial in terms of matter variables, metric components, and functions thereof. Estimating $\|\cdot\|_{C^k_\text{cusp}}$ for the whole collection reduces, via \hyperlink{lem:a.1}{Lemma~A.1}(i)-(ii), to estimating $\|\cdot\|_{C^k_\text{cusp}}$ for each individual variable. Let us begin with the matter variables. Here, it will not suffice to merely use, for example, $\|\sigma\|_{C^k_\text{cusp}}\leq B$, as, once again, our bounds must only depend on data. We therefore make use of our extra $\bm{u}$-derivative (see \hyperlink{def:2.5}{Definition~2.5}). \\ \\
\begin{minipage}{0.8\textwidth}
Taking $\sigma_+$ for concreteness (but the estimate works the same for any of $\sigma_\pm,\bm{u}^t_\pm,\bm{u}^x_\pm$), consider the integral curve of $\bm{u}_+$ passing through a fixed point $(t,\xi)$. Choose the parameter $\tau$ so that $\tau=0$ on the caustic, at, say $t=t_0$, and $\tau=\tau_1$ when the curve meets $(t,\xi)$. Then for each $i+j\leq k$, we have 
$$\partial^i_\xi\partial^j_t\sigma_+(t,\xi)=\partial^i_\xi\partial^j_t\sigma_+(t_0,0)+\int^{\tau_1}_0\bm{u}_+(\partial^i_\xi\partial^j_t\sigma_+)d\tau $$
Since $\ddot{x}(\tau)\geq\tfrac{1}{4}\delta_1$ holds for the curve, we have $x(\tau)\geq\tfrac{1}{8}\delta_1\tau^2$ and hence we can estimate
$$\tau_1\leq 4\delta^{-1/2}_1\varepsilon^{1/2} $$
\end{minipage}
\hfill
\begin{minipage}{0.18\textwidth}
\begin{figure}[H]
\begin{center}
\vspace{-1cm}
\begin{tikzpicture}[scale=1.5]
\draw [gray, thick, domain=0:2, samples=150] plot ({0.2*(\x)^(1.8)},{(\x)+1});
\draw [darkgray, thick] (0.0, 0.5) -- (0.0, 3.5);
\node[darkgray] at (0.73,3) [anchor = south] {$_{(t,\xi)}$};
\node[darkgray] at (0, 1) [anchor = east] {$_{(t_0,0)}$};
\node[darkgray] at (0.75, 2.95) [anchor = west] {$_{\tau=\tau_1}$};
\node[darkgray] at (0.24, 2.1) [anchor = west] {$_{\bm{u}_+}$};
\node[darkgray] at (0, 1) [anchor = west] {$_{\tau=0}$};
\node[darkgray] at (0.0, 2.6) [anchor = east] {$_\mathcal{C}$};
\draw [-stealth](0.24,2.1) -- (0.365,2.4);
\filldraw[color=black, fill=black](0,1) circle (0.03);
\filldraw[color=black, fill=black](0.7,3) circle (0.03);
\end{tikzpicture}
\end{center}
\end{figure}
\end{minipage}\\ \\
Taking suprema over $(t,\xi)$ and summing $i,j$, we now have 
$$\|\sigma_+\|_{C^k_\text{cusp}(\Sigma_\varepsilon)}\leq \sum_{i+j\leq k}\sup_{x=0}|\partial^i_\xi\partial^j_t\sigma_+|+4\delta^{-1/2}_1\varepsilon^{1/2}\sum_{i+j\leq k}\|\bm{u}_+(\partial^i_\xi\partial^j_t\sigma_+)\|_{C^0} $$
We next use \hyperlink{lem:a.2}{Lemma~A.2} to commute $\bm{u}_+$ with the coordinate derivatives:
\begin{align*}
\sum_{i+j\leq k}\|\bm{u}_+(\partial^i_\xi\partial^j_t\sigma_+)\|_{C^0}&\leq \|\bm{u}_+(\sigma_+)\|_{C^k_\text{cusp}}+\sum_{i+j\leq k}\|\partial^i_\xi\partial^j_t(\bm{u}_+(\sigma_+))-\bm{u}_+(\partial^i_\xi\partial^j_t\sigma_+)\|_{C_0}\\
&\leq \|\bm{u}_+(\sigma_+)\|_{C^k_\text{cusp}}+C_k\|\sigma_+\|_{C^k_\text{cusp}}\big(\|\bm{u}^t\|_{C^k_\text{cusp}}+\|\bm{u}^\xi\|_{C^k_\text{cusp}}\big)\\
&\leq B+2C_kB^2
\end{align*}
and we arrive at the desired bound:
$$\|\sigma_+\|_{C^k_\text{cusp}}\leq\sum_{i+j\leq k}\sup_{x=0}|\partial^i_\xi\partial^j_t\sigma_+| +4\delta^{-1/2}_1\varepsilon^{1/2}(B+2C_kB^2)$$
which, as intended, has one term related only to data, and the other controlled by $\varepsilon$. The bound holds in the same form for $\bm{u}^t_\pm$,$\bm{u}^\xi_\pm$: note that we have the extra $\bm{u}$-derivative for these because we already control them one level higher than $\sigma_\pm$.\\ \\
To derive a similar bound for the metric components $\log\Omega^2,r$, we can simply integrate from the boundary $x=0$, since they belong to a higher regularity class. For example, for each $i+j\leq k$,
$$\partial^i_\xi\partial^j_t(\partial_xr)(t,\xi)=\partial^i_\xi\partial^j_t(\partial_xr)(t,0)+\int^\xi_0\partial^{i+1}_\xi\partial^j_t(\partial_xr)d\xi $$
and, again taking the supremum over $(t,\xi)$ and summing $i,j$,
$$\|\partial_xr\|_{C^k_\text{cusp}}\leq\sum_{i+j\leq k}\sup_{x=0}|\partial^i_\xi\partial^j_t(\partial_xr)|+\varepsilon^{1/2}C $$
We can extend this bound to $r$ itself by first applying \hyperlink{lem:a.1}{Lemma~A.1}(i):
$$\|\partial_\xi \overline{r}\|_{C^k_\text{cusp}}=\|2\xi\partial_x\overline{r}\|_{C^k_\text{cusp}}\leq C_k\|2\xi\|_{C^k_\text{cusp}}\|\partial_x\overline{r}\|_{C^k_\text{cusp}}\leq 2C_k(1+\varepsilon^{1/2})C $$
and it again then follows by integration that 
$$\|r\|_{C^k_\text{cusp}}\leq\sum_{i+j\leq k}\sup_{x=0}|\partial^i_\xi\partial^j_tr|+2C_k\varepsilon^{1/2}(1+\varepsilon^{1/2})C $$
A similar observation is used to bound $\|\log\Omega^2\|_{C^k_\text{cusp}}$ in terms of $\|\Gamma_x\|_{C^k_\text{cusp}}$.\\ \\
Putting this together, we now compute $\|\cdot\|_{C^k_\text{cusp}}$ for the collection \textcircled{\footnotesize1}. Applying \hyperlink{lem:a.1}{Lemma~A.1}(i)-(ii) repeatedly, we can bound this in terms of norms of the individual variables. This includes, for instance, $r^{-1}$, which is bounded, through \hyperlink{lem:a.1}{Lemma~A.1}(ii), in terms of $\|r\|_{C^k_\text{cusp}}$ and $\delta_2$. As seen above, each bound has a part involving data, as well as a part proportional to $\varepsilon^{1/2}$. This extends to \textcircled{\footnotesize1} itself: in total, our estimate takes the form
$$\|\textcircled{\footnotesize1}\|_{C^k_\text{cusp}}\leq c_1(k,\delta_i,N_i)+c(k,\delta_i,N_i,C,B)\varepsilon^{1/2} $$
where $c,c_1>0$ are constants depending only on the indicated quantities. Note carefully that we are implicitly using the $k^\text{th}$ order boundary conditions here when we assert bounds in terms of data $\delta_i,N_i$ on the restrictions of the functions at $x=0$. \\ \\
\underline{Step 2: Estimates for \textcircled{\footnotesize2}:} Setting\vspace{-4mm} $$f_0(t):=(\Gamma_x)'_0(t)-\frac{m_0}{r_0(t)^3} $$
we first note that $\|f_0\|_{C^k[a,b]}$ may be bounded by a constant depending only on $\delta_i,N_i$ ($m_0$ is bounded in terms of $N_2$). Then, we need only apply \hyperlink{lem:a.3}{Lemma~A.3} to conclude
$$\|\textcircled{\footnotesize2}\|_{C^k_\text{cusp}}=\|f_0(v)\|_{C^k_\text{cusp}}\leq C_k\|f_0\|_{C^k[a,b]}\leq c_2(k,\delta_i,N_i) $$
for some constant $c_2(k,\delta_i,N_i)$.\\ \\
\underline{Step 3: Estimates for \textcircled{\footnotesize3}:} We intend to apply \hyperlink{lem:3.3}{Lemma~3.3} to handle this case, taking $\gamma=1$. The corresponding function $\psi\in C^\infty_\text{cusp}(\Sigma_\varepsilon)$ is then given by
$$\psi:=\xi\bigg[\partial_v\bigg(\frac{\Omega^2m}{r^3}\bigg)+\tfrac{1}{2}\Omega^4T^{uu}\partial_u\log r^2+\tfrac{1}{2}\Omega^4T^{uv}(\partial_v\log r^2-\Gamma_v)\bigg] $$
and so there are two claims to justify. The first is that $\|\psi\|_{C^k_\text{cusp}}$ depends only on $B,C$, so that the $\varepsilon$ obtained in the lemma also depends only on $B,C$. The second is that we can bound the quantity 
$$\sum_{i+j\leq k-1}\sup_{x=0}|\partial^i_\xi\partial^j_t\psi| \vspace{-2mm}$$
in terms of $\delta_i,N_i$. For the first, we observe that the prefactor $\xi$ makes $\partial_v$ into a `good' derivative\vspace{-2mm}
$$\xi\partial_v=\tfrac{1}{2}\xi\partial_t+\tfrac{1}{4}\partial_\xi \vspace{-1mm}$$
and moreover turns $\rho_\pm$ into $\sigma_\pm$ in the energy-momentum components. The first term sees first derivatives of $m$, but these are controlled to order $k$ in our assumptions. Hence $\psi$ is composed of terms whose $\|\cdot\|_{C^k_\text{cusp}}$ norm is bounded with respect to $B,C$, and so we once again apply \hyperlink{lem:a.1}{Lemma~A.1}(i)-(ii) repeatedly to estimate $\|\psi\|_{C^k_\text{cusp}}$ with respect to $B,C,\delta_2$ (the $\delta_2$ arising when we estimate the inverse factors $r^{-1}$).\vspace*{2mm} \\
For the second claim, our assumption that $(\Omega^2,r,\bm{u}_\pm,\sigma_\pm)$ obeys $k^\text{th}$ order boundary conditions tell us that $\partial^i_\xi\partial^j_t\psi|_{x=0}$ are given appropriately in terms of data. With $\psi$ involving second derivatives of $r$, as mentioned above, it is essential that the bound only sums over $i+j\leq k-1$, and thus we can indeed estimate this, in particular, in terms of $$N_2=\max\big\{\|r_0\|_{C^{k+2}},\|(\partial_xr)_0\|_{C^{k+1}}\big\}$$
Hence, we are justified in writing
$$\|\textcircled{\footnotesize3}\|_{C^k_\text{cusp}}\leq1+C_k\sum_{i+j\leq k-1}\sup_{x=0}|\partial^i_\xi\partial^j_t\psi|\leq c_3(k,\delta_i,N_i) $$
for small enough $\varepsilon=\varepsilon(B,C)>0$, where $c_3$ is a constant.\\ \\
\ul{Step 4: Combining the estimates:} Having now addressed the three cases \textcircled{\footnotesize1}-\textcircled{\footnotesize3}, we impose the condition that 
$$C>8\big(c_1(k,\delta_i,N_i)+c_2(k,\delta_i,N_i)+c_3(k,\delta_i,N_i)\big)$$
Recall that the same estimates are to be carried out also for $\partial_\xi\overline{\Gamma}_v$, $\partial_t\overline{\Gamma}_u$, $\partial_\xi\overline{\Gamma}_u$ and likewise for derivatives of $\overline{r}$. In each case, a similar condition is to be imposed on $C$. In addition, we need to ensure $C^0$ bounds for $\overline{\Gamma}_t,\overline{\Gamma}_x,\partial_t\overline{r},\partial_x\overline{r}$. To this end, we estimate
\begin{align*}
|\overline{\Gamma}_u(t,\xi)|&\leq |\overline{\Gamma}_u(t,0)|+\varepsilon^{1/2}\|\partial_\xi\overline{\Gamma}_u\|_{C^0}\leq \|(\Gamma_x)_0\|_{C^0}+\tfrac{1}{4}C
\end{align*}
using $\varepsilon\leq 1$ and the estimate below. If we impose that $C>4N_1$, then this gives us
$$\|\overline{\Gamma}_u\|_{C^0},\|\overline{\Gamma}_v\|_{C^0}\leq\tfrac{1}{2}C $$
Together with the analogous condition for $\overline{r}$, we then finally choose a single $C=C(k,\delta_i,N_i)>0$ meeting all of these conditions.\vspace*{2mm} \\
Having chosen such a $C$, we revisit our full estimate for $\partial_t\overline{\Gamma}_v$, which takes the form
\begin{align*}
\|\partial_t\overline{\Gamma}_v\|_{C^k_\text{cusp}}&\leq c_1(k,\delta_i,N_i)+c_2(k,\delta_i,N_i)+c_3(k,\delta_i,N_i)+c(k,\delta_i,N_i,C,B)\varepsilon^{1/2}\\
&\leq \tfrac{1}{8}C+c(k,\delta_i,N_i,C,B)\varepsilon^{1/2}
\end{align*}
and we require of $\varepsilon$ that the last term is at most $\tfrac{1}{8}C$. Revisiting similarly the estimates for $\partial_\xi\overline{\Gamma}_v$, $\partial_t\overline{\Gamma}_u$ etc$.$, we make the same requirement of $\varepsilon$ in each case. Finally, $\varepsilon$ must meet certain conditions, dependent only on $B,C$, in order to apply \hyperlink{lem:3.3}{Lemma~3.3} when controlling \textcircled{\footnotesize3}. We can thus make a single choice of $\varepsilon=\varepsilon(k,\delta_i,N_i,B,C)>0$ meeting all these requirements.\vspace*{2mm} \\
It is then a straightforward matter to recover the intended bounds for $\overline{\Gamma}_t$, $\overline{\Gamma}_x$, $\partial_t\overline{r}$, $\partial_x\overline{r}$. We have, for these choices of $C,\varepsilon$, \vspace{-2mm}
\begin{align*}
\|\overline{\Gamma}_t\|_{C^{k+1}_\text{cusp}},\|\overline{\Gamma}_x\|_{C^{k+1}_\text{cusp}}&\leq\tfrac{1}{2}\big(  \|\overline{\Gamma}_u\|_{C^{k+1}_\text{cusp}}+\|\overline{\Gamma}_v\|_{C^{k+1}_\text{cusp}}\big)\\
&\leq\tfrac{1}{2}\big( \|\partial_t\overline{\Gamma}_u\|_{C^k_\text{cusp}}+\|\partial_\xi\overline{\Gamma}_u\|_{C^k_\text{cusp}}+\|\partial_t\overline{\Gamma}_v\|_{C^k_\text{cusp}}+\|\partial_\xi\overline{\Gamma}_v\|_{C^k_\text{cusp}}+\|\overline{\Gamma}_u\|_{C^0}+\|\overline{\Gamma}_v\|_{C^0}\big)\\
&\leq C\vspace{-2mm}
\end{align*}
and likewise for $\partial_t\overline{r}$, $\partial_x\overline{r}$.\\ \\
\ul{Step 5: Further bounds for $\overline{\Gamma}_t,\overline{\Gamma}_x,\overline{r}$}: It remains only to show that the $C^0$ bounds for $\overline{\Gamma}_t$, $\overline{\Gamma}_x$, $\overline{r}$ can be achieved, which follows easily from what we have already done. We estimate
$$
|\overline{\Gamma}_t(t,\xi)|\leq \int^\xi_0|\partial_\xi\overline{\Gamma}_t|d\xi \leq\varepsilon^{1/2}C\vspace{-2mm}
$$
$$
|\overline{\Gamma}_x(t,\xi)|\geq |(\Gamma_x)_0(t)|-\int^\xi_0|\partial_\xi\overline{\Gamma}_x|d\xi \geq\delta_1-\varepsilon^{1/2}C
$$
$$
\overline{r}(t,x)\geq r_0(t)-\int^x_0|\partial_x\overline{r}|d\xi \geq \delta_2-\varepsilon C
$$
so that reducing $\varepsilon$, if needed, to satisfy 
$\varepsilon^{1/2}C<\tfrac{1}{2}\delta_1$ and $ \varepsilon C<\tfrac{1}{2}\delta_2 $ is sufficient.\\ \\
\ul{Step 6: Propagating $k^\text{th}$ order boundary conditions:} \vspace*{2mm} \\
The equations \hyperlink{eqn:3.3}{(3.3)}-\hyperlink{eqn:3.4}{(3.4)} satisfied by $\log\overline{\Omega}^2$ and $\overline{r}$ are of the form $\square\phi=\psi x^{-1/2}$, which was analysed in \hyperlink{sec:3.1}{Section~3.1}. For this equation, we have the identity (for $k\geq 2$)
\begin{align*}
\partial^{k+2}_\xi\phi=4&(k^2-1)\partial^2_t\partial^{k-2}_\xi\phi-4(k+1)\partial^{k-1}_\xi\psi\\
&+4\xi(2k+1)\partial^2_t\partial^{k-1}_\xi\phi-4\xi\partial^k_\xi\psi+4\xi^2\partial^2_t\partial^k_\xi\phi
\end{align*}
which is derived by differentiating equation \hyperlink{eqn:3.2}{(3.2)} $k-1$ times with $\partial_\xi$. Taking the limit $\xi\to0$, the second line of terms vanishes, leaving 
\begin{equation*}(\partial^{k+2}_\xi\phi)|_{\xi=0}=4(k^2-1)(\partial^2_t\partial^{k-2}_\xi\phi)|_{\xi=0}-4(k+1)(\partial^{k-1}_\xi\psi)|_{\xi=0}\tag{\hypertarget{eqn:3.10}{3.10}} \end{equation*}
In the present case, $\log\overline{\Omega}^2$ satisfies the same equation but with the function $\psi$ given by the expression 
$$\psi=\frac{2\xi}{r^2}\big(\Omega^2-(\partial_xr)^2+(\partial_tr)^2\big)-4\Omega^2(\sigma_++\sigma_-)\vspace{-2mm} $$
We now show by induction on $k$ that the $k^\text{th}$ order conditions are propagated to $\log\overline{\Omega}^2$.\\ \\
\hspace*{0.05\textwidth}\begin{minipage}{0.9\textwidth}\textsc{Base Cases.} The case $k=0$ is just the statement that $\overline{\Gamma}_t|_{\xi=0}=0$, $\overline{\Gamma}_x|_{\xi=0}=(\Gamma_x)_0$, which always holds because we solve for $\overline{\Omega^2}$ with the given $C^1$ data, regardless of what is on the right-hand side of \hyperlink{eqn:3.3}{(3.3)}. For the case $k=1$, evaluating equation \hyperlink{eqn:3.2}{(3.2)} at $\xi=0$ yields
$$\partial^3_\xi\phi|_{\xi=0}=-8\psi|_{\xi=0} $$
Assuming 1st order boundary conditions for $(\Omega^2,r,\bm{u}^t_\pm,\bm{u}^x_\pm,\sigma_\pm)$ is more than enough for $\psi|_{\xi=0}$ to take the correct value, namely, $-8\sigma_0$, so $\log\overline{\Omega^2}$ satisfies $3^\text{rd}$ order limits, and hence $\overline{\Gamma}_x=\tfrac{1}{2}\partial_x\log\overline{\Omega^2}=\tfrac{1}{\xi}\partial_\xi\log\overline{\Omega^2}$ satisfies $1^\text{st}$ order limits.
\\ \\ \end{minipage}
\\
\hspace*{0.05\textwidth}\begin{minipage}{0.9\textwidth}
\textsc{Inductive Step.} If $k\geq2$, we apply instead the formula \hyperlink{eqn:3.10}{(3.10)}. The assumption that $(\Omega^2,r,\bm{u}^t_\pm,\bm{u}^x_\pm,\sigma_\pm)$ satisfies $k^\text{th}$ order boundary conditions is more than enough to show that $\partial^{k-1}_\xi\psi|_{\xi=0}$ coincides with the correct values. And, by the induction hypothesis, the term in $\partial^2_t\partial^{k-2}_\xi\phi$ is also correct. Consequently, $\log\overline{\Omega^2}$ satisfies $(k+2)^\text{nd}$ order limits, so $\overline{\Gamma}_x=\tfrac{1}{2}\partial_x\log\overline{\Omega^2}=\tfrac{1}{\xi}\partial_\xi\log\overline{\Omega^2}$ satisfies $k^\text{th}$ order limits, completing the inductive step.\vspace*{0mm} \\
\end{minipage}\\
So, by induction, $\overline{\Omega}^2$ indeed satisfies $k^\text{th}$ order limits. The argument is the same for $\overline{r}$.
\hfill$\square$\\ \\ \\
\Large\hypertarget{sec:4}{\textbf{4\hspace{4mm}Analysis of the matter equations}}\label{4}\normalsize\\ \\
We now turn to the task of obtaining existence and bounds for the matter equations. In Section~3, it was relatively straightforward to prove the \textit{existence} of iterates for the metric components $\Omega^2,r$. As we will see presently, the proof of existence for iterates of $\bm{u},\sigma$ is more non-trivial. Since the Christoffel symbols $\Gamma_t,\Gamma_x$ are not Lipschitz on $\Sigma_\varepsilon$, geodesic curves must first be constructed by hand, using another iterative scheme (distinct from the one involved in \hyperlink{thm:1.1}{Theorem~1.1}!). Then, after carefully arguing that the curves form a foliation of $\Sigma_\varepsilon$, we finally obtain the desired vector field $\bm{u}$. Solving for the (renormalized) energy density $\sigma$ falls out easily after this task is completed.\\ \\
\hypertarget{pro:4.1}{\textbf{Proposition 4.1}} (Existence and bounds for the matter equations)\textbf{.} \textit{Fix $k\in\mathbb{N}$ and an admissible set of data $(r_0,\sigma_0,(\partial_xr)_0,(\Gamma_x)_0)$, along with associated constants $\delta_i$. Again define $$N_1=\|(\Gamma_x)_0\|_{C^{k+1}[a,b]}\hspace{0.5cm} N_2=\max\big\{\|r_0\|_{C^{k+2}[a,b]},\|(\partial_xr)_0\|_{C^{k+1}[a,b]}\big\}\hspace{0.5cm} N_3=\|\sigma_0\|_{C^k[a,b]}\vspace{2mm} $$ Then given $C>0$, there exist positive constants $\varepsilon=\varepsilon(k,\delta_i, N_i,C)$, $B=B(k,\delta_i,N_i,C)>0$  with the following property. Whenever $\log\Omega^2,r\in C^\infty_\text{cusp}(\Sigma_\varepsilon)$ are cusp-regular, satisfy $k^{\hspace{1pt}th}$ order boundary conditions, and obey the bounds}
$$\|\Gamma_x\|_{C^{k+1}_\text{cusp}},\hspace{4pt}\|\Gamma_t\|_{C^{k+1}_\text{cusp}},\hspace{4pt}\|\partial_xr\|_{C^{k+1}_\text{cusp}},\hspace{4pt}\|\partial_tr\|_{C^{k+1}_\text{cusp}}\leq C$$ $$ \inf_{\Sigma_\varepsilon}|\Gamma_x|\geq\tfrac{1}{2}\delta_1,\hspace{4pt}\sup_{\Sigma_\varepsilon}|\Gamma_t|\leq\tfrac{1}{2}\delta_1,\hspace{4pt}\inf_{\Sigma_\varepsilon} r\geq\tfrac{1}{2}\delta_2$$
\noindent\textit{there exist unique cusp-regular functions $\bm{u}^t_\pm, \bm{u}^x_\pm,\sigma_\pm\in C^\infty_\text{cusp}(\Sigma_\varepsilon)$, satisfying $k^{\hspace{1pt}th}$ order boundary conditions and
such that, writing $\bm{u}=\bm{u}^t\partial_t+\bm{u}^x\partial_x$, the geodesic and conservation equations are satisfied:}
$$\bm{u}^\nu\nabla_\nu\bm{u}^\mu=0 \qquad \nabla_\mu\big(\sigma x^{-1/2}\bm{u}^\mu\big)=0 $$
\textit{with $\bm{u}_\pm|_{x=0}=\partial_t$ and $\bm{u}^x_+,\partial_x\bm{u}^x_+>0$, (resp$.$ $\bm{u}^x_-,\partial_x\bm{u}^x_-<0$) on $\Sigma_\varepsilon$. The domain $\Sigma_\varepsilon$ is foliated by integral curves of $\bm{u}_+$ (resp. $\bm{u}_-$), each having past (resp. future) endpoint on $\{x=0\}\cap\overline{\Sigma}_\varepsilon$. Moreover, we have
$$\tfrac{1}{4}\delta_1\leq|\bm{u}_\pm(\bm{u}^x_\pm)|\leq 4C $$}
\textit{Finally, $\bm{u}_\pm,\sigma_\pm$ satisfy the following estimates.}
$$\|\sigma_\pm\|_{C^k_\text{cusp}},\hspace{1mm}\|\bm{u}_\pm(\sigma_\pm)\|_{C^k_\text{cusp}},\hspace{1mm}\|\bm{u}^t_\pm\|_{C^{k+1}_\text{cusp}},\hspace{1mm}\|\bm{u}^\xi_\pm\|_{C^{k+1}_\text{cusp}}\leq B $$
\textit{The statements above also hold if $\varepsilon$ is reduced further.}\\ \\
This proposition, which we return to prove after first establishing a number of lemmas, combines directly with the results of Section~3 to yield a well-defined iteration scheme for the matter and geometry variables. (See the discussion after \hyperlink{pro:3.2}{Proposition~3.2}.) 
\\ \\ \\
\large\hypertarget{sec:4.1}{\textbf{4.1\hspace{4mm}Estimates for the geodesic equation}}\label{4.1}\normalsize\\ \\
\hypertarget{def:4.1}{\textbf{Definition 4.1 }}(Comoving domain)\textbf{. }\textit{Let $\Theta\in\mathbb{R}^2$ be a subset of the $\tau$-$\chi$ plane. We call $\Theta$ a \ul{comoving domain} if 
\begin{enumerate}
\item $\Theta$ is an open subset of the upper half-plane $\{(\tau,\chi):\tau>0\}$.
\item for any $(\tau_0,\chi_0)\in\Theta$, we have $(0,\tau_0)\times\{\chi_0\}\subset\Theta $
\item the set of $\chi_0$ with $\mathbb{R}\times\{\chi_0\}\cap\Theta\neq\emptyset$ forms an open interval $I$ contained in $(a,b)$
\end{enumerate}
We use the notation $\Theta_\tau$ to denote the intersection $\Theta_\tau=\Theta\cap((0,\tau)\times\mathbb{R})$.} \\ \\
Our first task in obtaining \hyperlink{pro:4.1}{Proposition 4.1} is to solve for the geodesic curves generated by the vector field $\bm{u}$. Each of the ingoing and outgoing species is a one-parameter family of geodesics, specified by functions $t,x:\Theta\to\mathbb{R}$ defined on a suitable comoving domain $\Theta$: for each $\chi_0$, the curve $\tau\mapsto(t(\tau,\chi_0),x(\tau,\chi_0))$ is to be a geodesic in $\Sigma_\varepsilon$. The boundaries $\tau=0$ and $x=0$ are then identified---see Figure below\color{black}.\\ \\
Fix now a comoving domain $\Theta$. Suppose now that we have $\Gamma_t,\Gamma_x$ defined on some $\Sigma_\varepsilon$, with $\Gamma_t,\Gamma_x$ cusp-regular ($\in C^\infty_\text{cusp}(\Sigma_\varepsilon)$) and satisfying
$$\|\Gamma\|:=\|\Gamma_t\|_{C^0(\Sigma_\varepsilon)}+\|\Gamma_x\|_{C^0(\Sigma_\varepsilon)}<\infty\qquad\text{and}\qquad \Gamma_x\leq -\tfrac{1}{2}\delta_1 \vspace{-12mm}$$
\begin{minipage}{0.45\textwidth}
We consider smooth functions $t,x:\Theta\to\mathbb{R}$ with all derivatives extending continuously to $\{\tau=0\}$, and for which the image of the map \vspace{-2mm}
$$
\iota:\Theta\to\Sigma_\varepsilon\qquad
(\tau,\chi)\mapsto (t(\tau,\chi),x(\tau,\chi))
$$
is contained in $\Sigma_\varepsilon$. 
\end{minipage}
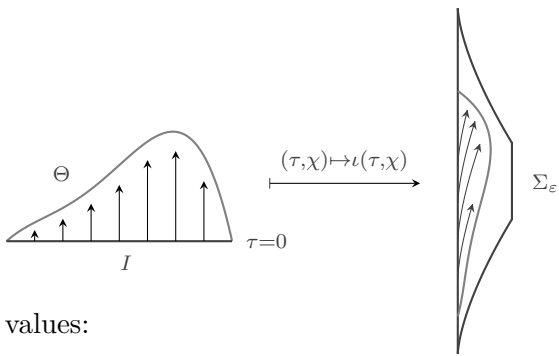
\begin{minipage}{0.55\textwidth}
\vspace{6mm}
\begin{figure}[H]
\begin{center}
\begin{tikzpicture}{scale=0.9}
\draw [darkgray, thick, domain=0:1.8, samples=150] plot ({0.3*(\x)^(1.5)},{\x});
\draw [darkgray, thick, domain=0:1.8, samples=150] plot ({0.3*(\x)^(1.5)},{4.6-\x});
\draw [darkgray, thick] (0.0, 0) -- (0.0, 4.6);
\draw [darkgray, thick] (0.72, 1.78) -- (0.72, 2.82);
\node[darkgray, align=left] at (1.5, 2.27) [anchor = east] {$_{\Sigma_\varepsilon}$};
\draw [gray, thick, domain=0:3, samples=150] plot ({-0.03*(\x)*(\x-3)*(2*(\x)^2-2*(\x)+3)},{(\x+0.5)});
\draw [darkgray, -stealth, domain=0:1, samples=150] plot ({0.2*(\x)^(1.5)},{\x+1});
\draw [darkgray, -stealth, domain=0:1.3, samples=150] plot ({0.2*(\x)^(1.5)},{\x+1.5});
\draw [darkgray, -stealth, domain=0:1.1, samples=150] plot ({0.2*(\x)^(1.5)},{\x+2.0});
\draw [darkgray, -stealth, domain=0:0.75, samples=150] plot ({0.2*(\x)^(1.5)},{\x+2.5});
\node[darkgray,align=left] at (-2.95,1.5) [anchor=west] {$_{\tau=0}$};
\node[darkgray,align=left] at (-4.4,1.45) [anchor=north] {$_{I}$};
\draw [gray, thick, domain=0:3, samples=150] plot ({(\x)-6},{1.5-0.1*(\x)*(\x-3)*(2*(\x)^2-2*(\x)+3)});
\draw [darkgray, thick] (-6, 1.5) -- (-3, 1.5);
\node[darkgray, align=left] at (-5, 2.4) [anchor = east] {$_{\Theta}$};
\draw [-stealth](-2.5,2.27) -- (-0.5,2.27);
\draw [-stealth](-3.375,1.5) -- (-3.375,2.3);
\draw [-stealth](-3.75,1.5) -- (-3.75,2.7);
\draw [-stealth](-4.125,1.5) -- (-4.125,2.58);
\draw [-stealth](-4.5,1.5) -- (-4.5,2.25);
\draw [-stealth](-4.875,1.5) -- (-4.875,2);
\draw [-stealth](-5.25,1.5) -- (-5.25,1.8);
\draw [-stealth](-5.625,1.5) -- (-5.625,1.65);
\draw [darkgray] (-2.5, 2.20) -- (-2.5, 2.34);
\node[darkgray,align=left] at (-1.5,2.3) [anchor=south] {$_{(\tau,\chi)\mapsto \iota(\tau,\chi)}$};
\end{tikzpicture}
\end{center}
\end{figure}
\end{minipage}
\vspace{-8mm} \\
We also suppose the following bounds and initial values:
\begin{equation*}
x|_{\tau=0}=\partial_\tau x|_{\tau=0}=0,\qquad t|_{\tau=0}=\chi\qquad \partial_\tau t|_{\tau=0}=1 \tag{\hypertarget{eqn:4.1}{4.1}}
\end{equation*}
\begin{equation*}
\partial^2_\tau x\geq\tfrac{1}{4}\delta_1\qquad\text{and}\qquad|\partial^2_\tau x|,|\partial^2_\tau t|\leq 2\|\Gamma\|\qquad\text{on }\Theta \tag{\hypertarget{eqn:4.2}{4.2}}
\end{equation*}
We note already that $\xi(\tau,\chi):=x(\tau,\chi)^{1/2}$ is also smooth with extension to $\tau=0$. To see this, we can write 
$$\xi(\tau,\chi)=\tau\bigg(\frac{x(\tau,\chi)}{\tau^2}\bigg)^{1/2} $$
Upper and lower bounds on $\partial^2_\tau x$ entail that $x(\tau,\chi)/\tau^2$ is also bounded above and below. Hence, after applying $(\cdot)^{1/2}$, it remains smooth with extension, since $(\cdot)^{1/2}$ is evaluated strictly away from zero.
\\ \\
As mentioned above, the pair $(t,x)$ represents a one-parameter family of geodesics, or an approximation to one. The next lemma gives estimates for an iteration scheme $(t,x)\mapsto(\overline{t},\overline{x})$ designed to improve such an approximation. Indeed, given $(t,x)$ above, we define $\overline{t},\overline{x}$ by solving
\begin{align*}
\partial^2_\tau\overline{t}&=-\Gamma_t\big((\partial_\tau t)^2+(\partial_\tau x)^2\big)-2\Gamma_x\hspace{1pt}\partial_\tau t\hspace{1pt}\partial_\tau x\\
\partial^2_\tau\overline{x}&=-\Gamma_x\big((\partial_\tau t)^2+(\partial_\tau x)^2\big)-2\Gamma_t\hspace{1pt}\partial_\tau t\hspace{1pt}\partial_\tau x
\end{align*}
subject to 
$$\overline{x}|_{\tau=0}=\partial_\tau \overline{x}|_{\tau=0}=0,\qquad \overline{t}|_{\tau=0}=\chi\qquad \partial_\tau \overline{t}|_{\tau=0}=1 $$
Here, $\Gamma_t=\Gamma_t(t(\tau,\chi),x(\tau,\chi)),\Gamma_x=\Gamma_x(t(\tau,\chi),x(\tau,\chi))$ are evaluated on $\Theta$ using the assumption above that $(t,x)$ maps into $\Sigma_\varepsilon$. Note that mapping into $\Sigma_\varepsilon$ is not assumed (or required) to hold for $(\overline{t},\overline{x})$ in the following lemma---that needs to be proved in applications of the lemma.
\\ \\ \\
\hypertarget{lem:4.2}{\textbf{Lemma 4.2 }}(Estimates for the geodesic equation)\textbf{. }\textit{For some $\tau_1\in(0,1)$ depending only on $\delta_1$, $\|\Gamma_t\|_{C^0}$, $\|\Gamma_x\|_{C^0}$, the bounds \hyperlink{eqn:4.2}{(4.2)} assumed for $\partial^2_\tau t,\partial^2_\tau x$ are recovered for $\overline{t},\overline{x}$ on $\Theta_{\tau_1}$:
$$\partial^2_\tau \overline{x}\geq\tfrac{1}{4}\delta_1\qquad\text{and}\qquad|\partial^2_\tau \overline{x}|,|\partial^2_\tau \overline{t}|\leq 2\|\Gamma\| $$
Moreover, suppose that $(t_n,x_n)^\infty_{n=1}$ is a sequence of smooth functions on $\Theta_{\tau_1}$ obtained by iteration:
$$(t_{n+1},x_{n+1}):=(\overline{t}_n,\overline{x}_n) $$
with $(t_1,x_1)$ satisfying \hyperlink{eqn:4.1}{(4.1)}-\hyperlink{eqn:4.2}{(4.2)}. Then for each $k\in\mathbb{N}_0$, the sequence is bounded in the $C^k$-norm on $\Theta_{\tau_1}$:
$$\sup_n\|t_n\|_{C^k(\Theta_{\tau_1})},\hspace{4pt}\sup_n\|x_n\|_{C^k(\Theta_{\tau_1})}<\infty $$
Finally, if already $(\overline{t},\overline{x})=(t,x)$, with $(t,x)$ satisfying \hyperlink{eqn:4.1}{(4.1)}-\hyperlink{eqn:4.2}{(4.2)} then we have the estimate
\begin{align*}&\|t\|_{C^{k+1}(\Theta_{\tau_1})},\hspace{2pt}\|\partial_\tau t\|_{C^{k+1}(\Theta_{\tau_1})},\hspace{2pt}\|\partial^2_\tau t\|_{C^{k+1}(\Theta_{\tau_1})},\|x\|_{C^{k+1}(\Theta_{\tau_1})},\hspace{2pt}\|\partial_\tau x\|_{C^{k+1}(\Theta_{\tau_1})},\hspace{2pt}\|\partial^2_\tau x\|_{C^{k+1}(\Theta_{\tau_1})},\\
&\qquad\qquad\|\xi\|_{C^{k+1}(\Theta_{\tau_1})},\hspace{2pt}\|\partial_\tau \xi\|_{C^{k+1}(\Theta_{\tau_1})}\leq C(k,\delta_1,\|\Gamma_t\|_{C^{k+1}_\text{cusp}(\Sigma_\varepsilon)},\|\Gamma_x\|_{C^{k+1}_\text{cusp}(\Sigma_\varepsilon)}) \tag{\hypertarget{eqn:4.3}{4.3}}
\end{align*}
where the constant $C>0$ depends only on the indicated quantities.}\\ \\
\textit{Proof.} \ul{Step 1: Bounds for $\partial^2_\tau t,\partial^2_\tau x$.} We assume already that $|\partial^2_\tau t|,|\partial^2_\tau x|\leq 2\|\Gamma\|$, so we begin by integrating once. For all $(\tau,\chi)\in\Theta$,
$$1-2\|\Gamma\|\tau\leq\partial_\tau t(\tau,\chi)\leq 1+2\|\Gamma\|\tau\qquad\text{and}\qquad|\partial_\tau x(\tau,\chi)|\leq 2\|\Gamma\|\tau $$
Now if $\tau_1$ satisfies $\tau_1\leq \tfrac{1}{8}\|\Gamma\|^{-1}$, then we have
\begin{align*}
|\partial^2_\tau\overline{t}|&\leq |\Gamma_t|\big((1+2\|\Gamma\|\tau)^2+4\|\Gamma\|^2\tau^2\big)+2|\Gamma_x|(1+2\|\Gamma\|\tau)\cdot2\|\Gamma\|\tau\\
&<2\|\Gamma\|
\end{align*}
and similarly for $\partial^2_\tau\overline{x}$ (by swapping $\Gamma_t,\Gamma_x$). For the lower bound on $\partial^2_\tau x$, we have simply
\begin{align*}
\partial^2_\tau\overline{x}&\geq \inf|\Gamma_x|(1-2\|\Gamma\|\tau)^2-2\|\Gamma\|(1+2\|\Gamma\|\tau)\cdot2\|\Gamma\|\tau\\
&\geq\tfrac{1}{2}\delta_1(\tfrac{3}{4})^2-4\|\Gamma\|^2\tau(\tfrac{5}{4})
\end{align*}
and so, imposing further that $\tau_1<\tfrac{1}{160}\|\Gamma\|^{-2}\delta_1$, we achieve the bound $\partial^2_\tau\overline{x}\geq\tfrac{1}{4}\delta_1$ on $\Theta_{\tau_1}$. Adjusting if necessary to arrange that $\tau_1<1$, we're done.
\\ \\
\ul{Step 2: Uniform bounds for $(t_n,x_n)$.} \hspace{2mm}We will prove the uniform $\|\cdot\|_{C^k}$ bounds by induction on $k$. It is helpful to formulate the inductive step as follows.\\ \\ 
\hspace*{0.05\textwidth}\begin{minipage}{0.9\textwidth}\textsc{Claim.} \textit{Fix a non-negative integer $k\in\mathbb{N}_0$. Suppose, for some $\tilde{C}>0$, that
\begin{align*}&\|t\|_{C^{k}(\Theta_{\tau_1})},\hspace{2pt}\|\partial_\tau t\|_{C^{k}(\Theta_{\tau_1})},\hspace{2pt}\|\partial^2_\tau t\|_{C^{k}(\Theta_{\tau_1})},\|x\|_{C^{k}(\Theta_{\tau_1})},\hspace{2pt}\|\partial_\tau x\|_{C^{k}(\Theta_{\tau_1})},\hspace{2pt}\|\partial^2_\tau x\|_{C^{k}(\Theta_{\tau_1})}\leq \tilde{C}
\end{align*}
Then there exists a positive constant $C>0$, depending only on $\tilde{C},k,\delta_1,\|\Gamma_t\|_{C^{k+1}_\text{cusp}(\Sigma_\varepsilon)}$ and $\|\Gamma_x\|_{C^{k+1}_\text{cusp}(\Sigma_\varepsilon)}$, with the following property. Let $i+j\in\mathbb{N}_0$ satisfy $i+j=k+1$. If, for all $\tau\in(0,\tau_1)$,
\begin{equation*}|\partial^2_\tau(\partial^i_\tau\partial^j_\chi t)|(\tau,\chi),\hspace{1mm}|\partial^2_\tau(\partial^i_\tau\partial^j_\chi x)|(\tau,\chi)\leq Ce^{C\tau} \tag{\hypertarget{eqn:4.4}{4.4}}\end{equation*} then $\overline{t},\overline{x}$ satisfy these same bounds \hyperlink{eqn:4.4}{(4.4)}. Alternatively, if already $(\overline{t},\overline{x})=(t,x)$, then \hyperlink{eqn:4.4}{(4.4)} holds.\vspace*{2mm}}
\end{minipage}\\ \\
The reason for postulating the upper bound $Ce^{C\tau}$ is that, if we merely bounded above by $C$, then we would have to reduce $\tau_1$ depending on the value of $k$, in order to recover the bound. Having a fixed domain $\Theta_{\tau_1}$ for all $k$ allows the iteration scheme to converge to a smooth limit on a single domain.\\ \\
\begin{mdframed}
\textit{Proof of Claim} (enclosed by the vertical line). Studying $\overline{t}$ first, we apply the derivative $\partial^i_\tau\partial^j_\chi$ to both sides of
$$\partial^2_\tau\overline{t}=-\Gamma_t\big((\partial_\tau t)^2+(\partial_\tau x)^2\big)-2\Gamma_x\hspace{1pt}\partial_\tau t\hspace{1pt}\partial_\tau x$$
Due to the cusp-regularity of $\Gamma_t,\Gamma_x$ (see \hyperlink{def:2.3}{Definition~2.3}), we think of 
$$\Gamma_t=\Gamma_t(t(\tau,\chi),\xi(\tau,\chi))\qquad \Gamma_x=\Gamma_x(t(\tau,\chi),\xi(\tau,\chi)) $$
so that, when we apply the chain rule, we extract $\partial_\xi$-derivatives, which are controlled by $\|\cdot\|_{C^{k+1}_\text{cusp}}$, rather than $\partial_x$. This means that derivatives of $\xi(\tau,\chi)$ appear, which we control by means of \hyperlink{lem:a.4}{Lemmas~A.4} and \hyperlink{lem:a.5}{A.5}. Applying now the Leibniz and chain rule repeatedly, the derivative $\partial^2_\tau(\partial^i_\tau\partial^j_\chi \overline{t})$ becomes a polynomial expression containing the following terms:
\begin{itemize}
\item[$\circ$] $\Gamma_t$, $\Gamma_x$ and cusp-derivatives up to order $k+1$
\item[$\circ$] derivatives $\partial^a_\tau\partial^b_\chi t$ with $0\leq a\leq i+1$, $0\leq b\leq j$
\item[$\circ$] derivatives $\partial^a_\tau\partial^b_\chi x$ with $1\leq a\leq i+1$, $0\leq b\leq j$
\item[$\circ$] derivatives $\partial^a_\tau\partial^b_\chi \xi $ with $0\leq a\leq i$, $0\leq b\leq j$
\end{itemize}
All the terms $\partial^a_\tau\partial^b_\chi t$, $\partial^a_\tau\partial^b_\chi x$ above are bounded above by $\tilde{C}$, except possibly for the highest order derivatives $\partial^i_\tau\partial^j_\chi t$, $\partial^{i+1}_\tau\partial^j_\chi t$, $\partial^i_\tau\partial^j_\chi x$, $\partial^{i+1}_\tau\partial^j_\chi x$. When $i\geq 1$, we can still use $\tilde{C}$ to bound these terms: 
$$|\partial^i_\tau\partial^j_\chi t|=|\partial^{i-1}_\tau\partial^j_\chi(\partial_\tau t)|\leq\|\partial_\tau t\|_{C^k(\Theta)}\leq\tilde{C} $$
$$|\partial^{i+1}_\tau\partial^j_\chi t|=|\partial^{i-1}_\tau\partial^j_\chi(\partial^2_\tau t)|\leq\|\partial^2_\tau t\|_{C^k(\Theta)}\leq\tilde{C} $$
and similarly for $x$. We therefore treat two cases separately:\\ \\
\ul{Case $i\geq 1$.} Applying \hyperlink{lem:a.4}{Lemma~A.4}, we see that every derivative $\partial^a_\tau\partial^b_\chi\xi$ appearing in the polynomial is bounded by a constant depending only on $k,\delta_1,\tilde C$. This includes the top-order derivative $\partial^i_\tau\partial^j_\chi\xi$, because \hyperlink{lem:a.4}{Lemma~A.4} gives us an estimate for $\|\partial_t\xi\|_{C^k(\Theta)}$. It follows that we have
\begin{align*}\big|\partial^2_\tau(\partial^i_\tau\partial^j_\chi \overline{t})\big|(\tau,\chi)&= \Big|\partial^i_\tau\partial^j_\chi\Big[-\Gamma_x\big((\partial_\tau t)^2+(\partial_\tau x)^2\big)-2\Gamma_t\partial_\tau t\partial_\tau x\Big]\Big|\\
&\leq A\big(k,\delta_1,\|\Gamma_t\|_{C^{k+1}_\text{cusp}},\|\Gamma_x\|_{C^{k+1}_\text{cusp}},\tilde C\big) 
\end{align*}
for some constant $A>0$.
Now, whenever $C$ is at least $2A$, we have
$$\big|\partial^2_\tau(\partial^i_\tau\partial^j_\chi \overline{t})\big|(\tau,\chi)\leq \tfrac{1}{2}Ce^{C\tau}<Ce^{C\tau} $$
(The extra factors of 2 are introduced to give room to close a bootstrap argument below.) Repeating the analysis for $\overline{x}$, which is a simple matter of exchanging $\Gamma_t,\Gamma_x$, we obtain one further condition on $C$. Hence the bound \hyperlink{eqn:4.4}{(4.4)} does indeed propagate from $(t,x)$ to $(\overline{t},\overline{x})$ for sufficiently large $C$, depending only on the constants given, concluding the proof of the Claim in this case.\\ \\
\ul{Case $i=0$.} We now need to actually use \hyperlink{eqn:4.4}{(4.4)}. By differentiating initial conditions for $t,x$, we have
$$\partial^{k+1}_\chi t|_{\tau=0}=\partial_\tau\partial^{k+1}_\chi t|_{\tau=0}=\partial^{k+1}_\chi x|_{\tau=0}=\partial_\tau\partial^{k+1}_\chi x|_{\tau=0}=0 $$
So we have 
$$|\partial_\tau\partial^{k+1}_\chi t(\tau,\chi)|\leq\bigg|\int^\tau_0\partial^2_\tau\partial^{k+1}_\chi t(\tilde{\tau},\chi)d\tilde{\tau}\bigg|\leq\int^\tau_0Ce^{C\tilde{\tau}}d\tilde{\tau}\leq e^{C\tau} $$
$$|\partial^{k+1}_\chi t(\tau,\chi)|\leq\bigg|\int^\tau_0(\tau-\tilde{\tau})\partial^2_\tau\partial^{k+1}_\chi t(\tilde{\tau},\chi)d\tilde{\tau}\bigg|\leq\tau_1\int^\tau_0Ce^{C\tilde{\tau}}d\tilde{\tau}\leq e^{C\tau} $$
and similarly for $x$. We underline the importance of being able to remove the $C$ factor by integration.\\ \\ 
As in the previous case, we observe that all terms appearing in the polynomial expression are bounded above by one of $\|\Gamma_t\|_{C^{k+1}_\text{cusp}}$, $\|\Gamma_x\|_{C^{k+1}_\text{cusp}}$, $\tilde{C}$, $\delta_1$ or functions thereof, except this time for the top-order derivatives, which are $\partial^{k+1}_\chi\xi$, $\partial^{k+1}_\chi t$, $\partial_\tau\partial^{k+1}_\chi t$, $\partial^{k+1}_\chi x$, $\partial_\tau\partial^{k+1}_\chi x$. For the derivative $\partial^{k+1}_\chi\xi$, we apply \hyperlink{lem:a.5}{Lemma~A.5}. Since these top-order derivatives appear \textit{linearly} in the polynomial, we can deduce that, in total, we have
$$|\partial^{2}_\tau(\partial^{k+1}_\chi\overline t)|(\tau,\chi)\leq A_1\big(k,\delta_1,\|\Gamma_t\|_{C^{k+1}_\text{cusp}},\|\Gamma_x\|_{C^{k+1}_\text{cusp}},\tilde C\big) e^{C\tau}+A_2\big(k,\delta_1,\|\Gamma_t\|_{C^{k+1}_\text{cusp}},\|\Gamma_x\|_{C^{k+1}_\text{cusp}},\tilde C\big) $$
for constants $A_1,A_2>0$. We now impose that $C>2(A_1+A_2)$, which gives the same conclusion as in the previous case (achieving \hyperlink{eqn:4.4}{(4.4)} with an extra factor of $\tfrac{1}{2}$). Considering $\overline{x}$ similarly, we thus choose a single $C$ which enables the bound \hyperlink{eqn:4.4}{(4.4)} to hold for $\overline{t},\overline{x}$. This concludes the proof of the Claim, except for the final part.\\ \\
If already $(\overline{t},\overline{x})=(t,x)$, then we observe that, with the same $C$ just chosen, we have, in either case $i=0$ or $i\geq1$:
$$|\partial^2_\tau(\partial^i_\tau\partial^j_\chi t)|(0,\chi),\hspace{2pt}|\partial^2_\tau(\partial^i_\tau\partial^j_\chi x)|(0,\chi)\leq A_1+A_2\leq\tfrac{1}{2}C $$
To see this, note that each of the terms $\partial^i_\tau\partial^j_\chi t$, $\partial^{i+1}_\tau\partial^j_\chi t$, $\partial^i_\tau\partial^j_\chi x$, $\partial^{i+1}_\tau\partial^j_\chi x$ vanishes at $\tau=0$, as mentioned above, while in $\partial^i_\tau\partial^j_\chi\xi$, we are allowed to `evaluate at $\tau=0$', according to the last statement in \hyperlink{lem:a.5}{Lemma~A.5}.\\ \\
We now establish \hyperlink{eqn:4.4}{(4.4)} by a bootstrap argument, carried out (for reasons that will become clear momentarily) along each individual geodesic. That is, fix $\chi_0$ such that $\big(\mathbb{R}\times\{\chi_0\}\big)\cap\Theta\neq\emptyset$, and suppose
$$\big(\mathbb{R}\times\{\chi_0\}\big)\cap\Theta_{\tau_1}=(0,\tau_0)\times\{\chi_0\} $$
where possibly $\tau_0=\tau_1$. We claim that $A\subset[0,\tau_0)$ given by
$$A:=\Big\{a\in[0,\tau_1):(\forall\tau\in[0,a])\hspace{3pt}|\partial^2_\tau(\partial^i_\tau\partial^j_\chi t)|(\tau,\chi_0),\hspace{2pt}|\partial^2_\tau(\partial^i_\tau\partial^j_\chi x)|(\tau,\chi_0)\leq Ce^{C\tau}\Big\} $$
is nonempty, open and closed, which would complete the proof. The smoothness of $t,x$ with extension to $\tau=0$ is what enables us to make sense of this at $\tau=0$, and indeed we have shown above that $0\in A$. Closedness follows immediately from continuity of the functions
$$\tau\mapsto\partial^2_\tau(\partial^i_\tau\partial^j_\chi t)(\tau,\chi_0)\qquad \tau\mapsto\partial^2_\tau(\partial^i_\tau\partial^j_\chi x)(\tau,\chi_0)$$
To prove openness, let $a\in A$. We can now apply our earlier bound, and since we left room in our estimate, we actually have
$$|\partial^2_\tau(\partial^i_\tau\partial^j_\chi t)|(a,\chi_0),\hspace{2pt}|\partial^2_\tau(\partial^i_\tau\partial^j_\chi x)|(a,\chi_0)\leq \tfrac{1}{2}Ce^{C\tau} $$
Hence, by continuity again, we deduce $a+\delta\in A$ for some $\delta>0$. Note that we need to work with individual geodesics in order to obtain a positive $\delta$. Hence $A=(0,\tau_0)$, and since this holds for every $\chi_0$, we conclude that \hyperlink{eqn:4.4}{(4.4)} holds on $\Theta_{\tau_1}$.
\end{mdframed}
\vspace{3mm}We now apply the Claim to demonstrate uniform bounds for a sequence $(t_n,x_n)^\infty_{n=1}$. By the calculation in step~1, we have uniform bounds of the form 
$$\|t_n\|_{C^k},\|x_n\|_{C^k},\|\partial_\tau t_n\|_{C^k},\|\partial_\tau x_n\|_{C^k},\|\partial^2_\tau t_n\|_{C^k},\|\partial^2_\tau x_n\|_{C^k}\leq \tilde{C}_k $$
in the case $k=0$. We now show by induction that these bounds hold for every $k$.\\ \\
Indeed, suppose there exists $\tilde{C}_k$ such that, for all $n\in\mathbb{N}$, the above bounds hold. By \hyperlink{lem:4.2}{Lemma~4.2}, we obtain a constant $C=C(k,\delta_1,\|\Gamma_t\|_{C^{k+1}_\text{cusp}},\|\Gamma_x\|_{C^{k+1}_\text{cusp}})>0$ such that, if $i+j=k+1$ and 
$$|\partial^2_\tau(\partial^i_\tau\partial^j_\chi t_n)|(\tau,\chi),\hspace{2pt}|\partial^2_\tau(\partial^i_\tau\partial^j_\chi x_n)|(\tau,\chi)\leq Ce^{C\tau} $$
holds on $(0,\tau_1)\times I$, then it also holds for $t_{n+1},x_{n+1}$. (The new $C$ does not truly depend on $\tilde{C}_k$, since this, inductively, depends only on $k,\delta_1,\|\Gamma_t\|_{C^{k+1}_\text{cusp}},\|\Gamma_x\|_{C^{k+1}_\text{cusp}}$.) After possibly increasing $C$ to make the above true for $t_1,x_1$, we deduce that, for all $n$,
$$\|\partial^2_\tau(\partial^i_\tau\partial^j_\chi t_n)\|_{C^0},\hspace{2pt}\|\partial^2_\tau(\partial^i_\tau\partial^j_\chi x_n)\|_{C^0}\leq Ce^{C\tau_1} $$
In particular, 
$$\|\partial^2_\tau t_n\|_{C^{k+1}}=\|\partial^2_\tau t_n\|_{C^{k}}+\sum_{i+j=k+1}\|\partial^2_\tau(\partial^i_\tau\partial^j_\chi t_n)\|_{C^0} $$
is uniformly bounded.\\ \\
This immediately gives bounds for $\partial^k_\chi\partial_\tau t_n$, $\partial^k_\chi t_n$. Noting, as in the proof of \hyperlink{lem:4.2}{Lemma~4.2}, that these quantities vanish at $\tau=0$, we have
$$\|\partial^k_\chi\partial_\tau t_n\|_{C^0}\leq \tau_1\|\partial^2_\tau t_n\|_{C^k},\qquad \|\partial^k_\chi t_n\|_{C^0}\leq \tfrac{1}{2}\tau^2_1\|\partial^2_\tau t_n\|_{C^k}$$
and hence $\|\partial_\tau t_n\|_{C^{k+1}},\| t_n\|_{C^{k+1}}$ are uniformly bounded via
$$\|\partial_\tau t_n\|_{C^{k+1}}\leq\|\partial^2_\tau t_n\|_{C^{k+1}}+\|\partial_\tau t_n\|_{C^k}+\|\partial^k_\chi\partial_\tau t_n\|_{C^0} $$
$$\| t_n\|_{C^{k+1}}\leq\|\partial_\tau t_n\|_{C^{k+1}}+\|t_n\|_{C^k}+\|\partial^k_\chi t_n\|_{C^0} $$
Identical reasoning applies to $x_n$. We conclude that some $\tilde{C}_{k+1}>0$ indeed exists, and hence by induction, all derivatives of $t_n,x_n$ are uniformly bounded.\\ \\
\ul{Step 3: If already $(\overline{t},\overline{x})=(t,x)$.} We apply the Claim again, choosing a constant $\tilde{C}_k$ inductively in $k$ as above---except we do not need to increase $C$ again this time. This gives bounds for $t,x$, and we use \hyperlink{lem:a.4}{Lemma~A.4} to obtain the bounds for $\xi$. After renaming the final overall constant $C$, we establish \hyperlink{eqn:4.3}{(4.3)}.\hfill $\square$\\ \\ \\
\large\hypertarget{sec:4.2}{\textbf{4.2\hspace{4mm}Obtaining a smooth one-parameter family of geodesic tangents}}\label{4.2}\normalsize
\\ \\
\hypertarget{lem:4.3}{\textbf{Lemma 4.3 }}(Solving on a maximal comoving domain $\Theta$)\textbf{. }\textit{Let $\varepsilon, \delta_1>0$ be positive constants. Then whenever $\Omega^2\in C^\infty_\text{cusp}(\Sigma_\varepsilon)$ satisfies $\Gamma_x\leq -\tfrac{1}{2}\delta_1 $ on $\Sigma_\varepsilon$, there exists a unique comoving domain $\Theta$ and functions $t,x\in C^\infty(\Theta)$ such that, for each $\chi\in(a,b)$, the curve
$$\tau\mapsto (t(\tau,\chi),x(\tau,\chi)) $$
is an inextendible geodesic curve in $\Sigma_\varepsilon$, tangent to $\partial_t$ at $(\chi,0)$ in the limiting sense at $\tau=0$.}
\\ \\
\textit{Proof of Lemma 4.3. }Our strategy will be to apply an iteration scheme, with the previous lemma taking care of the hard work. The main obstacle to overcome is that the comoving domain $\Theta$ is unknown at the start, but we must commit to \textit{some} domain on which the iterates of $t,x$ live. If too large a domain $\Theta$ is chosen, then the curves $\tau\mapsto (t(\tau,\chi),x(\tau,\chi))$ may leave the domain $\Sigma_\varepsilon$ on which $\Omega^2$ is defined. On the other hand, once each geodesic is solved for up to a small proper time, we are now in the interior of $\Sigma_\varepsilon$, on which $\Omega^2$ is smooth. We first obtain, by solving the geodesic equation on small comoving rectangles, a domain $\Theta$ which at least initializes each geodesic, and we may then apply standard theorems on flows of vector fields to extend each geodesic as far as possible.\\ \\
\ul{Step 1: Choose comoving rectangles.} First, fix $\chi_0\in(a,b)$. We begin by observing that we can choose $\tau_0=\tau_0(\varepsilon,\delta_1,\|\Gamma_t\|_{C^0},\|\Gamma_x\|_{C^0},|a-\chi_0|,|b-\chi_0|)>0$ so that any curve $(t(\tau),x(\tau))$ satisfying
$$t(0)=\chi_0,\quad \dot{t}(0)=1,\quad x(0)=\dot{x}(0)=0 $$
$$|\ddot{x}(\tau)|,\hspace{2pt}|\ddot{t}(\tau)|\leq 2(\|\Gamma_t\|_{C^0}+\|\Gamma_x\|_{C^0}),\qquad\ddot{x}(\tau)\geq\tfrac{1}{4}\delta_1 $$
remains in $\Sigma_\varepsilon$ while $\tau\in(0,\tau_0)$. This can be established by elementary calculus, involving some casework for the multiple components of the boundary of $\Sigma_\varepsilon$. We shrink $\tau_0$ even further if necessary to make it less than the $\tau_1=\tau_1(\|\Gamma_t\|_{C^0},\|\Gamma_x\|_{C^0},\delta_1)$ in \hyperlink{lem:4.2}{Lemma~4.2}. Now, given an open interval $I$ whose endpoints lie in $(a,b)$, we can compute $\tau_0$ for the two endpoints, and it follows straightforwardly that the smaller of the two $\tau_0$'s works for all $\chi_0\in I$. Hence fix the domain $(0,\tau_0)\times I$.
\\ \\
\ul{Step 2: Initialising a convergent iteration scheme.} Having chosen $(0,\tau_0)\times I$, define 
$$t_1(\tau,\chi)=\chi+\tau,\qquad x_1(\tau,\chi)=(\|\Gamma_t\|_{C^0}+\|\Gamma_x\|_{C^0})\tau^2$$
This satisfies conditions \hyperlink{eqn:4.1}{(4.1)}-\hyperlink{eqn:4.2}{(4.2)}, and so we can define a sequence $(t_n,x_n)^\infty_{n=1}$ of smooth functions on $(0,\tau_0)\times I$ by setting 
$$(t_{n+1},x_{n+1})=(\overline{t}_n,\overline{x}_n) $$
with $\overline{t}_n,\overline{x}_n$ as described above. Step~1 and \hyperlink{lem:4.2}{Lemma~4.2} together show that this sequence is well-defined, and moreover bounded with respect to the norm $\|\cdot\|_{C^k((0,\tau_0)\times I)}$ for all $k\in\mathbb{N}$.
\\ \\
We now show that, by possibly shrinking $\tau_0$ (relative to $\delta_1,\|\Gamma_t\|_{C^1_\text{cusp}},\|\Gamma_x\|_{C^1_\text{cusp}}$), we have for each $n$,
$$\|\partial^2_\tau t_{n+2}-\partial^2_\tau t_{n+1}\|_{C^0}+\|\partial^2_\tau x_{n+2}-\partial^2_\tau x_{n+1}\|_{C^0}\leq\tfrac{1}{2}\Big(\|\partial^2_\tau t_{n+1}-\partial^2_\tau t_{n}\|_{C^0}+\|\partial^2_\tau x_{n+1}-\partial^2_\tau x_{n}\|_{C^0}\Big) $$
We have
\begin{align*}
\partial^2_\tau t_{n+2}-\partial^2_\tau t_{n+1}&=\Big[-\Gamma_t(t_{n+1},x_{n+1})\big((\partial_\tau t_{n+1})^2+(\partial_\tau x_{n+1})^2\big)-2\Gamma_x(t_{n+1},x_{n+1})\partial_\tau t_{n+1}\partial_\tau x_{n+1}\Big]\\
&\hspace{1cm}-\Big[-\Gamma_t(t_{n},x_{n})\big((\partial_\tau t_{n})^2+(\partial_\tau x_{n})^2\big)-2\Gamma_x(t_{n},x_{n})\partial_\tau t_{n}\partial_\tau x_{n}\Big]
\end{align*}
The tricky terms to control come from the differences between $\Gamma_t,\Gamma_x$. For this, we use the estimate
\begin{align*}|x^{1/2}_{n+1}(\tau)-x^{1/2}_n(\tau)|&=\frac{|x_{n+1}(\tau)-x_n(\tau)|}{|x^{1/2}_{n+1}(\tau)+x^{1/2}_n(\tau)|}\\
&\leq\frac{\tfrac{1}{2}\|\partial^2_\tau x_{n+1}-\partial^2_\tau x_{n}\|_{C^0}\tau^2}{2\big(\tfrac{1}{8}\delta_1\tau^2\big)^{1/2}}\\
&=\tfrac{1}{4}\|\partial^2_\tau x_{n+1}-\partial^2_\tau x_{n}\|_{C^0}\big(\tfrac{1}{8}\delta_1\big)^{-1/2}\cdot\tau
\end{align*}
so that, taking $\Gamma_x$ for example, we have 
\begin{align*}
|\Gamma_x(t_{n+1},x_{n+1})-\Gamma_x(t_{n},x_{n})|&\leq |\Gamma_x(t_{n+1},x_{n+1})-\Gamma_x(t_{n+1},x_{n})|+|\Gamma_x(t_{n+1},x_{n})-\Gamma_x(t_{n},x_{n})|\\
&\leq \sup_{\Sigma_\varepsilon}|\partial_\xi\Gamma_x|\cdot|x^{1/2}_{n+1}-x^{1/2}_{n}|+\sup_{\Sigma_\varepsilon}|\partial_t\Gamma_x|\cdot|t_{n+1}-t_{n}|\\
&\leq \|\Gamma_x\|_{C^1_\text{cusp}}\cdot\Big[\tfrac{1}{4}\|\partial^2_\tau x_{n+1}-\partial^2_\tau x_{n}\|_{C^0}\big(\tfrac{1}{8}\delta_1\big)^{-1/2}\tau+\tfrac{1}{2}\|\partial^2_\tau t_{n+1}-\partial^2_\tau t_n\|_{C^0}\tau^2\Big]\\
&\leq \tau\cdot C(\delta_1,\|\Gamma_x\|_{C^1_\text{cusp}})\Big(\|\partial^2_\tau t_{n+1}-\partial^2_\tau t_{n}\|_{C^0}+\|\partial^2_\tau x_{n+1}-\partial^2_\tau x_{n}\|_{C^0}\Big)
\end{align*}
for some constant $C=C(\delta_1,\|\Gamma_x\|_{C^1_\text{cusp}})$.\\ \\
Note that we have uniform bounds on the terms in $t_n$, $x_n$ themselves:
$$\|x_n\|_{C^0},\|t_n\|_{C^0}\leq (\|\Gamma_t\|_{C^0}+\|\Gamma_x\|_{C^0})\tau^2\leq (\|\Gamma_t\|_{C^0}+\|\Gamma_x\|_{C^0})$$
$$\|\partial_\tau x_n\|_{C^0},\|\partial_\tau t_n\|_{C^0}\leq 2(\|\Gamma_t\|_{C^0}+\|\Gamma_x\|_{C^0})\tau\leq 2(\|\Gamma_t\|_{C^0}+\|\Gamma_x\|_{C^0})$$
Meanwhile, difference in these terms are easier to compute. For example:
$$|\partial_\tau t_{n+1}-\partial_\tau t_{n}|\leq\|\partial^2_\tau t_{n+1}-\partial^2_\tau t_n\|_{C^0}\tau $$
Putting these bounds together, we find that, for some constant $C=C(\delta_1,\|\Gamma_x\|_{C^1_\text{cusp}},\|\Gamma_t\|_{C^1_\text{cusp}})$, we have
\begin{align*}
\|\partial^2_\tau t_{n+2}-\partial^2_\tau t_{n+1}\|_{C^0}+&\|\partial^2_\tau x_{n+2}-\partial^2_\tau x_{n+1}\|_{C^0}\\
&\leq\tau C(\delta_1,\|\Gamma_x\|_{C^1_\text{cusp}},\|\Gamma_t\|_{C^1_\text{cusp}})\Big(\|\partial^2_\tau t_{n+1}-\partial^2_\tau t_{n}\|_{C^0}+\|\partial^2_\tau x_{n+1}-\partial^2_\tau x_{n}\|_{C^0}\Big)
\end{align*}
and hence a small enough choice of $\tau_0$, relative to $\delta_1,\|\Gamma_x\|_{C^1_\text{cusp}},\|\Gamma_t\|_{C^1_\text{cusp}}$ achieves the desired contraction.\\ \\
It follows that the derivatives $\partial^2_\tau t_{n+1}, \partial^2_\tau x_{n+1}$ form a Cauchy sequence for the $\|\cdot\|_{C^0}$ norm, and this extends straightforwardly also to $t_n,x_n$ and the first derivatives $\partial_\tau t_n,\partial_\tau x_n$. Hence $(t_n,x_n)$ converges uniformly to a limit $(t,x)$ defined on $(0,\tau_0)\times I$ which satisfies exactly 
$$\partial^2_\tau t+\Gamma_t\big((\partial_\tau t)^2+(\partial_\tau x)^2\big)+2\Gamma_x\partial_\tau t\partial_\tau x =0$$
$$\partial^2_\tau x+\Gamma_x\big((\partial_\tau t)^2+(\partial_\tau x)^2\big)+2\Gamma_t\partial_\tau t\partial_\tau x =0$$
We know that $t,x$ admits first and second $\partial_\tau$ derivatives, so that in particular this equation makes sense. In fact, the limiting functions are smooth, with extension to $\tau=0$. This follows immediately by a standard application of the Arzela-Ascoli theorem, since the limiting functions $t_n,x_n$ are bounded with respect to each norm $\|\cdot\|_{C^k((0,\tau_0)\times I)}$. Moreover, since all derivatives of $t_n,x_n$ extend continuously to $\tau=0$, this property also holds for $t,x$.\\ \\
\ul{Step 3: Extending the geodesics maximally.} So far, we have obtained a smooth one-parameter family of geodesic curves in $\Sigma_\varepsilon$, parametrized by the open rectangle $(0,\tau_0)\times I$. We would like to maximally extend the geodesics in this family. To this end, note that $\Omega^2$ is smooth on $\Sigma_\varepsilon$, and hence the geodesic spray is a smooth vector field on $T\Sigma_\varepsilon$. By the Fundamental Theorem on Flows (see $[$\hyperlink{Lee13}{Lee13}, Theorem~9.12$]$), there is a unique maximal flow domain $\mathcal{D}\subset\mathbb{R}\times T\Sigma_\varepsilon$ and smooth flow $\Phi:\mathcal{D}\to T\Sigma_\epsilon$ whose integral curves are (lifts of) geodesic curves in $\Sigma_\varepsilon$.\\ \\
We now let $\gamma:I\to T\Sigma_\varepsilon$ be the smooth curve defined by saying that $\gamma(\chi)$ is the vector $$\partial_\tau t\big(\tfrac{1}{2}\tau_0,\chi\big)\partial_\tau+\partial_\tau x\big(\tfrac{1}{2}\tau_0,\chi\big)\partial_x $$
based at the point $(t\big(\tfrac{1}{2}\tau_0,\chi\big),x\big(\tfrac{1}{2}\tau_0,\chi\big))$.\\ \\
The maximal flow domain $\mathcal{D}$ allows us to immediately read off the desired extension of $(0,\tau_0)\times I$. Indeed, let 
$$\Theta_I:=\{(\tau,\chi)\in\mathbb{R}_{>0}\times I:\big(\tau-\tfrac{1}{2}\tau_0,\gamma(\chi)\big)\in\mathcal{D}\} $$
This is the pre-image of an open subset $\mathcal{D}\subset\mathbb{R}\times T\Sigma_\varepsilon$ under a continuous map, so is open. Moreover, $\Theta_I$ contains $(0,\tau_0)\times I$. Now, if $\pi_t,\pi_x$ are the coordinate functionals on $T\Sigma_\varepsilon$ for $t,x$, we extend $t,x\in C^\infty((0,\tau_0)\times I)$ to $\Theta_I$ by setting
$$t(\tau,\chi)=\pi_t\circ\Phi\big(\tau-\tfrac{1}{2}\tau_0,\gamma(\chi)\big),\qquad x(\tau,\chi)=\pi_x\circ\Phi\big(\tau-\tfrac{1}{2}\tau_0,\gamma(\chi)\big)$$
By smoothness of $\gamma$, these $t,x$ are themselves smooth, and by uniqueness of integral curves, they agree on $(0,\tau_0)\times I$ with the $t,x$ defined in Step~2 (because they agree at $\tau=\tfrac{1}{2}\tau_0$). Hence $\Theta_I$ parametrizes a family of smooth \ul{inextendible} geodesics, extending the previously obtained family. Note that the $\Theta_I$ just obtained may be an unbounded subset of $\mathbb{R}^2$---that is, the geodesics may not terminate after a finite parameter interval.
\\ \\
\ul{Step 4: Gluing together all the $\Theta_I$.} For each open interval $I$ with endpoints contained in $(a,b)$, the $\Theta_I$ constructed above, and the $t,x$ defined on it, are unique. Indeed, up to parameter $\tau_0$, the functions $t,x$ arose as fixed points of a contractive self-map, and were hence unique, while their subsequent continuation began in the interior of $\Sigma_\varepsilon$ and was therefore also uniquely determined, including in particular the (possibly unbounded) domain $\Theta_I$ on which they were maximally defined.\\ \\
We now define
$$\Theta:=\cup\{\Theta_I:\text{open interval }I\text{ with endpoints in }(a,b)\} $$
Note that we could not have simply taken $I=(a,b)$ itself, as we would not be able to obtain a positive $\tau_0$. Whenever the domains overlap, the different $t,x$ define on each domain coincide. Indeed, if $\Theta_{I_1}\cap\Theta_{I_2}\neq\emptyset$, then the $t,x$ defined on either domain would be the fixed point of a contractive self-map defined on some small comoving rectangle with base $I_1\cap I_2$. Hence they coincide on this rectangle, and are thereafter uniquely continued as smooth geodesic curves in $\Sigma_\varepsilon$. Hence we finally obtain a single comoving domain $\Theta$ and smooth functions $t,x$ satisfying all the desired properties.~\hfill $\square$
\\ \\ \\
\large\hypertarget{sec:4.3}{\textbf{4.3\hspace{4mm}Forming a foliation of $\bm{\Sigma_\varepsilon}$}}\label{4.3}\normalsize\\ \\
With no further smallness assumption on $\varepsilon$, the smooth one-parameter family obtained above may \textit{a priori} involve geodesics that intersect in $\Sigma_\varepsilon$. Our next task is to obtain (quantitatively) a small enough $\varepsilon$ to ensure that this does not occur, and that in fact the geodesics smoothly foliate $\Sigma_\varepsilon$. It is here that we make essential use of the shape of the domain $\Sigma_\varepsilon$ (see \hyperlink{def:2.2}{Definition~2.2}) to ensure that every point lies on some geodesic, leaving no `voids'.\\ \\
For sufficiently small $\varepsilon$, the derivative of the map $(\tau,\chi)\mapsto (t,\xi)$ is invertible, and the Inverse Function Theorem applies to give a \textit{local} diffeomorphism. Recall that the proof of the Inverse Function Theorem applies the Mean Value Inequality to a suitable function. We would like  to follow the same reasoning to upgrade to a \textit{global} diffeomorphism, but the domain $\Theta$ may be non-convex. The next lemma addresses this difficulty.\\ \\
\hypertarget{lem:4.4}{\textbf{Lemma 4.4. }}\textit{Suppose $\Theta$ is a comoving domain and $\tilde{t},\tilde{\xi}\in C^\infty(\Theta)$ are such that 
$$\iota:\Theta\to\Sigma_\varepsilon\qquad (\tau,\chi)\mapsto (\tilde{t}(\tau,\chi),\tilde{\xi}^2(\tau,\chi)) $$
(that is, $\iota$ maps into $\Sigma_\varepsilon$.) Moreover, suppose that $\tilde{t}(0,\chi)=\chi$, $\tilde{\xi}(0,\chi)=0$ in the limiting sense, that each curve $\chi\mapsto (t(\tau,\chi),\xi^2(\tau,\chi))$ is inextendible, and that the following bounds are satisfied on $\Theta$.
$$|\partial_\tau \tilde{t}|<\tfrac{1}{2},\qquad |\partial_\tau \tilde{\xi}-1|<\tfrac{1}{2},\qquad |\partial_\chi \tilde{t}-1|<\tfrac{1}{2},\qquad |\partial_\chi \tilde{\xi}|<\tfrac{1}{2} $$
Then $\iota$ is injective.}\\ \\
\textit{Proof. }Any comoving domain (see \hyperlink{def:4.1}{Definition~4.1}) has four boundary components: $\{\tau=0\}$, two vertical line segments $\{\chi=\text{const.}\}$ (possibly empty) and, due to openness, the graph $\{(\tau_*(\chi),\chi):\chi\in I\}$ of a positive, lower-semicontinuous function $\tau_*(\chi)$ defined on $I$. Now suppose that $\iota$ fails to be injective. Thus some pair of curves $\iota(\{\chi=\chi_1\})$, $\iota(\{\chi=\chi_2\})$ intersect, with $\chi_1<\chi_2$, say. (Since $\partial_\tau\tilde{\xi}>0$, no curve cannot intersect itself.) By lower-semicontinuity, $\tau_*(\chi)$ attains a (positive) minimum at some $\chi=\chi_\text{min}$ with $\chi_\text{min}\in(\chi_1,\chi_2)$, say. (If $\chi_\text{min}$ equals an endpoint, the argument to follow still goes through.) We now claim that the curves $\iota(\{\chi=\chi_1\})$ and $\iota(\{\chi=\chi_\text{min}\})$ do not intersect, and similarly with $\iota(\{\chi=\chi_2\})$. If this claim is true, then the proof is complete, because by inextendability, $\iota(\{\chi=\chi_\text{min}\})$ disconnects $\Sigma_\varepsilon$, with $\iota(\{\chi=\chi_1\})$ and $\iota(\{\chi=\chi_2\})$ living in distinct connected components, a contradiction.\\ \\
\begin{minipage}{0.75\textwidth}
Suppose then that $\iota(\tau_\text{min},\chi_\text{min})=\iota(\tau_1,\chi_1)$. We now define on $\Theta$
$$\hat{\iota}:(\tau,\chi)\mapsto\big(\tilde{\xi}(\tau,\chi)-\tau,\tilde{t}(\tau,\chi)-\chi\big)$$
and apply the Mean Value Inequality along a suitable path to obtain a contradiction. On the one hand, we have
$$\|\hat{\iota}(\tau_\text{min},\chi_\text{min})-\hat{\iota}(\tau_1,\chi_1)\|_{l_2(\mathbb{R}^2)} =\big(|\tau_\text{min}-\tau_1|^2+|\chi_\text{min}-\chi_1|^2\big)^{1/2}$$
since $\iota(\tau_\text{min},\chi_\text{min})=\iota(\tau_1,\chi_1)$.\\ \\
\end{minipage}
\hfill
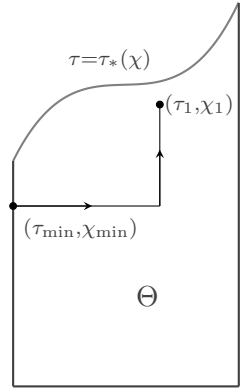
\begin{minipage}{0.23\textwidth}
\begin{figure}[H]
\begin{center}
\vspace{-1cm}
\begin{tikzpicture}[scale=1.5]
\draw [gray, thick, domain=0:2, samples=150] plot ({(\x)},{2+2*(\x)-2*(\x)^2+0.676*(\x)^3});
\draw [darkgray, thick] (0.0, 0) -- (0.0, 2);
\draw [darkgray, thick] (2.0, 0) -- (2.0, 3.4);
\draw [darkgray, thick] (0.0, 0) -- (2.0, 0);
\node[darkgray] at (1.25,2.5) [anchor = west] {$_{(\tau_1,\chi_1)}$};
\node[darkgray] at (0, 1.4) [anchor = west] {$_{(\tau_{\min},\chi_{\min})}$};
\node[darkgray] at (1, 0.8) [anchor = west] {$\Theta$};
\node[darkgray] at (0.4, 2.9) [anchor = west] {$_{\tau=\tau_*(\chi)}$};
\draw [black] (0.0, 1.6) -- (1.3,1.6);
\draw [black] (1.3, 1.6) -- (1.3,2.5);
\draw [-stealth](0,1.6) -- (0.7,1.6);
\draw [-stealth](1.3,1.6) -- (1.3,2.1);
\filldraw[color=black, fill=black](0,1.6) circle (0.03);
\filldraw[color=black, fill=black](1.3,2.5) circle (0.03);
\end{tikzpicture}
\end{center}
\end{figure}
\end{minipage}
\\ \\
On the other hand, by minimality of $\tau_*(\chi_\text{min})$, the straight-line path depicted in the Figure is contained in $\Theta$. Hence, applying the Mean Value Inequality to $\hat{\iota}$, we have
\begin{align*}
\|\hat{\iota}(\tau_\text{min},\chi_\text{min})-\hat{\iota}(\tau_1,\chi_1)\|_{l_2(\mathbb{R}^2)} &\leq \|\hat{\iota}(\tau_\text{min},\chi_\text{min})-\hat{\iota}(\tau_\text{min},\chi_1)\|_{l_2(\mathbb{R}^2)} +\|\hat{\iota}(\tau_\text{min},\chi_1)-\hat{\iota}(\tau_1,\chi_1)\|_{l_2(\mathbb{R}^2)} \\
&\leq \sup_\Theta\|D\hat{\iota}(\partial_\chi)\|_{l^2(\mathbb{R}^2)}|\chi_\text{min}-\chi_1|+\sup_\Theta\|D\hat{\iota}(\partial_\tau)\|_{l^2(\mathbb{R}^2)}|\tau_\text{min}-\tau_1|\\
&<\tfrac{1}{\sqrt{2}}|\chi_\text{min}-\chi_1|+\tfrac{1}{\sqrt{2}}|\tau_\text{min}-\tau_1|\\
&\leq\big(|\tau_\text{min}-\tau_1|^2+|\chi_\text{min}-\chi_1|^2\big)^{1/2}
\end{align*}
which is overall a strict inequality, contradicting the equality above.\hfill$\square$
\\ \\
To apply this result, we need to rescale the coordinates on $\Theta$ and $\Sigma_\varepsilon$ suitably so that the derivative bounds are satisfied. (This is why the previous lemma was written with $\tilde{\hspace{4pt}}$.) This in turn requires a dissection into subdomains, as detailed in the next result.\\ \\
\hypertarget{lem:4.5}{\textbf{Lemma 4.5 }}(Forming a foliation of $\Sigma_\varepsilon$)\textbf{. }\textit{Let $\delta_1,C>0$ be positive constants. There exists $\varepsilon=\varepsilon(\delta_1,C)>0$ with the following property. Whenever $\Omega^2\in C^\infty_\text{cusp}(\Sigma_\varepsilon)$ satisfies
$$\Gamma_x\leq-\tfrac{1}{2}\delta_1,\hspace{4pt}\sup|\Gamma_t|\leq\tfrac{1}{2}\delta_1\quad\text{and}\quad \|\Gamma_x\|_{C^2_\text{cusp}(\Sigma_\varepsilon)}, \|\Gamma_t\|_{C^2_\text{cusp}(\Sigma_\varepsilon)}\leq C $$
then if we take the maximal comoving domain $\Theta$ given by \hyperlink{lem:4.3}{Lemma~4.3}, the map $$
\iota:\Theta\to\Sigma_\varepsilon\qquad
(\tau,\chi)\mapsto (t(\tau,\chi),x(\tau,\chi))
$$
is a diffeomorphism $\Theta\leftrightarrow\Sigma_\varepsilon$ (in particular bijective), with the inextendible geodesics forming a foliation of $\Sigma_\varepsilon$. Moreover, coordinatizing $\Sigma_\varepsilon$ with $(t,\xi)$, the Jacobian determinant $\bm{J}:=\partial_\tau t\partial_\chi\xi-\partial_\tau \xi\partial_\chi t$ satisfies $|\bm{J}|\geq\tfrac{1}{16}\delta^{1/2}_1$ on $\Theta$.}\\ \\
\textit{Proof. }\ul{Step 1: Injectivity.} We would like to rescale the coordinates on $\Theta$ and $\Sigma_\varepsilon$ to arrange that the hypotheses of \hyperlink{lem:4.4}{Lemma~4.4} hold. The most difficult one is that $|\partial_\tau\xi-1|<\tfrac{1}{2}$, since even on the caustic $\tau=0$ itself, the values of $\partial_\tau\xi$ may vary within a large range. We write $$\bm{u}^\xi_0:=\partial_\tau\xi|_{\tau=0}\quad\implies\quad\bm{u}^\xi_0(t)=\big(-\tfrac{1}{2}(\Gamma_x)_0(t)\big)^{1/2} $$
To manage the possible variance of $\bm{u}^\xi_0$, we consider the dissection
$$a=\chi_0<\chi_1<\chi_2<\dots<\chi_N=b $$
where $\chi_i=a+\tfrac{i}{N}(b-a)$. We choose $N$ to be any integer satisfying 
$$N>4(b-a)\sup_{[a,b]}|(\bm{u}^\xi_0)'|\big(\inf_{[a,b]}\bm{u}^\xi_0\big)^{-1} $$
Our claim is that, subject to a further condition on $\varepsilon$ which will now be stated, the restriction of $\iota$ to each subdomain
$$\Theta_i:=\Theta\cap\{\chi_{i-1}<\chi<\chi_{i+1}\} $$
is injective. First, it follows from applying \hyperlink{lem:4.2}{Lemma~4.2} to $\Theta$ (in the $(\overline{t},\overline{x})=(t,x)$ case) that we have bounds on $t,\xi\in C^\infty(\Theta)$ in terms of $C,\delta_1$. Namely, for some $\tilde{C}=\tilde{C}(C,\delta_1)$, we have 
$$\|t\|_{C^2(\Theta)},\hspace{2pt}\|\xi\|_{C^2(\Theta)}\leq \tilde{C} $$
So we now make the assumption 
$$\varepsilon^{1/2}<\tfrac{1}{16}\big(\tfrac{1}{8}\delta_1\big)^{1/2}\tilde{C}^{-1}\min\{1,\inf_{[a,b]}\bm{u}^\xi_0\} $$
Since we have $\partial^2_\tau x>\tfrac{1}{4}\delta_1$ on $\Theta$, integration shows that $x(\tau,\chi)>\tfrac{1}{8}\delta_1\tau^2$, and so the above assumption on $\varepsilon$ means that on $\Theta$ we have $$\tau<\tfrac{1}{16}\tilde{C}^{-1}\min\{1,\inf_{[a,b]}\bm{u}^\xi_0\} $$
Fix now a subdomain $\Theta_i$. Then first, by our choice of $N$, we have that
$$|\bm{u}^\xi_0(\chi)-\bm{u}^\xi_0(\chi_i)|<\bigg(\frac{b-a}{N}\bigg)\sup_{[a,b]}|(\bm{u}^\xi_0)'|<\tfrac{1}{4}\inf_{[a,b]}\bm{u}^\xi_0 $$
and so at each point $(\tau,\chi)$ on $\Theta_i$ we have, by our assumption on $\varepsilon$,
\begin{align*}
|\partial_\tau\xi(\tau,\chi)-\bm{u}^\xi_0(\chi_i)|&\leq |\partial_\tau\xi(\tau,\chi)-\bm{u}^\xi_0(\chi)|+|\bm{u}^\xi_0(\chi)-\bm{u}^\xi_0(\chi_i)|\\
&<\tau\sup_\Theta|\partial^2_\tau\xi|+\tfrac{1}{4}\inf_{[a,b]}\bm{u}^\xi_0 <\tfrac{1}{2}\inf_{[a,b]}\bm{u}^\xi_0
\end{align*}
Since $\partial_\chi t|_{\tau=0}=\partial_\tau t|_{\tau=0}=1$, we have
$$|\partial_\chi t-1|\leq\tau\sup_\Theta|\partial_\tau\partial_\chi t|<\tfrac{1}{2} $$
and
$$|\partial_\tau t|\leq 1+|\partial_\tau t-1|\leq 1+\tau\sup_\Theta|\partial^2_\tau t|<2 $$
Similarly since $\partial_\chi\xi|_{\tau=0}=0$, we have
$$|\partial_\chi\xi(\tau,\chi)|\leq\tau\sup_\Theta|\partial_\tau\partial_\chi\xi|<\tfrac{1}{16}\inf_{[a,b]}\bm{u}^\xi_0 $$
Now that the setup is complete, we rescale our coordinates as follows:
$$\tilde{\tau}=\tau,\quad\tilde{\chi}=\tfrac{1}{4}\chi,\quad\tilde{t}=\tfrac{1}{4}\tau,\quad\tilde{\xi}=\bm{u}^\xi_0(\chi_i)\xi $$
The bounds obtained above give precisely that $\tilde{t}(\tilde{\tau},\tilde{\chi})$, $\tilde{\xi}(\tilde{\tau},\tilde{\chi})$ satisfy
$$|\partial_{\tilde{\tau}}\tilde{t}|<\tfrac{1}{2},\qquad|\partial_{\tilde{\tau}}\tilde{\xi}-1|<\tfrac{1}{2},\qquad|\partial_{\tilde{\chi}}\tilde{t}-1|<\tfrac{1}{2},\qquad|\partial_{\tilde{\chi}}\tilde{\xi}|<\tfrac{1}{2} $$
and so \hyperlink{lem:4.4}{Lemma~4.4} applies, yielding that $\iota|_{\Theta_i}$ is injective.\\ \\
It remains only to ensure that two curves \textit{not} belonging to a common subdomain $\Theta_i$ may not intersect. This is straightforward by further imposing that $$\varepsilon^{1/2}<\big(\tfrac{1}{16}\delta_1\big)^{1/2}\bigg(\frac{b-a}{N}\bigg) $$
with $N$ as selected above. Indeed, if two curves do not belong to a common domain $\Theta_i$, then their initial $t$-values are at least $\tfrac{2}{N}(b-a)$ apart. However, in Step~1 we showed that each curve satisfies $d\xi/dt>(\tfrac{1}{16}\delta_1)^{1/2}$, so the condition above on $\varepsilon$ makes the domain $\Sigma_\varepsilon$ too narrow for any such pair of curves to intersect.
\\ \\
\ul{Step 2: Lower bounds on $|\bm{J}|$.} Using the bounds we established above, we have immediately that
\begin{align*}|\bm{J}(\tau,\chi)|&\geq |\partial_\tau\xi||\partial_\chi t|-|\partial_\tau t||\partial_\chi\xi| \\
&\geq \tfrac{1}{2}\inf_{[a,b]}\bm{u}^\xi_0\cdot\tfrac{1}{2}-2\cdot\tfrac{1}{16}\inf_{[a,b]}\bm{u}^\xi_0\\
&=\tfrac{1}{8}\inf_{[a,b]}(-\tfrac{1}{2}(\Gamma_x))_0)^{1/2}>\tfrac{1}{16}\delta^{1/2}_1
\end{align*}
on $\Theta$, as required.\\ \\
\ul{Step 3: Lower bounds on $d\xi/dt$.} We will now show that the shape of $\Sigma_\varepsilon$ is correctly chosen. We first impose that $\varepsilon<\tfrac{1}{4}\delta_1\tau^2_1$, where $\tau_1=\tau_1(\delta_1,\|\Gamma_t\|_{C^0},\|\Gamma_x\|_{C^0})$ is as in \hyperlink{lem:4.2}{Lemma~4.2}. This $\tau_1$ itself was chosen small enough so that $$|\partial_\tau x|\leq\tfrac{1}{4}\quad\text{and}\quad \partial_\tau t\geq\tfrac{3}{4},\quad\text{so}\quad\bigg|\frac{dx}{dt}\bigg|=\bigg|\frac{\partial_\tau x}{\partial_\tau t}\bigg|\leq\tfrac{1}{2}$$
This in turn gives us control of $dx^2/dt^2$. Using the geodesic equations, we have the identity
$$\frac{dx^2}{dt^2}=\bigg(-\Gamma_x-\bigg(\frac{dx}{dt}\bigg)^2\Gamma_t\bigg)\bigg(1-\bigg(\frac{dx}{dt}\bigg)^2\bigg) $$
and so the assumptions on $\Gamma_t$, $\Gamma_x$ give the lower bound
$$\frac{d^2x}{dt^2}\geq\big(\tfrac{1}{2}\delta_1-\tfrac{1}{8}\delta_1\big)\big(1-\big(\tfrac{1}{2}\big)^2\big)>\tfrac{1}{8}\delta_1 $$
Differentiating $\xi^2$ twice, we now have that
$$\frac{d}{dt}\bigg(2\xi\frac{d\xi}{dt}\bigg)>\tfrac{1}{8}\delta_1,\quad\text{so}\quad 2\xi\frac{d\xi}{dt}>\tfrac{1}{8}\delta_1 t $$
Meanwhile we have $x(t)>\tfrac{1}{16}\delta_1 t^2$, so it follows that
$$\frac{d\xi}{dt}>\xi^{-1}\big(\tfrac{1}{16}\delta_1 t\big)>\big(\tfrac{1}{16}\delta_1 t^2\big)^{-1/2}\big(\tfrac{1}{16}\delta_1 t\big)=\big(\tfrac{1}{16}\delta_1\big)^{1/2} $$
This should be compared with the quadratic curves which bound $\Sigma_\varepsilon$: in $(t,\xi)$ coordinates, they satisfy exactly $d\xi/dt=\big(\tfrac{1}{16}\delta_1\big)^{1/2}$.\\ \\
\ul{Step 4: Surjectivity.} Showing surjectivity is now geometrically intuitive. Given any point $(t_0,\xi_0)$ in $\Sigma_\varepsilon$, let $\gamma$ be a smooth curve in $\Sigma_\varepsilon$ of the form $t\mapsto (t,\xi_*(t))$, with $\xi_*(t)$ defined on $(a,b)$ and satisfying $d\xi_*/dt<(\tfrac{1}{16}\delta_1)^{1/2}$. In particular, such a curve joins the point $(0,a)$ and $(0,b)$. It then follows from the fact that the geodesics are inextendible, terminate on the part of the boundary with $\xi>0$, and satisfy $d\xi/dt>(\tfrac{1}{16}\delta_1)^{1/2}$, that each one intersects $\gamma$ precisely once, at time $\tau=\tau_*(\chi)$, say.
\vspace*{2mm} \\
\begin{minipage}{0.5\textwidth}
By smoothness of $\iota$, and the invertibility of $D\iota$ everywhere (see Step~2), the Pre-Image Theorem (see [\hyperlink{Lee13}{Lee13}, Theorem~5.12]) entails that $\iota^{-1}(\gamma)$ is a smooth curve in $\Theta$. In particular, $\tau_*=\tau_*(\chi)$ is continuous. Now $\chi\mapsto t(\tau_*(\chi),\chi)$ varies continuously between $a$ and $b$, so by the Intermediate Value Theorem, takes the value $t_0$ somewhere, say $\chi=\chi_0$. So the geodesic labelled $\chi_0$ passes through $(t_0,\xi_0)$. That is, $(t_0,\xi_0)$ lies in the image of $\iota$.\hfill$\square$
\end{minipage}
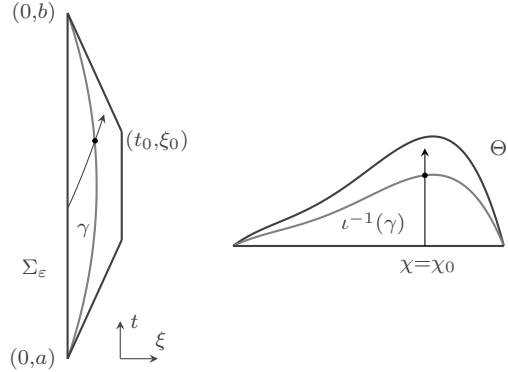
\begin{minipage}{0.5\textwidth}
\vspace{-2mm}
\begin{figure}[H]
\begin{center}
\begin{tikzpicture}{scale=0.9}
\draw [gray, thick, domain=0:4.6, samples=150] plot ({0.073*(4.6*\x)-0.073*(\x)^(2)},{\x});
\draw [darkgray, thick] (0.0, 0) -- (0.0, 4.6);
\draw [darkgray, thick] (0.72, 1.58) -- (0.72, 3.02);
\draw [darkgray, thick] (0,0) -- (0.72, 1.58);
\draw [darkgray, thick] (0,4.6) -- (0.72, 3.02);
\node[darkgray, align=left] at (-0.1, 1.2) [anchor = east] {$_{\Sigma_\varepsilon}$};
\node[darkgray, align=left] at (0.45, 1.7) [anchor = east] {$_{\gamma}$};
\draw [darkgray, -stealth, domain=0:1.25, samples=150] plot ({0.5*(\x)-0.1*(\x)^(1.5)},{\x+2.0});
\draw [darkgray, -stealth] (0.7,0) -- (1.2,0);
\draw [darkgray, -stealth] (0.7,0) -- (0.7,0.5);
\filldraw[color=black, fill=black](0.365,2.9) circle (0.03);
\node[darkgray,align=left] at (3.5,1.8) [anchor=west] {$_{\iota^{-1}(\gamma)}$};
\node[darkgray,align=left] at (0.7,0.5) [anchor=west] {$_{t}$};
\node[darkgray,align=left] at (1.25,0.5) [anchor=north] {$_{\xi}$};
\node[darkgray, align=left] at (4.8, 1.45) [anchor = north] {$_{\chi=\chi_0}$};
\draw [darkgray, thick, domain=0:3, samples=150] plot ({1.2*(\x)+2.2},{1.5-0.1*(\x)*(\x-3)*(2*(\x)^2-2*(\x)+3)});
\draw [gray, thick, domain=0:3, samples=150] plot ({1.2*(\x)+2.2},{1.5+0.65*(-0.1*(\x)*(\x-3)*(2*(\x)^2-2*(\x)+3))});
\filldraw[color=black, fill=black](4.75,2.44) circle (0.03);
\draw [darkgray, thick] (2.2, 1.5) -- (5.8, 1.5);
\node[darkgray, align=left] at (6.0, 2.8) [anchor = east] {$_{\Theta}$};
\node[darkgray, align=left] at (0.0, 4.6) [anchor = east] {$_{(0,b)}$};
\node[darkgray, align=left] at (0.0, 0) [anchor = east] {$_{(0,a)}$};
\draw [-stealth](4.75,1.5) -- (4.75,2.8);
\node[darkgray,align=left] at (1.2,2.62) [anchor=south] {$_{(t_0,\xi_0)}$};
\end{tikzpicture}
\end{center}
\end{figure}
\end{minipage}\\ \\ \\
\large\hypertarget{sec:4.4}
{\textbf{4.4\hspace{4mm}Existence and bounds for $\bm{\sigma}$}}\label{4.4}\normalsize\\ \\
\hypertarget{lem:4.6}{\textbf{Lemma 4.6}} (Comparing comoving- and cusp-derivatives)\textbf{.} \textit{Let $\Theta$ be a comoving domain, and  let $t,\xi\in C^\infty(\Theta)$. Moreover, suppose the map$$
\iota:\Theta\to\Sigma_\varepsilon\qquad
(\tau,\chi)\mapsto (t(\tau,\chi),\xi^2(\tau,\chi))
$$
is a diffeomorphism, with the Jacobian determinant
$\bm{J}:=\partial_\tau t\hspace{1pt}\partial_\chi\xi-\partial_\tau \xi\hspace{1pt}\partial_\chi t$ 
satisfying $|\bm{J}|\geq \tfrac{1}{16}\delta_1^{1/2}$ on $\Theta$. Then for each $k\in\mathbb{N}_0$, there exists $C=C(k,\delta_1,\|t\|_{C^k(\Theta)},\|\xi\|_{C^k(\Theta)})>0$ such that, for all $f\in$~$C^\infty_\text{cusp}(\Sigma_\varepsilon)$, the inequality holds}
$$\|f\|_{C^k_\text{cusp}(\Sigma_\varepsilon)}\leq C\|f\|_{C^k(\Theta)} $$
\textit{whenever the norms exist. (Here, the latter quantity is really $\|\iota^*f\|_{C^k(\Theta)}$, but we suppress the notation $\iota^*f$, viewing $(\tau,\chi)$, $(t,\xi)$ as two choices of smooth chart on $\Sigma_\varepsilon$.)}\\ \\
\textit{Proof. }The coordinate bases vectors of the two smooth charts $(\tau,\chi)$, $(t,\xi)$ on $\Sigma_\varepsilon$ are related by 
$$\partial_\tau=\partial_\tau t\hspace{2pt}\partial_t+\partial_\tau \xi\hspace{2pt}\partial_\xi \qquad \partial_\chi=\partial_\chi t\hspace{2pt}\partial_t+\partial_\chi \xi\hspace{2pt}\partial_\xi $$
After a $2\times 2$ matrix inversion, we have
$$\partial_t=\bm{J}^{-1}\big(\partial_\chi\xi\hspace{2pt}\partial_\tau-\partial_\chi\xi\hspace{2pt}\partial_\chi\big) \qquad \partial_\xi=\bm{J}^{-1}\big(-\partial_\chi t\hspace{2pt}\partial_\tau+\partial_\tau t\hspace{2pt}\partial_\chi\big)$$
Now suppose indices $i,j$ are chosen with $i+j\leq k$. We have
$$\partial^i_t\partial^j_\xi f=\Big[\bm{J}^{-1}\big(\partial_\chi\xi\hspace{2pt}\partial_\tau-\partial_\chi\xi\hspace{2pt}\partial_\chi\big)\Big]^i\Big[\bm{J}^{-1}\big(-\partial_\chi t\hspace{2pt}\partial_\tau+\partial_\tau t\hspace{2pt}\partial_\chi\big)\Big]^jf $$
which expands, via the Leibniz rule, to a polynomial expression containing derivatives of $t,\xi,f$ (of order $\leq k$) and $\bm{J}^{-1}$ (of order $\leq k-1$). Moreover, the derivatives of $f$ appear linearly in the polynomial, while each derivative of $\bm{J}^{-1}$ takes the form 
$$\partial^a_\tau\partial^b_\chi(\bm{J}^{-1})=\bm{J}^{-(a+b+1)}\times\{\text{polynomial in order }\leq (a+b+1)\text{ derivatives of }t,\xi\} $$
So $\partial^i_t\partial^j_\chi f$ is really a polynomial in $\bm{J}^{-1}$, and order $\leq k$ derivatives of $t,\xi,f$. Hence, using $|\bm{J}|\geq \tfrac{1}{4}\delta^{1/2}_1$, and bounding all derivatives of $t,\xi$ by $\|t\|_{C^k(\Theta))}$, $\|\xi\|_{C^k(\Theta))}$, we finally sum over $i+j\leq k$ to obtain a constant
$C=C(k,\delta_1,\|t\|_{C^k(\Theta))}, \|\xi\|_{C^k(\Theta))})>0$ with the stated property.\hfill$\square$
\\ \\
\hypertarget{lem:4.7}{\textbf{Lemma 4.7}} (Existence and bounds for $\sigma$)\textbf{.} \textit{Fix $k\in\mathbb{N}$ and an admissible set of data $(r_0,\sigma_0$, $(\partial_xr)_0,(\Gamma_x)_0)$, along with associated constants $\delta_i,N_i$. Let $\log\Omega^2,r\in C^\infty_\text{cusp}(\Sigma_\varepsilon)$ 
satisfy $k^{\hspace{1pt}th}$ order limits at $x=0$, and 
obey the bounds
$$\|\Gamma_x\|_{C^{k+1}_\text{cusp}},\hspace{4pt}\|\Gamma_t\|_{C^{k+1}_\text{cusp}},\hspace{4pt}\|\partial_xr\|_{C^{k+1}_\text{cusp}},\hspace{4pt}\|\partial_tr\|_{C^{k+1}_\text{cusp}}\leq C \qquad \textit{and}\qquad \Gamma_x\leq-\tfrac{1}{2}\delta_1,\hspace{4pt}\inf r\geq\tfrac{1}{2}\delta_2$$
Let $\Theta$ be a comoving domain, and  let $t,\xi\in C^\infty(\Theta)$ satisfy $k^{\hspace{1pt}th}$ order limits at $\tau=0$. Moreover, suppose
\begin{enumerate}
\item The map $(\tau,\chi)\mapsto (t(\tau,\chi),\xi(\tau,\chi)^2)$ is a diffeomorphism $\Theta\to\Sigma_\varepsilon$, and the Jacobian determinant
$\bm{J}:=\partial_\tau t\hspace{1pt}\partial_\chi\xi-\partial_\tau \xi\hspace{1pt}\partial_\chi t $
satisfies $$|\bm{J}|\geq \tfrac{1}{16}\delta_1^{1/2} \quad\text{on }\Theta\vspace{-2mm}$$
\item $t,\xi$ satisfy the bounds
$$\|t\|_{C^{k+1}(\Theta)},\hspace{4pt}\|\partial_\tau t\|_{C^{k+1}(\Theta)},\hspace{4pt}\|\xi\|_{C^{k+1}(\Theta)},\hspace{4pt}\|\partial_\tau\xi\|_{C^{k+1}(\Theta)} \leq B$$
\end{enumerate}
Then there exists a unique $\sigma\in C^\infty(\Theta)$ satisfying}
\begin{equation*}\partial_\tau\big(\sigma \bm{J}\hspace{1pt}\Omega^2 r^2\big)=0,\qquad\sigma\big|_{\tau=0}=\sigma_0\tag{\hypertarget{eqn:4.5}{4.5}}
\end{equation*}
\textit{where $\Omega^2,r$ are understood to be defined on $\Theta$ through the identification $\Theta\leftrightarrow\Sigma_\varepsilon$. Finally, under the same identification, we have the bounds}
$$\|\sigma\|_{C^k_\text{cusp}(\Sigma_\varepsilon)}, \|\partial_\tau\sigma\|_{C^k_\text{cusp}(\Sigma_\varepsilon)}\leq\tilde{B} $$
\textit{where $\tilde{B}=\tilde{B}(k,\delta_i,N_i,C,B)>0$ is a constant.}
\\ \\
\textit{Proof. }For every $(\tau,\chi)\in\Theta$, the line segment $(0,\tau]\times\{\chi\}$ is contained in $\Theta$. Integrating equation \hyperlink{eqn:4.5}{(4.5)} along this line segment, we immediately obtain 
\begin{equation*}
\sigma(\tau,\chi)|\bm{J}(\tau,\chi)|\hspace{2pt}\Omega^2(\tau,\chi)\hspace{2pt}r(\tau,\chi)^2=\sigma_0(\chi)\bm{u}^\xi_0(\chi)r^2_0(\chi)
\end{equation*}
We can now use lower bounds on $\bm{J}$ (which is everywhere negative) to safely rearrange
\begin{equation*}
\sigma(\tau,\chi)=|\bm{J}(\tau,\chi)|^{-1}\exp\big(-2\log\Omega^2(\tau,\chi)\big)r(\tau,\chi)^{-2}\sigma_0(\chi)\bm{u}^\xi_0(\chi)r^2_0(\chi) \tag{\hypertarget{eqn:4.6}{4.6}}
\end{equation*}
All the constituent factors are well-defined and smooth. So existence, uniqueness and smoothness of $\sigma$ is established. We turn to writing down bounds for $\sigma$, $\partial_\tau\sigma$.\\ \\
For $\sigma$, we repeatedly apply \hyperlink{lem:a.1}{Lemma~A.1}(i)-(ii) to write $\|\sigma\|_{C^k_\text{cusp}}$ in terms of the norms of its factors, and to handle compositions of functions. We also apply \hyperlink{lem:4.6}{Lemma~4.6} to convert $\|\cdot\|_{C^k_\text{cusp}}$ into $\|\cdot\|_{C^k(\Theta)}$ for everything except $\Omega^2,r$ terms which naturally live on $\Sigma_\varepsilon$. The resulting combination of constants we write as $C=C(k,\delta_1,\|t\|_{C^k(\Theta)},\|\xi\|_{C^k(\Theta)})>0$. Altogether we have
\begin{align*}
\|\sigma\|_{C^k_\text{cusp}}\leq&\hspace{2pt}C(k,\delta_1,\|t\|_{C^k(\Theta)},\|\xi\|_{C^k(\Theta)})\times\|(\cdot)^{-1}\|_{C^k\big[\tfrac{1}{16}\delta^{1/2},\infty\big)}\\
&\times\Big(1+\|\bm{J}\|_{C^k(\Theta)}\Big)^k\times\|\exp(-2(\cdot))\|_{C^k[-\|\log\Omega^2\|_{C^0},\|\log\Omega^2\|_{C^0}]}\times\Big(1+\|\log\Omega^2\|_{C^k_\text{cusp}}\Big)^k\\
&\times\|(\cdot)^{-2}\|_{C^k\big[\tfrac{1}{2}\delta_2,\infty\big)}\times\Big(1+\|r\|_{C^k_\text{cusp}}\Big)^k\times \|\sigma_0\|_{C^k[a,b]}\times \|u^\xi_0\|_{C^k[a,b]}\times \|r_0\|^2_{C^k[a,b]}
\end{align*}
Now note that 
\begin{align*}
\|\bm{J}\|_{C^k(\Theta)}&\lesssim_k \|\partial_\tau t\|_{C^k(\Theta)}\|\partial_\chi \xi\|_{C^k(\Theta)}+\|\partial_\chi t\|_{C^k(\Theta)}\|\partial_\tau \xi\|_{C^k(\Theta)}\\
&\leq 2\|t\|_{C^{k+1}(\Theta)}\|\xi\|_{C^{k+1}(\Theta)}\leq 2B^2
\end{align*}
and $\|\log\Omega^2\|_{C^0}\leq\varepsilon\|\Gamma_x\|_{C^0}\leq C$, say, so that we have a bound like
$$\|\exp(-2(\cdot))\|_{C^k[-\|\log\Omega^2\|_{C^0},\|\log\Omega^2\|_{C^0}]}\leq 2^{k+1}e^C $$
Everything else is straightforwardly bounded through our constants. Hence a suitable algebraic combination $\tilde{B}=\tilde{B}(k,\delta_i,N_i,C,B)>0$ offers
$$\|\sigma\|_{C^k_\text{cusp}}\leq \tilde{B} $$
For $\partial_\tau\sigma$, we differentiate \hyperlink{eqn:4.6}{(4.6)}, apply the Leibniz rule and rearrange, yielding
\begin{align*}
\partial_\tau\sigma&=-\sigma\big[\partial_\tau\log\Omega^2+\bm{J}^{-1}\partial_\tau \bm{J}+2r^{-1}\partial_\tau r\big]\\
&=-\sigma\big[\Gamma_t\partial_\tau t+2\xi\Gamma_x\partial_\tau\xi+ \bm{J}^{-1}\partial_\tau \bm{J}+2r^{-1}\partial_\tau t\partial_t r+4\xi r^{-1}\partial_\tau\xi\partial_x r\big]
\end{align*}
The estimates are identical, applying \hyperlink{lem:a.1}{Lemma~A.1}(i)-(ii) to bound $\|\partial_\tau\sigma\|_{C^k_\text{cusp}}$ in terms of the norms of the quantities on the right-hand side. We apply \hyperlink{lem:4.4}{Lemma~4.4} to convert $\|\cdot\|_{C^k_\text{cusp}}$ into $\|\cdot\|_{C^k(\Theta)}$ for $\bm{J},t,\xi$ terms, but not $\Omega^2,r$ terms, as before. The only additional information used is the extra $\partial_\tau$-derivative for $t,\xi$. For example,
\begin{align*}
\|\partial_\tau \bm{J}\|_{C^k_\text{cusp}}&\lesssim_k \|\partial_\tau t\|_{C^{k+1}(\Theta)}\|\xi\|_{C^{k+1}(\Theta)}+\|\partial_\tau\xi \|_{C^{k+1}(\Theta)}\|t\|_{C^{k+1}(\Theta)}\\
&\leq4B^2
\end{align*}
Hence, after possibly increasing $\tilde{B}=\tilde{B}(k,\delta_i,N_i,C,B)>0$, we conclude
$$\|\sigma\|_{C^k_\text{cusp}(\Sigma_\varepsilon)},\hspace{4pt}\|\partial_\tau\sigma\|_{C^k_\text{cusp}(\Sigma_\varepsilon)}\leq \tilde{B} \vspace{-0.8cm}$$
\hfill$\square$\\ \\ \\
\large\hypertarget{sec:4.5}{\textbf{4.5\hspace{4mm}Proof of Proposition 4.1}}\label{4.5}\normalsize\\ \\
We are now in a position to combine all the foregoing lemmas to complete the proof of \hyperlink{pro:4.1}{Proposition~4.1}.\\ \\
\textit{Proof of Proposition 4.1.} By \hyperlink{lem:4.5}{Lemma~4.5}, there exists $\varepsilon=\varepsilon(\delta_1,\|\Gamma_t\|_{C^2_\text{cusp}},\|\Gamma_x\|_{C^2_\text{cusp}})>0$ and a diffeomorphism
$$\iota:\Theta\to\Sigma_\varepsilon\qquad(\tau,\chi)\mapsto (t(\tau,\chi),x(\tau,\chi))$$
with the property that, for each $\chi\in(a,b)$, the curve $\tau\mapsto(t(\tau,\chi),x(\tau,\chi))$ is an inextendible geodesic curve, tangent to $\partial_t$ at $(t,x)=(\chi,0)$ in the limiting sense as $\tau\to0$. The pushforward $\bm{u}=\iota_*\partial_\tau$ is a smooth unit timelike vector field on $\Sigma_\varepsilon$ with
$$\tfrac{1}{4}\delta_1\leq\bm{u}(\bm{u}^x)\leq 2(\|\Gamma_t\|_{C^0}+\|\Gamma_x\|_{C^0})\leq 4C $$
Moreover the Jacobian determinant $\bm{J}$ satisfies $|\bm{J}|\geq\tfrac{1}{16}\delta^{1/2}_1$. We will not need to reduce $\varepsilon$ further, and we fix it now.\\ \\
Next, plugging these $t,x$ into \hyperlink{lem:4.2}{Lemma~4.2} (in the $(\overline{t},\overline{x})=(t,x)$ case), we have bounds
\begin{align*}
\|t\|_{C^{k+1}(\Theta)}&,\hspace{2pt}\|\partial_\tau t\|_{C^{k+1}(\Theta)},\hspace{2pt},\|\xi\|_{C^{k+1}(\Theta)},\hspace{2pt}\|\partial_\tau \xi\|_{C^{k+1}(\Theta)}\leq B(k,\delta_1,\|\Gamma_t\|_{C^{k+1}_\text{cusp}},\|\Gamma_x\|_{C^{k+1}_\text{cusp}})
\end{align*}
We proceed to consider $\sigma$. Applying \hyperlink{lem:4.7}{Lemma~4.7} yields a smooth $\sigma\in C^\infty(\Theta)$. When interpreted on $\Sigma_\varepsilon$, the defining equation \hyperlink{eqn:4.5}{(4.5)} says that
$$\nabla_\mu(\sigma x^{-1/2}\bm{u}^\mu)=0 $$
We moreover obtain bounds, which after renaming $B$, read
$$\|\sigma\|_{C^k_\text{cusp}},\|\bm{u}(\sigma)\|_{C^k_\text{cusp}}\leq B(k,\delta_1,N_i,C) $$
Smoothness on $\Theta$ (with extension to $\tau=0$) is equivalent to cusp-regularity on $\Sigma_\varepsilon$. So the collection of variables obtained above, which we now rename to $(\bm{u}^t_+,\bm{u}^x_+,\sigma_+)$, belongs to $C^\infty_\text{cusp}$. We also check that $\partial_x\bm{u}^x>0$:
$$\bm{u}^x\partial_x\bm{u}^x\equiv \bm{u}(\bm{u}^x)-\bm{u}^t\partial_t\bm{u}^x\geq \tfrac{1}{4}\delta_1-B\varepsilon^{1/2}\|\partial_t\partial_\xi\bm{u}^x\|_{C^0}\geq \tfrac{1}{4}\delta_1-B^2\varepsilon^{1/2} $$
and the final quantity is positive for sufficiently small $\varepsilon>0$ relative to $B,\delta_1$. Since $\bm{u}^x>0$, we conclude that, for such $\varepsilon$, $\partial_x\bm{u}^x>0$ on $\Sigma_\varepsilon$. We now perform the same construction in the `ingoing' case, which follows from identical arguments to the above, and satisfies the same properties and bounds.\\ \\
It remains to verify, again by induction as in Section~3, that the collection $(\bm{u}^t_\pm,\bm{u}^x_\pm,\sigma_\pm)$ satisfies $k^\text{th}$ order boundary conditions. We prove further, in fact, that if $(\Omega^2,r)$ satisfies $k^\text{th}$ order boundary conditions, then we also have correct limits for $\partial^i_\xi\bm{u}^\xi|_{\xi=0}$ for $i=0,\dots,k$.\\ \\
\hspace*{0.05\textwidth}\begin{minipage}{0.9\textwidth}\textsc{Base Case.} The case $k=0$ is just the statement that $\bm{u}^t|_{\xi=0}=1$, $\bm{u}^x|_{\xi=0}=0$ and $\sigma|_{\xi=0}=\sigma_0$, which are simply the $C^0$ boundary conditions imposed for the matter equations. We verify also that $\bm{u}^\xi_\pm|_{\xi=0}=\pm(-2(\Gamma_x)_0)^{1/2}$, which relies on $0^\text{th}$ order boundary conditions for $(\Omega^2,r)$. Indeed, in the outgoing case, the geodesic equation \hyperlink{eqn:2.8}{(2.8)} tells us that, along an integral curve of $\bm{u}_+$, we have $d^2x/d\tau^2|_{\tau=0}=-(\Gamma_x)_0$. It follows that $d\xi/d\tau|_{\tau=0}=(-2(\Gamma_x)_0)^{1/2}$ 
\\
\end{minipage}\\
\hspace*{0.05\textwidth}\begin{minipage}{0.9\textwidth}
\textsc{Inductive Step.} We begin by establishing the $k^\text{th}$ order condition for $\bm{u}^\xi_\pm$, omitting the $\pm$ signs in the sequel. First, apply $\bm{u}=\bm{u}^t\partial_t+\bm{u}^\xi\partial_\xi$ successively, $k$ times in total, to the equation for $\bm{u}^x$, giving an equation for $\bm{u}(\bm{u}(\dots(\bm{u}^x)\dots)$. Where the $\bm{u}$ operator is otherwise applied to $\bm{u}^t,\bm{u}^x$, we may use the geodesic equations \hyperlink{eqn:2.7}{(2.7)}-\hyperlink{eqn:2.8}{(2.8)} to rewrite it in terms of Christoffel symbols. \end{minipage}\\
\hspace*{0.05\textwidth}\begin{minipage}{0.9\textwidth}Thus the derived equation involves up to $k$ (cusp-) derivatives of $\Gamma_t,\Gamma_x$ (which by assumption satisfy the correct limits at $\xi=0$), but no other derivatives of $\bm{u}^t,\bm{u}^x$. We can therefore take the limit as $\xi=0$. Considering now $x(\tau)$ along a fixed integral curve of $\bm{u}$, we have correct limits for $d^{k+2}x/d\tau^{k+2}$ at $\tau=0$, which allow us to deduce correct limits for $d^{k+1}\xi/d\tau^{k+1}$ at $\tau=0$. Expanding $\bm{u}(\bm{u}(\dots(\bm{u}^\xi)\dots)$ (where in total $k-1$ $\bm{u}$ derivatives are applied to $\bm{u}^\xi$) via $\bm{u}=\bm{u}^t\partial_t+\bm{u}^\xi\partial_\xi$, we have correct limits for everything except the term in $\partial^k_\xi\bm{u}^\xi$, so we can isolate this and read off correct limits for $\partial^k_\xi\bm{u}^\xi$. With $k^\text{th}$ order limits for $\bm{u}^\xi_\pm$ in hand, it is simple to get the claimed limits for $\bm{u}^t_\pm,\bm{u}^x_\pm,\sigma_\pm$. By the induction hypothesis, we have the correct $i^\text{th}$ order limits for $i=0,\dots,k-1$. Deducing the correct $k^\text{th}$ order limit for $\bm{u}^x_\pm$ is immediate because $\bm{u}^x_\pm=2\xi\bm{u}^\xi_\pm$. Meanwhile, $\bm{u}^t_\pm$ (resp$.$ $\sigma_\pm$) obeys a transport equation along $\bm{u}=\bm{u}^t\partial_t+\bm{u}^\xi\partial_\xi$. Applying $\partial^{k-1}_\xi$ to both sides of the equation, we can rearrange to make $\partial^k_{\xi}\bm{u}^t_\pm$ (resp$.$ $\partial^k_{\xi}\sigma_\pm$) the subject. Every other term is known to satisfy the correct limits at $\xi=0$, and so this also applies to $\partial^k_\xi\bm{u}^t_\pm$ (resp$.$ $\partial^k_\xi\sigma_\pm$) concluding the proof. Note that we needed this to hold for $\bm{u}^\xi_\pm$ first, because the equation for $\sigma_\pm$ involves $\partial_\xi\bm{u}^\xi_\pm$.\vspace*{2mm} \\
\end{minipage}\\
So $(\bm{u}^t_\pm,\bm{u}^x_\pm,\sigma_\pm)$ have the desired properties, with $\varepsilon,B$ indeed depending only on $k,\delta_i,N_i,C$.\hfill$\square$
\\ \\ \\
\Large\hypertarget{sec:5}{\textbf{5\hspace{4mm}Contraction bounds}}\label{5}\normalsize\\ \\
The aim of this section is to show that the iteration map defined in the foregoing sections satisfies a contraction estimate with respect to a suitable metric. It turns out that the correct metric in which to estimate differences is defined by regular $\partial_x,\partial_t$ coordinate derivatives, rather than the $\partial_\xi,\partial_t$ derivatives which define the $\|\cdot\|_{C^k_\text{cusp}}$ norms used in the previous sections. Though $\partial_x$ derivatives of our variables are usually unbounded, the $\partial_x$ derivative of their \textit{differences} may be bounded, and are indeed the right choice. We must also estimate differences in $\Omega^2,r$ to the $C^2$ level, because the energy density $\sigma_\pm$ lives two levels of differentiability below $\Omega^2,r$ (see \hyperlink{def:2.5}{Definition 2.5})\color{black}.\\ \\
To that end, let $\Omega^2_1,\Omega^2_2,r_1,r_2\in C^\infty_\text{cusp}(\Sigma_\varepsilon)$. We define the metric
$$d((\Omega^2_1,r_1),(\Omega^2_2,r_2)):=\|\hspace{-1pt}\log\Omega^2_1-\log\Omega^2_2\|_{C^2(\Sigma_\varepsilon)}+\|r_1-r_2\|_{C^2(\Sigma_\varepsilon)}$$
(whenever it exists), where we emphasize again that the norms appearing above are the usual $\|\cdot\|_{C^2}$ norms associated to the chart $(t,x)$, and not the $\|\cdot\|_{C^2_\text{cusp}}$ norms  appearing in the rest of the paper. For brevity, we often use the abbreviation $d(\cdots)$ in the sequel to refer to the quantity above.\\ \\ \\
\large\hypertarget{sec:5.1}{\textbf{5.1\hspace{4mm}Difference estimates for $\bm{u,\sigma}$}}\label{5.1}\normalsize\\ \\
We first address the matter equations, obtaining control of differences in $\bm{u},\sigma$ in the norms $\|\cdot\|_{C^1}$, $\|\cdot\|_{C^0}$ respectively. \\ \\
\hypertarget{lem:5.1}{\textbf{Lemma 5.1}} (Difference estimates for $\bm{u},\sigma$)\textbf{.} \textit{Let $\varepsilon\in(0,1]$. Let $C,\delta_1,\delta_2>0$ be positive constants. Then there exist $C_1,C_2,C_3,C_4>0$, depending only on $C,\delta_1,\delta_2$, with the following property. Let $\Omega^2_1,\Omega^2_2,r_1,r_2\in C^\infty_\text{cusp}(\Sigma_\varepsilon)$ be such that $d((\Omega^2_1,r_1),(\Omega_2^2,r_2))<\infty $ and
$$\log\Omega^2_i|_{x=0}=0,\quad \partial_x\log\Omega^2_i|_{x=0}=2(\Gamma_x)_0, \qquad r_i|_{x=0}=r_0,\quad \partial_xr_i|_{x=0}=(\partial_xr)_0\vspace{1mm}$$and let $\bm{u}_i,\sigma_i$ solve exactly the matter equations on each respective background
$$\bm{u}^\nu_i(\nabla_i)_\nu\bm{u}^\mu_i=0\qquad(\nabla_i)_\mu(\sigma_ix^{-1/2}\bm{u}^\mu_i)=0 $$
Moreover, suppose that
\begin{enumerate}
\item For each $i=1,2$, it holds that $\bm{u}^x_i,\partial_x\bm{u}^x_i>0$ on $\Sigma_\varepsilon$  
\item For each $i=1,2$, the integral curves of $\bm{u}_i$ terminate on $x=0$, where $\bm{u}^x_i=0$, and we have lower and upper bounds
$$\tfrac{1}{4}\delta_1\leq \bm{u}_i(\bm{u}^x_i)\leq 4C $$  
\item The following bounds hold in terms of the constants $C,\delta_1,\delta_2$ $$\|\bm{u}^t_i\|_{C^2_\text{cusp}},\|\bm{u}^x_i\|_{C^2_\text{cusp}},\|(\Gamma_t)_i\|_{C^1_\text{cusp}},\|(\Gamma_x)_i\|_{C^1_\text{cusp}},\|\partial_tr_i\|_{C^0},\|\partial_xr_i\|_{C^0},\|\sigma_i\|_{C^1_\text{cusp}}\leq C $$
$$\inf_{\Sigma_\varepsilon}|(\Gamma_x)_i|\geq\tfrac{1}{2}\delta_1,\quad\inf_{\Sigma_\varepsilon} r_i\geq\tfrac{1}{2}\delta_2\vspace{-2mm}$$
\end{enumerate}
Then the following difference estimates hold for the matter variables on $\Sigma_\varepsilon$:}
$$|\bm{u}^t_1-\bm{u}^t_2|(t,x),\hspace{2pt} |\bm{u}^x_1-\bm{u}^x_2|(t,x) \leq C_1x^{3/2}\cdot d((\Omega^2_1,r_1),(\Omega^2_2,r_2))\vspace{-4mm}$$
\begin{align*}
|\partial_t\bm{u}^t_1-&\partial_t\bm{u}^t_2|(t,x),\hspace{2pt}|\partial_x\bm{u}^t_1-\partial_x\bm{u}^t_2|(t,x),\\&|\partial_t\bm{u}^x_1-\partial_t\bm{u}^x_2|(t,x),
|\partial_x\bm{u}^x_1-\partial_x\bm{u}^x_2|(t,x) \leq C_2x^{1/2}\cdot d((\Omega^2_1,r_1),(\Omega^2_2,r_2))\hspace{2.5cm}
\end{align*}
\vspace{-4mm}
$$\hspace{2.35cm}|\sigma_1-\sigma_2|(t,x)\leq C_3x\cdot d((\Omega^2_1,r_1),(\Omega^2_2,r_2)) $$
$$|T^{tt}_1-T^{tt}_2|(t,x),\hspace{2pt}|T^{xx}_1-T^{xx}_2|(t,x),\hspace{2pt}|T^{tx}_1-T^{tx}_2|(t,x)\leq C_4x^{1/2}\cdot d((\Omega^2_1,r_1),(\Omega^2_2,r_2)) \vspace{2mm}$$
\textit{where here the energy-momentum components $T^{\mu\nu}=x^{-1/2}\sigma\bm{u}^\mu\bm{u}^\nu$ are those of a single dust species.}\\ \\ 
\textit{Proof.} We introduce square-brackets notation $[\hspace{2pt}\cdot\hspace{2pt}]$ to denote the difference between $i=1,2$, so that eg.
$[\sigma]:=\sigma_1-\sigma_2 $. We also introduce the notation
$$[\bm{u}]:=\big([\bm{u}^t]^2+[\bm{u}^x]^2\big)^{1/2}\qquad [\partial\bm{u}]:=\big([\partial_t\bm{u}^t]^2+[\partial_x\bm{u}^t]^2+[\partial_t\bm{u}^x]^2+[\partial_x\bm{u}^x]^2\big)^{1/2} $$
and we write $\tau$ to denote proper time computed along any integral curve of $\bm{u}_1$, with $\tau=0$ on the caustic. (Then $\tfrac{d}{d\tau}:=\bm{u}_1$.) Our strategy is to take the equation defining each quantity for $i=1,2$, and subtract. After rearranging, we obtain an estimate for $\tfrac{d}{d\tau}([\hspace{2pt}\cdot\hspace{2pt}])$ to which Gronwall's inequality applies. \\ \\
\ul{Step 1: $C^0$ differences in $\bm{u}$.} First, we seek an estimate of the form 
\begin{equation*}\tfrac{d}{d\tau}[\bm{u}]\lesssim_{C,\delta_1}[\bm{u}]+x\cdot d(\cdots)\tag{\hypertarget{eqn:5.1}{5.1}} \end{equation*}
Taking equation \hyperlink{eqn:2.8}{(2.8)} for $\bm{u}^x$
$$\bm{u}^t_i\partial_t\bm{u}^x_i+\bm{u}^x_i\partial_x\bm{u}^x_i+(\Gamma_x)_i((\bm{u}^t_i)^2+(\bm{u}^x_i)^2)+2(\Gamma_t)_i\bm{u}^t_i\bm{u}^x_i =0$$
and subtracting $i=2$ from $i=1$ yields
$$\tfrac{d}{d\tau}[\bm{u}^x]+[\bm{u}^t]\partial_t\bm{u}^x_2+[\bm{u}^x]\partial_x\bm{u}^x_2+\big[\Gamma_x((\bm{u}^t)^2+(\bm{u}^x)^2)+2\Gamma_t\bm{u}^t\bm{u}^x\big]=0 $$
The term $[\bm{u}^x]\partial_x\bm{u}^x_2$ is problematic because $\partial_x\bm{u}^x_2$ is unbounded. However, since it has a favourable sign ($\partial_x\bm{u}^x_2>0$), it can be ignored when we estimate $\tfrac{d}{d\tau}[\bm{u}^x]$. Combining with the equation for $\bm{u}^t$, we estimate
$$\big|\tfrac{d}{d\tau}[\bm{u}]\big|\leq [\bm{u}]\big(|\partial_t\bm{u}^t_2|+|\partial_x\bm{u}^t_2|+|\partial_t\bm{u}^x_2|\big)\hspace{2pt}+\hspace{2pt}\text{terms involving }\Gamma_t,\hspace{2pt}\Gamma_x $$
To control the term $|\partial_x\bm{u}^t_2|$, which \textit{a priori} could be unbounded, we note that evaluating 
$$\bm{u}(\bm{u}^t)=\bm{u}^t\partial_t\bm{u}^t+\bm{u}^\xi\partial_\xi\bm{u}^t$$
at $x=0$ reveals that $\partial_\xi\bm{u}^t|_{x=0}=0$. So in fact $|\partial_x\bm{u}^t_2|$ is bounded, and we can write
$$|\partial_x\bm{u}^t_2|=\bigg|\frac{\partial_\xi\bm{u}^t_2}{2\xi}\bigg|\leq\tfrac{1}{2}\|\partial^2_\xi\bm{u}^t_2\|_{C^0}\leq \tfrac{1}{2}\|\bm{u}^t_2\|_{C^2_\text{cusp}}\leq \tfrac{1}{2}C $$Turning to the terms involving $\Gamma_t,\Gamma_x$, we manipulate these into terms proportional to $[\Gamma_t],[\Gamma_x]$ and terms proportional to $[\bm{u}^t],[\bm{u}^x]$. The latter become `Gronwall terms', while for the former we use the fact that $[\Gamma_t]\big|_{x=0}=[\Gamma_x]\big|_{x=0}=0$ to write 
$$|[\Gamma_t]|\leq \tfrac{1}{2}\int^x_0|[\partial_t\partial_x\log\Omega^2]|dx\leq \tfrac{1}{2}x\cdot d(\cdots) $$
and similarly $|[\Gamma_x]|\leq \tfrac{1}{2}x\cdot d(\cdots)$. Combining these observations yields \hyperlink{eqn:5.1}{(5.1)}. With \hyperlink{eqn:5.1}{(5.1)} in hand, observe that along each fixed integral curve of $\bm{u}_1$, the value of $x$ is sandwiched between multiples of $\tau^2$:
$$\tfrac{1}{8}\delta_1\tau^2\leq x\leq 2C\tau^2 $$
Hence our estimate says
$$ \tfrac{d}{d\tau}[\bm{u}]\lesssim_{C,\delta_1}[\bm{u}]+\tau^2\cdot d(\cdots)$$
Since $[\bm{u}]=0$ at $\tau=0$, we apply Gronwall's inequality to obtain 
$$[\bm{u}](\tau)\lesssim_{C,\delta_1}\tau^3\cdot d(\cdots)e^{(\cdots)_{C,\delta_1}\tau} $$
where $(\cdots)_{C,\delta_1}$ is another constant in $C,\delta_1$. Converting the $\tau$ back into $x^{1/2}$:
$$[\bm{u}](t,x)\lesssim_{C,\delta_1}x^{3/2}\cdot d(\cdots)e^{(\cdots)_{C,\delta_1}x^{1/2}}$$
and since $x^{1/2}< \varepsilon^{1/2}\leq 1$, the latter term is another constant in $C,\delta_1$. So we have the desired bound.\\ \\
\ul{Step 2: $C^1$ differences in $\bm{u}$.} Using the conclusion of Step~1, we now seek an estimate of the form 
\begin{equation*}\tfrac{d}{d\tau}[\partial\bm{u}]\lesssim_{C,\delta_1}[\partial\bm{u}]+d(\cdots) \tag{\hypertarget{eqn:5.2}{5.2}} \end{equation*}
We differentiate each geodesic equation \hyperlink{eqn:2.7}{(2.7)}-\hyperlink{eqn:2.8}{(2.8)} and commute with the $\bm{u}$ transport operator. So, to estimate $[\partial_x\bm{u}^x]$, for instance, we study
$$\bm{u}_i(\partial_x\bm{u}^x_i)+\partial_t\bm{u}^x_i\partial_x\bm{u}^t_i+(\partial_x\bm{u}^x_i)^2+\partial_x\Big((\Gamma_x)_i((\bm{u}^t_i)^2+(\bm{u}^x_i)^2)+2(\Gamma_t)_i\bm{u}^t_i\bm{u}^x_i\Big) =0$$
The evolution equation for $[\partial_x\bm{u}^x]$ then reads
\begin{align*}\tfrac{d}{d\tau}[\partial_x\bm{u}^x]+[\partial_t\bm{u}^x]\partial_x\bm{u}^t_1 +[\partial_x\bm{u}^t]&\partial_t\bm{u}^x_2 + [\partial_x\bm{u}^x](\partial_x\bm{u}^x_1+\partial_x\bm{u}^x_2)\\&+\Big[\partial_x\Big(\Gamma_x((\bm{u}^t)^2+(\bm{u}^x)^2)+2\Gamma_t\bm{u}^t\bm{u}^x\Big)\Big] =0 
\end{align*}
Here again, the term $[\partial_x\bm{u}^x](\partial_x\bm{u}^x_1+\partial_x\bm{u}^x_2)$ has a benevolent sign, and so may be ignored when we estimate $\tfrac{d}{d\tau}[\partial\bm{u}]$. We next need to establish that
$$\Big[\partial_x\Big(\Gamma_x((\bm{u}^t)^2+(\bm{u}^x)^2)+2\Gamma_t\bm{u}^t\bm{u}^x\Big)\Big] \lesssim_{C,\delta_1}[\partial\bm{u}]+d(\cdots)$$
Expanding both the $\partial_x$ and $[\hspace{2pt}\cdot\hspace{2pt}]$ yields a large number of terms which may be examined schematically. With terms of the schematic form $[\Gamma](\partial\bm{u})\bm{u}$, we have as in Step~1 that $[\Gamma]\leq x\cdot d(\cdots)$ which is more than enough to balance the unbounded contribution $|\partial\bm{u}|\lesssim_C x^{-1/2}$. Terms of the form $\Gamma[\partial\bm{u}]\bm{u}$ are easily seen to be $\lesssim_{C} [\partial\bm{u}]$. Meanwhile, with terms of the form $\partial\Gamma[\bm{u}]\bm{u}$ or $\Gamma(\partial\bm{u})[\bm{u}]$, we use the result of Step~1 that $[\bm{u}]\lesssim_{C,\delta_1}x^{3/2}\cdot d(\cdots)$, which is more enough to balance the unbounded terms $|\partial\Gamma|\lesssim_C x^{-1/2}$ or $|\partial\bm{u}|\lesssim_C x^{-1/2}$. So the terms involving $\Gamma_t,\Gamma_x$ are bounded as claimed.\\ \\
Combining with similarly manipulated equations for $[\partial_t\bm{u}^t]$, $[\partial_x\bm{u}^t]$ and $[\partial_t\bm{u}^x]$, we conclude estimate \hyperlink{eqn:5.2}{(5.2)}. With \hyperlink{eqn:5.2}{(5.2)} in hand, we now argue as in Step~1, converting between $\tau$ and $x^{1/2}$ and applying Gronwall, to arrive at the desired conclusion.\\ \\
\ul{Step 3: $C^0$ differences in $\sigma$.} In this Step, we seek an estimate of the form
\begin{equation*}\big|\tfrac{d}{d\tau}[\sigma]\big|\lesssim_{C,\delta_1}[\sigma]+x^{1/2}\cdot d(\cdots)\tag{\hypertarget{eqn:5.3}{5.3}} \end{equation*}
We write equation \hyperlink{eqn:2.9}{(2.9)} for $\sigma$ in the form
$$\bm{u}(\sigma)+\sigma\big(\partial_\xi\bm{u}^\xi+\partial_t\bm{u}^t+\bm{u}(\log\Omega^2)+\bm{u}(\log r^2)\big)=0 $$
Taking this equation for $i=1,2$ and subtracting, we obtain an equation for $\tfrac{d}{d\tau}[\sigma]$:
\begin{align*}
\tfrac{d}{d\tau}[\sigma]+[\bm{u}^t]\partial_t\sigma_2+
[\bm{u}^x]\partial_x\sigma_2+\sigma_1\Big[\partial_\xi\bm{u}^\xi&+\partial_t\bm{u}^t+\bm{u}(\log\Omega^2)+\bm{u}(\log r^2)\Big]\\+[\sigma]&\big(\partial_\xi\bm{u}^\xi_2+\partial_t\bm{u}^t_2+\bm{u}_2(\log\Omega^2_2)+\bm{u}_2(\log r^2_2)\big)=0
\end{align*}
The latter terms, proportional to $[\sigma]$, are readily controlled by a constant depending only on $C,\delta_1,\delta_2$. Meanwhile, the terms involving $[\bm{u}^t],[\bm{u}^x]$ can be estimated using the result of Step~1, using in particular that
\begin{align*}
|[\bm{u}^x]\partial_x\sigma_2|\lesssim_{C,\delta_1} x^{3/2}\cdot d(\cdots)\tfrac{1}{2x^{1/2}}\|\partial_\xi\sigma_2\|_{C^0}\lesssim_{C,\delta_1} x\cdot d(\cdots)
\end{align*}
The remaining terms above, with a factor $\sigma_1$, are estimated using the $C^1$ difference bounds for $\bm{u}$ obtained in Step~2. The only one not already discussed is $[\partial_\xi\bm{u}^\xi]$, but this may be controlled by $[\partial_x\bm{u}^x]$. Indeed,
\begin{align*}
[\partial_\xi\bm{u}^\xi](t,x)& \leq [\partial_x\bm{u}^x](t,x)+|\tfrac{1}{2x}[\bm{u}^x](t,x)|\\
&\leq [\partial_x\bm{u}^x](t,x)+\tfrac{1}{2}\sup_{[0,x]}|[\partial_x\bm{u}^x]|\lesssim_{C,\delta_1} x^{1/2}\cdot d(\cdots)
\end{align*}
Putting this together, we obtain \hyperlink{eqn:5.3}{(5.3)}, which, after 
again arguing as in Steps 1-2, leads us to the desired $C^0$ difference estimate for $\sigma$.\\ \\
\ul{Step 4: Differences in $T^{\mu\nu}$.} \hspace{2pt}Immediate by combining the results of Steps 1-3. 
\hfill$\square$\\ \\ \\
\large\hypertarget{sec:5.2}{\textbf{5.2\hspace{4mm}Contraction estimates for $\bm{\Omega^2,r}$}}\label{5.2}\normalsize\\ \\
We turn now to the Einstein equations, stating our bounds in terms of energy-momentum components $T^{\mu\nu}$ only.\newpage
\hypertarget{lem:5.2}{\textbf{Lemma 5.2}} (Contraction estimates for $\Omega^2,r$)\textbf{.} \textit{Let $C_1,C_2,\delta_1,\delta_2>0$ be positive constants. Then there exists $\varepsilon=\varepsilon(C_1,C_2,\delta_1,\delta_2)>0$ with the following property. Let $\Omega^2_1,\Omega^2_2,r_1,r_2\in C^\infty_\text{cusp}(\Sigma_\varepsilon)$ be such that $d((\Omega^2_1,r_1),(\Omega^2_2,r_2))<\infty $ and
$$\log\Omega^2_i|_{x=0}=0,\quad \partial_x\log\Omega^2_i|_{x=0}=2(\Gamma_x)_0, \qquad r_i|_{x=0}=r_0,\quad \partial_xr_i|_{x=0}=(\partial_xr)_0\vspace{1mm}$$Let also $T^{uu}_i,T^{vv}_i,T^{uv}_i\in C^\infty(\Sigma_\varepsilon)$ $(i=1,2)$ be components of an exactly conserved energy momentum tensor on each background, i$.$e$.$ equations \hyperlink{eqn:3.7}{(3.7)}-\hyperlink{eqn:3.8}{(3.8)} hold. Suppose that the following bounds hold:
\begin{align*}\|x^{1/2}T^{uu}_i\|_{C^0}&,\hspace{2pt}\|x^{1/2}T^{vv}_i\|_{C^0},\hspace{2pt}\|x^{1/2}T^{uv}_i\|_{C^0}, \\
&\|\hspace{-1pt}\log\Omega^2_i\|_{C^0},\hspace{2pt}\|(\Gamma_t)_i\|_{C^1_\text{cusp}},\hspace{2pt}\|(\Gamma_x)_i\|_{C^1_\text{cusp}},\hspace{2pt}\|r_i\|_{C^0},\hspace{2pt}\|\partial_tr_i\|_{C^1_\text{cusp}},\hspace{2pt}\|\partial_xr_i\|_{C^1_\text{cusp}}\leq C_1\end{align*}
$$\inf_{\Sigma_\varepsilon}|(\Gamma_x)_i|\geq\tfrac{1}{2}\delta_1,\quad\inf_{\Sigma_\varepsilon} r_i\geq\tfrac{1}{2}\delta_2$$
Moreover, suppose the energy-momentum components $T^{\mu\nu}$ satisfy
$$|T^{uu}_1-T^{uu}_2|(t,x),\hspace{2pt}|T^{vv}_1-T^{vv}_2|(t,x),\hspace{2pt}|T^{uv}_1-T^{uv}_2|(t,x)\leq C_2x^{1/2} \cdot d((\Omega^2_1,r_1),(\Omega^2_2,r_2))$$
Then the solutions $(\overline{\Omega}^2_1,\overline{r}_1),(\overline{\Omega}^2_2,\overline{r}_2)\in C^\infty_\text{cusp}(\Sigma_\varepsilon)$ to the linear wave equations
\begin{align*}
\partial_u\partial_v\log\overline{\Omega}^2_i=&\frac{\Omega^2_im_i}{r^3_i}-\Omega^4_iT^{uv}_i\\
\partial_u\partial_v\overline{r}_i=-&\frac{\Omega^2_im_i}{2r^2_i}+\tfrac{1}{2}r_i\Omega^4_iT^{uv}_i
\end{align*}
\vspace{-4mm}
$$\log\overline{\Omega}^2_i|_{x=0}=0,\quad \partial_x\log\overline{\Omega}^2_i|_{x=0}=2(\Gamma_x)_0, \qquad \overline{r}_i|_{x=0}=r_0,\quad \partial_x\overline{r}_i|_{x=0}=(\partial_xr)_0\vspace{1mm}$$
satisfy}
$$d\big((\overline{\Omega}^2_1,\overline{r}_1),(\overline{\Omega}^2_2,\overline{r}_2)\big)\leq\tfrac{1}{2}d((\Omega^2_1,r_1),(\Omega^2_2,r_2)) \vspace{3mm}$$
\textit{Proof.} Because $[\hspace{1pt}\log\Omega^2]|_{x=0}=[\partial_x\log\Omega^2]|_{x=0}=[r]|_{x=0}=[\partial_xr]|_{x=0}=0$, we can integrate to show that
$$|[\hspace{1pt}\log\Omega^2]|(t,x),\hspace{2pt}|[r]|(t,x)\leq\tfrac{1}{2}x^2\cdot d(\cdots) $$
$$|[\partial_t\log\Omega^2]|(t,x),\hspace{2pt}|[\partial_x\log\Omega^2]|(t,x),\hspace{2pt}|[\partial_t r]|(t,x),\hspace{2pt}|[\partial_x r]|(t,x)\leq x\cdot d(\cdots) $$
Consequently, functions $f(\Omega^2,r)$, where $f$ is smooth (for example $\Omega^4$) obey bounds of the form
$$|[f(\Omega^2,r)]|(t,x)\lesssim_{C_1,C_2,\delta_1,\delta_2}x^2\cdot d(\cdots) $$
and functions $g(\Omega^2,r,\partial\Omega^2,\partial r)$ with $g$ smooth (for example $m$) obey bounds of the form
$$|[g(\Omega^2,r,\partial\Omega^2,\partial r)]|(t,x)\lesssim_{C_1,C_2,\delta_1,\delta_2}x\cdot d(\cdots) $$
\ul{Step 1: Bounds for $\partial_u\partial_v$ derivatives.} We seek an estimate of the form
\begin{equation*}
|[\partial_u\partial_v\log\overline{\Omega}^2]|(t,x),\hspace{2pt}|[\partial_u\partial_v\overline{r}]|(t,x)\lesssim_{C_1,C_2,\delta_1,\delta_2} x^{1/2}\cdot d(\cdots)\tag{\hypertarget{eqn:5.4}{5.4}}
\end{equation*}
The equations \hyperlink{eqn:3.3}{(3.3)}-\hyperlink{eqn:3.4}{(3.4)} defining $\overline{\Omega}^2,\overline{r}$ are given in terms of the derivative $\partial_u\partial_v$, so we can simply subtract the defining equation from itself in the case $i=1,2$ to obtain, in the case of $\Omega^2$,
$$[\partial_u\partial_v\log\overline{\Omega}^2]=\bigg[\frac{\Omega^2m}{r^3}\bigg]-[\tfrac{1}{2}\Omega^4T^{uv}] $$
Using the above comments, and the hypotheses given, we estimate the terms on the right-hand side to immediately yield \hyperlink{eqn:5.4}{(5.4)}. The argument is identical for $\overline{r}$.
\\ \\
\ul{Step 2: Bounds for $\partial^2_u$, $\partial^2_v$ derivatives.} Likewise we seek estimates of the form
\begin{equation*}
|[\partial^2_u\log\overline{\Omega}^2]|(t,x),\hspace{2pt}|[\partial^2_v\log\overline{\Omega}^2]|(t,x),\hspace{2pt}|[\partial^2_u\overline{r}]|(t,x),\hspace{2pt}|[\partial^2_v\overline{r}]|(t,x)\lesssim_{C_1,C_2,\delta_1,\delta_2} x^{1/2}\cdot d(\cdots)\tag{\hypertarget{eqn:5.5}{5.5}}
\end{equation*}Here we can apply the same strategy as in the proof of \hyperlink{pro:3.2}{Proposition~3.2}, exploiting the conservation of the energy-momentum tensor to write the $\partial^2_u,\partial^2_v$ derivatives of $\log\overline{\Omega}^2,\overline{r}$ in terms of undifferentiated $T^{\mu\nu}$ components. Indeed, applying equation \hyperlink{eqn:3.9}{(3.9)} to $\Omega^2_1,\Omega^2_2$ and subtracting, the terms written explicitly in terms of data cancel, giving us
$$[\partial^2_v\log\overline{\Omega^2}](u,v)=[\tfrac{1}{2}\Omega^4T^{uu}](u,v)+\int^v_u\Big[\partial_v\bigg(\frac{\Omega^2m}{r^3}\bigg)+ \Omega^4T^{uu}r^{-1}\partial_ur+\Omega^4T^{uv}(r^{-1}\partial_vr-\tfrac{1}{2}\partial_v\log\Omega^2)\Big](\tilde{u},v)du $$
Bounding the first term is the same as in Step~1. Meanwhile, we claim that the integrand is bounded above by a multiple of $d(\cdots)$ (that is, with a constant depending only on $C_1,C_2,\delta_2,\delta_2$). If true, then the integral itself gives us an additional factor of $v-u=2x$, which will be more than enough for \hyperlink{eqn:5.5}{(5.5)}. But the integrand is indeed thus bounded, using again the comments at the start and the hypotheses given.\\ \\
This establishes \hyperlink{eqn:5.5}{(5.5)} for $\partial^2_v\log\overline{\Omega}^2$ only, and the corresponding arguments for $\partial^2_u\log\overline{\Omega^2}$, $\partial^2_u\overline{r}$, $\partial^2_v\overline{r}$ are identical and omitted.\\ \\
\ul{Step 3: Obtaining the contraction.} After rotating into $(t,x)$ coordinates, Steps 1-2 together give
$$|[\partial^2_x\log\overline{\Omega}^2]|(t,x),\hspace{2pt}|[\partial_x\partial_t\log\overline{\Omega}^2]|(t,x),\hspace{2pt}|[\partial^2_t\log\overline{\Omega}^2]|(t,x),\qquad\qquad\qquad\qquad\vspace{-1mm} $$
$$\qquad\qquad\qquad|[\partial^2_x\overline{r}]|(t,x),\hspace{2pt}|[\partial_x\partial_t\overline{r}]|(t,x),\hspace{2pt}|[\partial^2_t\overline{r}]|(t,x)\lesssim_{C_1,C_2,\delta_1,\delta_2} x^{1/2}\cdot d((\Omega^2_1,r_1),(\Omega^2_2,r_2)) $$
Applying the comments at the start of the proof to $\overline{\Omega}^2,\overline{r}$, we may also bound $[\hspace{1pt}\log\overline{\Omega}^2],[\overline{r}]$ etc$.$ in terms of the left-hand side, and so in fact we have
\begin{align*}
d\big((\overline{\Omega}^2_1,\overline{r}_1),(\overline{\Omega}^2_2,\overline{r}_2)\big)&\lesssim_{C_1,C_2,\delta_1,\delta_2}x^{1/2}\cdot d((\Omega^2_1,r_1),(\Omega^2_2,r_2))\\
&\lesssim_{C_1,C_2,\delta_1,\delta_2}\varepsilon^{1/2}\cdot d((\Omega^2_1,r_1),(\Omega^2_2,r_2))
\end{align*}
and hence we can choose $\varepsilon=\varepsilon(C_1,C_2,\delta_1,\delta_2)>0$ so that the final constant is at most $\tfrac{1}{2}$, completing the proof.
\hfill$\square$
\\ \\ \\
\Large\hypertarget{sec:6}{\textbf{6\hspace{4mm}Proof of Theorem 1.1}}\label{6}\normalsize\\ \\
\large\hypertarget{sec:6.1}{\textbf{6.1\hspace{4mm}Preparatory lemmas}}\label{6.1}\normalsize\\ \\
We begin by combining all the work of Sections~3-5 to identify a contractive self-map which forms the basis of our iterative scheme.\\ \\
\hypertarget{lem:6.1}{\textbf{Lemma 6.1}} (Existence of contraction map)\textbf{.} \textit{Fix $k\in\mathbb{N}$ and an admissible set $(r_0,\sigma_0,(\partial_xr)_0,(\Gamma_x)_0)$ of data. Define the constants $\delta_i,N_i$ controlling the data at the k$^{\hspace{1pt}th}$ level:}
$$\delta_1=\inf_{[a,b]}|(\Gamma_x)_0|>0\qquad\delta_2=\inf_{[a,b]}r_0>0$$
$$N_1=\|(\Gamma_x)_0\|_{C^{k+1}[a,b]}\hspace{0.5cm} N_2=\max\big\{\|r_0\|_{C^{k+2}[a,b]},\|(\partial_xr)_0\|_{C^{k+1}[a,b]}\big\}\hspace{0.5cm} N_3=\|\sigma_0\|_{C^k[a,b]}\vspace{2mm} $$
\textit{There exist constants $\varepsilon,B,C$, depending only on $k,\delta_i,N_i$, and a self-map
$$\Phi:\mathcal{A}(k,\varepsilon,B,C)\to \mathcal{A}(k,\varepsilon,B,C),\quad (\Omega^2,r,\bm{u}_\pm,\sigma_\pm)\mapsto(\overline{\Omega}^2,\overline{r},\overline{\bm{u}}_\pm,\overline{\sigma}_\pm)$$
with the property that $\overline{\Omega}^2,\overline{r}$ satisfy}
\begin{align*}
(-\partial^2_x+\partial^2_t)(\log\overline{\Omega}^2)=&\frac{4\Omega^2m}{r^3}-4\Omega^2(\sigma_++\sigma_-) x^{-1/2}\\
(-\partial^2_x+\partial^2_t)\hspace{1pt}\overline{r}=-&\frac{2\Omega^2m}{r^2}+2r\Omega^2(\sigma_++\sigma_-) x^{-1/2}
\end{align*}
\vspace{-4mm}
$$\log\overline{\Omega}^2|_{x=0}=0,\quad \partial_x\log\overline{\Omega}^2|_{x=0}=2(\Gamma_x)_0, \qquad \overline{r}|_{x=0}=r_0,\quad \partial_x\overline{r}|_{x=0}=(\partial_xr)_0\vspace{1mm}$$
\textit{(That $\overline{\bm{u}}_\pm, \overline{\sigma}_\pm$ satisfy the matter equations is included in the definition of $\mathcal{A}(k,\varepsilon,B,C)$.)\vspace*{2mm} \\
Moreover, if $(\Omega^2_i,r_i,(\bm{u}_\pm)_i,(\sigma_\pm)_i)\in \mathcal{A}(k,\varepsilon,B,C)$ ($i=1,2$) are such that $$d((\Omega^2_1,r_1),(\Omega^2_2,r_2)):=\|\hspace{-1pt}\log\Omega^2_1-\log\Omega^2_2\|_{C^2(\Sigma_\varepsilon)}+\|r_1-r_2\|_{C^2(\Sigma_\varepsilon)}$$ 
exists, then we have
$$d\big((\overline{\Omega}^2_1,\overline{r}_1),(\overline{\Omega}^2_2,\overline{r}_2)\big)\leq\tfrac{1}{2}d((\Omega^2_1,r_1),(\Omega^2_2,r_2)) $$}
\hspace{-1mm}\textit{Finally, we can decrease $\varepsilon$ if needed, and the above continues to hold with the same $B,C$. Alternatively, we can increase $C$ if needed, and the above continues to hold with some possibly larger $B=B(k, \delta_i, N_i, C)$, smaller $\varepsilon=\varepsilon(k,\delta_i,N_i,C)$.}\\ \\
\textit{Proof.} With $k\in\mathbb{N},\delta_i,N_i$ fixed, we obtain $C=C(k,\delta_i,N_i)>0$ from \hyperlink{pro:3.2}{Proposition~3.2}. With $C$ in hand, we then obtain $\varepsilon=\varepsilon(k,\delta_i,N_i,C)>0,B=B(k,\delta_i,N_i,C)>0$ from \hyperlink{pro:4.1}{Proposition~4.1}. With this $B$ in hand, we finally obtain a possibly smaller $\varepsilon=\varepsilon(k,\delta_i,N_i,C,B)>0$ from \hyperlink{pro:3.2}{Proposition~3.2}. This chooses the constants $\varepsilon,B,C$, which manifestly depend only on $k,\delta_i,N_i$. Moreover, we are free to increase $C$ and/or decrease $\varepsilon$, and the propositions still apply, but we see that our larger choice of $C$ may affect the $B,\varepsilon$ yielded by \hyperlink{pro:4.1}{Proposition~4.1}, and the $\varepsilon$ yielded by \hyperlink{pro:3.2}{Proposition~3.2}. We can now return to the statements of these propositions and see what this choice of constants gives us.\\ \\
Suppose now that we have $(\Omega^2,r,\bm{u}^t_\pm,\bm{u}^x_\pm,\sigma_\pm)\in \mathcal{A}(k,\varepsilon,B,C)$. The statement of \hyperlink{pro:3.2}{Proposition~3.2} gives us $\overline{\Omega}^2,\overline{r}\in C^\infty_\text{cusp}(\Sigma_\varepsilon)$ satisfying equations \hyperlink{eqn:3.3}{(3.3)}-\hyperlink{eqn:3.4}{(3.4)}, and possessing the bounds stipulated in \hyperlink{def:2.5}{Definition~2.5}. Feeding this into \hyperlink{pro:4.1}{Proposition~4.1} yields $\overline{\bm{u}}^t_\pm,\overline{\bm{u}}^x_\pm,\overline{\sigma}_\pm\in C^\infty(\Sigma_\varepsilon)$ satisfying exactly the matter equations, bounds and other properties required in \hyperlink{def:2.5}{Definition~2.5}. Hence the collection $(\overline{\Omega}^2,\overline{r},\overline{\bm{u}}^t_\pm,\overline{\bm{u}}^x_\pm,\overline{\sigma}_\pm)$ belongs to $\mathcal{A}(k,\varepsilon,B,C)$, and so we indeed have a self-map $\Phi$ as desired.\\ \\
It remains to argue that, after possibly reducing $\varepsilon$ further (relative to $B,C$ only, and thus relative to $k,\delta_i,N_i$ only), the contraction property holds. Indeed, \hyperlink{lem:5.2}{Lemma~5.2} gives us such an $\varepsilon$, depending on certain constants $C_1,C_2$, as well as $\delta_1,\delta_2$. We see that, even at the lowest possible level $k=1$, $C_1$ manifestly depends on the bounds $B,C$ already achieved for $(\Omega^2_i,r_i,(\bm{u}_\pm)_i,(\sigma_\pm)_i)$. Meanwhile, $C_2$ depends on differences in the energy-momentum components induced by $(\bm{u}_\pm)_i,(\sigma_\pm)_i$. Turning to \hyperlink{lem:5.1}{Lemma~5.1}, we have that this $C_2$ (called $C_4$ in that lemma) depends once again only on the bounds $B,C$. (Note that, to apply \hyperlink{lem:5.1}{Lemma~5.1}, we may have to reduce $\varepsilon$ again relative to $B,C$.) In conclusion, \hyperlink{lem:5.1}{Lemmas 5.1} and \hyperlink{lem:5.2}{5.2} yield an $\varepsilon$, depending only on $B,C,\delta_1,\delta_2$, such that the desired contraction property indeed holds for the corresponding self-map $\Phi$.\hfill$\square$\\ \\
\hypertarget{lem:6.2}{\textbf{Lemma~6.2}} (Extension of higher regularity solutions)\textbf{.} \textit{Let $k\geq 2$ be an integer and $s>0$. Let $(\Omega^2,r,\bm{u}_\pm,\sigma_\pm)$ be a solution to the joint Einstein-matter system \hyperlink{eqn:2.5}{(2.5)}-\hyperlink{eqn:2.9}{(2.9)} on the domain $\Sigma_s$ with $\Omega^2,r\in C^{k+1}(\Sigma_s)$, $\bm{u}_\pm\in C^k(\Sigma_s)$, $\sigma_\pm\in C^{k-1}(\Sigma_s)$ and suppose that each variable converges, along with all its derivatives (i$.$e$.$ up to order $k+1$ derivatives of $\Omega^2,r$, etc$.$), as $x\to s$. Then there exist $\delta>0$ such that the solution may be uniquely extended at the same level of differentiability to $\Sigma_{s+\delta}$. Moreover, if (the extended) $\Omega^2,r$ remains bounded in $C^2$, $\bm{u}_\pm$ in $C^1$ and $\sigma_\pm$ in $C^0$, then each variable converges, along with all its derivatives, as $x\to s+\delta$.}\\ \\
\textit{Proof.} This is a statement of local existence for the Einstein 2-dust system, in the spacelike direction, at the level of regularity $C^{k+1}\times C^k\times C^{k-1}$, but with data taken on a hypersurface $x=s$ to which the fluid velocities are \textit{not} tangent. This is substantially easier than the corresponding result of this paper, proved in Steps 1-3 of the proof of \hyperlink{thm:1.1}{Theorem~1.1}. In particular, it does not involve blowup of variables, loss of derivatives, or require any recourse to cusp-regularity etc$.$, and the proof is omitted.\hfill $\square$
\\ \\ \\
\large\hypertarget{sec:6.2}{\textbf{6.2\hspace{4mm}Proof of Theorem 1.1}}\label{6.2}\normalsize\\ \\
With these lemmas in hand, we are ready to establish \hyperlink{thm:1.1}{Theorem~1.1}.\\ \\
\textit{Proof of Theorem 1.1.} We first outline the steps of the proof before launching into the detail. We begin by obtaining the 2-dust region $\{x>0\}$ as a fixed point of the self-map $\Phi$ defined above. Before we can do this, we need to justify that the function spaces $\mathcal{A}(k,\varepsilon,B,C)$ are non-empty (after possibly reducing $\varepsilon$ further), so that we can initialise the iteration scheme. Then the contraction property gives a limiting object $(\Omega^2,r,\bm{u}_\pm,\sigma_\pm)$ where $\Omega^2,r\in C^2(\Sigma_\varepsilon)$, $\bm{u}_\pm\in C^1(\Sigma_\varepsilon)$ and $\sigma_\pm\in C^0(\Sigma_\varepsilon)$. \\ \\
After that, we upgrade the regularity. Since we obtained bounds at the $k^\text{th}$ level of differentiability, a standard Arzel\'{a}-Ascoli argument deduces that $(\Omega^2,r,\bm{u}_\pm,\sigma_\pm)$ admit higher derivatives. However, this only applies on $\Sigma_{\varepsilon(k)}$ where $\varepsilon(k)\to 0$ possibly. We thus rely on the black-box continuation result \hyperlink{lem:6.2}{Lemma~6.2} to propagate higher regularity to the whole of the original $\Sigma_\varepsilon$. During this argument, we show that $\Omega^2,r$ satisfy the Einstein equations, that $\bm{u}_\pm$ are geodesic, and that $\sigma_\pm$ satisfies the conservation equation, (This cannot be done earlier since the convergence is only $C^0$ for the latter variable.) We then show that the constraint equations \hyperlink{eqn:2.3}{(2.3)}-\hyperlink{eqn:2.4}{(2.4)}, and thus the full Einstein equations, hold on $\Sigma_\varepsilon$. We finally solve the Einstein equations in the vacuum region (which is trivial), and avoid $r=0$ by possibly reducing $\varepsilon$ further. We conclude the proof by verifying the conditions needed for a weak solution to Einstein's equation outlined in Appendix B.\\ \\
\ul{Step 1: Non-emptiness of $\mathcal{A}(k,\varepsilon,B,C)$.} Fix $k\in\mathbb{N}$ and an admissible set of data $(r_0,\sigma_0,(\partial_x r)_0,(\Gamma_x)_0)$, together with associated constants $\delta_i,N_i$ as in \hyperlink{lem:6.1}{Lemma~6.1}. Applying that lemma, we have constants $\varepsilon,B,C$ and a self-map $\Phi:\mathcal{A}(k,\varepsilon,B,C)\to\mathcal{A}(k,\varepsilon,B,C)$ which is contractive for the metric $d$, i$.$e$.$ for the $C^2$ norm computed for the metric components $(\Omega^2, r)$ only.\\ \\
We can pick metric components, which we name $(\Omega^2_1,r_1)$, given by suitable polynomials in $\xi$ of degree $k+2$, with coefficients depending on $t$, so that $k^\text{th}$ order boundary conditions are satisfied. The leading terms would be
$$\log\Omega^2_1(t,x)=2(\Gamma_x)_0(t)\xi^2+\cdots \qquad r_1(t,x)= r_0(t)+\cdots$$
These belong to $C^\infty_\text{cusp}(\Sigma_\varepsilon)$, but may fail to satisfy the bounds in \hyperlink{def:2.5}{Definition~2.5}. For this reason, we enlarge $C>0$ and reduce $\varepsilon>0$, if needed, so that $(\Omega^2_1,r_1)$ satisfy
$$ \inf_{\Sigma_\varepsilon}|\Gamma_x|\geq\tfrac{1}{2}\delta_1,\hspace{4pt}\sup_{\Sigma_\varepsilon}|\Gamma_t|\leq\tfrac{1}{2}\delta_1,\hspace{4pt}\inf_{\Sigma_\varepsilon} r\geq\tfrac{1}{2}\delta_2$$
and so that the upper bounds hold in terms of $C$. This is at the cost of changing the values of $B,\varepsilon$, as mentioned at the end of the statement of \hyperlink{lem:6.1}{Lemma~6.1}. It now follows from the way $B,\varepsilon$ are chosen in \hyperlink{lem:6.1}{Lemma~6.1} that the induced $(\bm{u}_\pm)_1,(\sigma_\pm)_1$ obey the other properties of \hyperlink{def:2.5}{Definition~2.5}, including the bounds in terms of $B$, and so, for this new choice of constants, we have an element $(\Omega^2_1,r_1,(\bm{u}_\pm)_1,(\sigma_\pm)_1)$ of $\mathcal{A}(k,\varepsilon,B,C)$.\\ \\
\ul{Step 2: Applying the iteration scheme.} Define now a sequence $(\Omega^2_n,r_n,(\bm{u}_\pm)_n,(\sigma_\pm)_n)^\infty_{n=1}$ of elements of $\mathcal{A}(k,\varepsilon,B,C)$ through the self-map $\Phi$, namely
$$ (\Omega^2_{n+1},r_{n+1},(\bm{u}_\pm)_{n+1},(\sigma_\pm)_{n+1}):=\Phi\big((\Omega^2_n,r_n,(\bm{u}_\pm)_n,(\sigma_\pm)_n)\big)$$ for all $n\geq 1$. In order to apply the contraction property, we need to know that $d$ is defined for pairs of consecutive elements of the sequence. This follows from the fact that the definition of $\mathcal{A}(k,\varepsilon,B,C)$ includes $k^\text{th}$ order boundary conditions, which already at the level $k=1$ are sufficient to ensure that the possibly unbounded differences $\partial^2_x\log\Omega^2_1-\partial^2_x\log\Omega^2_2$ and $\partial^2_xr_1-\partial^2_xr_2$ are actually bounded.
\\ \\
Applying the contraction property, we now have inductively that, for all $n\geq 2$, 
$$d((\Omega^2_{n+2},r_{n+2}),(\Omega^2_{n+1},r_{n+1})) \leq \tfrac{1}{2}d((\Omega^2_{n+1},r_{n+1}),(\Omega^2_{n},r_{n}))$$
and so $(\Omega^2_n,r_n)$ is Cauchy in the $C^2$ norm. Applying \hyperlink{lem:5.1}{Lemma~5.1}, we have immediately that $(\bm{u}_\pm)_n, (\sigma_\pm)_n$ are also Cauchy, in the $C^1$ and $C^0$ norm respectively. (Of course, $(\Omega^2_n,r_n)$ itself does not have finite $C^2$ norm. If the reader is concerned about this use of language, one may say instead that e$.$g$.$ $(\Omega^2_n-\Omega^2_1,r_n-r_1)$ is Cauchy in the $C^2$ norm.) We conclude that the variables converge to a limit $(\Omega^2, r, \bm{u}_\pm, \sigma_\pm)$ with $\Omega^2,r\in C^2(\Sigma_\varepsilon)$, $\bm{u}_\pm\in C^1(\Sigma_\varepsilon)$, $\sigma_\pm\in C^0(\Sigma_\varepsilon)$.\\ \\
\ul{Step 3: Propagation of regularity.} The analysis in Steps 1 and 2 applied to an arbitrary $k\in\mathbb{N}$. The constants $\varepsilon,B,C$ involved in the self-map will depend on $k$, and we write $\varepsilon_k$ as a notational reminder of this dependence. The sequences of iterates $(\Omega^2_n,r_n,(\bm{u}_\pm)_n,(\sigma_\pm)_n)^\infty_{n=1}$ in each case are defined on different domains $\Sigma_{\varepsilon_k}$, but we observe that the iterates coincide on the intersection of the domains. This is because they are obtained as unique solutions to PDE with data taken at $x=0$. Consequently, the limit $(\Omega^2,r,\bm{u}_\pm,\sigma_\pm)$ is unique, not depending on $k$.\\ \\
We fix now $k=2$ (for reasons to be explained momentarily), and write $\varepsilon:=\varepsilon_2$, which will be our final value of $\varepsilon$. In Step~2 above, we obtained $(\Omega^2, r, \bm{u}_\pm, \sigma_\pm)$ which was the limit of iterates $(\Omega^2_n,r_n,(\bm{u}_\pm)_n,(\sigma_\pm)_n)$ satisfying the equations
\begin{align*}
(-\partial^2_x+\partial^2_t)(\log\Omega^2_{n+1})=&\frac{4\Omega^2_nm_n}{r^3_n}-4\Omega^2_n((\sigma_+)_n+(\sigma_-)_n) x^{-1/2}\\
(-\partial^2_x+\partial^2_t)\hspace{1pt}r_{n+1}=-&\frac{2\Omega^2_nm_n}{r^2_n}+2r_n\Omega^2_n((\sigma_+)_n+(\sigma_-)_n) x^{-1/2}
\end{align*}
Since we have $C^2$ convergence of $\Omega^2_n,r_n$, and $C^0$ convergence of $(\sigma_\pm)_n$, we can take pointwise limits in these equations and deduce that $(\Omega^2,r)$ satisfy the Einstein equations 
\begin{align*}
(-\partial^2_x+\partial^2_t)(\log\Omega^2)=&\frac{4\Omega^2m}{r^3}-4\Omega^2(\sigma_++\sigma_-) x^{-1/2}\\
(-\partial^2_x+\partial^2_t)\hspace{1pt}r=-&\frac{2\Omega^2m}{r^2}+2r\Omega^2(\sigma_++\sigma_-) x^{-1/2}
\end{align*}
Similarly, we can use $C^2$ convergence of $\Omega^2_n,r_n$ and $C^1$ convergence of $(\bm{u}^t_\pm)_n$, $(\bm{u}^x_\pm)_n$ to take pointwise limits on both sides of 
\begin{align*}
(\bm{u}_\pm(\bm{u}^t_\pm))_n+(\Gamma_t)_n\big((\bm{u}^t_\pm)^2_n+(\bm{u}^x_\pm)^2_n\big)+2(\Gamma_x)_n(\bm{u}^t_\pm)_n(\bm{u}^x_\pm)_n&=0\\
(\bm{u}_\pm(\bm{u}^x_\pm))_n+(\Gamma_x)_n\big((\bm{u}^t_\pm)^2_n+(\bm{u}^x_\pm)^2_n\big)+2(\Gamma_t)_n(\bm{u}^t_\pm)_n(\bm{u}^x_\pm)_n&=0
\end{align*}
and conclude that $\bm{u}_\pm$ satisfies
\begin{align*}
\bm{u}_\pm(\bm{u}^t_\pm)+\Gamma_t\big((\bm{u}^t_\pm)^2+(\bm{u}^x_\pm)^2\big)+2\Gamma_x\bm{u}^t_\pm\bm{u}^x_\pm&=0\\
\bm{u}_\pm(\bm{u}^x_\pm)+\Gamma_x\big((\bm{u}^t_\pm)^2+(\bm{u}^x_\pm)^2\big)+2\Gamma_t\bm{u}^t_\pm\bm{u}^x_\pm&=0
\end{align*}
The situation is not so immediate for $\sigma_\pm$, because the convergence is only $C^0$ for that variable. However, because we have taken $k=2$, we have uniform bounds for $\|(\sigma_\pm)_n\|_{C^2_\text{cusp}}$. By Arzel\'{a}-Ascoli, there is a subsequence $\phi(n)$ with the property that $(\sigma_\pm)_{\phi(n)}$ converges in the $C^1_\text{cusp}$ norm. This must coincide with the $C^0$ limit which we named $\sigma_\pm$. We can now take pointwise limits in the equation
$$(\bm{u}_\pm(\sigma_\pm))_{\phi(n)}+(\sigma_\pm)_{\phi(n)}\big(\partial_\xi(\bm{u}^\xi_\pm)_{\phi(n)}+\partial_t(\bm{u}^t_\pm)_{\phi(n)}+(\bm{u}_\pm(\log\Omega^2))_{\phi(n)}+(\bm{u}_\pm(\log r^2))_{\phi(n)}\big)=0 $$
to conclude that 
$$\bm{u}_\pm(\sigma_\pm)+\sigma_\pm\big(\partial_\xi\bm{u}^\xi_\pm+\partial_t\bm{u}^t_\pm+\bm{u}_\pm(\log\Omega^2)+\bm{u}_\pm(\log r^2)\big)=0 $$
So the collection $(\Omega^2, r, \bm{u}_\pm, \sigma_\pm)$ solves the Einstein and matter equations on $\Sigma_\varepsilon$.\\ \\
We will now argue along similar lines that $(\Omega^2, r, \bm{u}_\pm, \sigma_\pm)$ actually possesses the regularity $C^\infty_\text{cusp}$. By applying Arzel\'{a}-Ascoli in exactly the above manner, we may conclude that, on the domain $\Sigma_{\varepsilon_k}$, the metric components $\Omega^2,r$ belong to $C^{k+1}_\text{cusp}$, the fluid velocities $\bm{u}_\pm$ belong to $C^k_\text{cusp}$, and the energy densities $\sigma_\pm$ belong to $C^{k-1}_\text{cusp}$. For each fixed $k$, we may now uniquely extend $(\Omega^2, r, \bm{u}_\pm, \sigma_\pm)$ as a $C^{k+1}\times C^k\times C^{k-1}$ solution to the joint Einstein-matter equations from $\Sigma_{\varepsilon_k}$ to $\Sigma_\varepsilon$ by means of \hyperlink{lem:6.2}{Lemma~6.2}. Here we use the known existence of the $C^2\times C^1\times C^0$ solution on $\Sigma_\varepsilon$ (which must coincide with any solution of higher regularity) to show that the blowup criterion is avoided. Hence $(\Omega^2, r, \bm{u}_\pm, \sigma_\pm)$ actually possesses the regularity $C^\infty_\text{cusp}$ as advertised.
\\ \\
\ul{Step 4: Propagation of constraints.} So far, we have a collection $(\Omega^2,r,\bm{u}_\pm,\sigma_\pm)$, defined on a domain $\Sigma_\varepsilon$, with regularity $C^\infty_\text{cusp}$ and which satisfy equations \hyperlink{eqn:2.5}{(2.5)}-\hyperlink{eqn:2.9}{(2.9)}. In particular, the energy-momentum tensor $T^{\mu\nu}$ induced by $\bm{u}_\pm,\sigma_\pm$ is covariantly conserved. We now wish to show that the constraint equations \hyperlink{eqn:2.3}{(2.3)}-\hyperlink{eqn:2.4}{(2.4)} also hold on $\Sigma_\varepsilon$ under these conditions, if they already hold (in the limiting sense) at $x=0$, and therefore that the full set of Einstein equations is satisfied.
\\ \\
A standard (yet tedious) calculation reveals that, if \hyperlink{eqn:2.5}{(2.5)}-\hyperlink{eqn:2.6}{(2.6)} hold, together with $\nabla^\mu T_{\mu\nu}=0$, then 
$$\partial_x\Big(r\partial^2_tr-r\Gamma_t\partial_xr-r\Gamma_x\partial_tr+r^2T_{tx}\Big)=0 $$
$$\partial_x\Big(r\partial_x\partial_tr-r\Gamma_t\partial_tr-r\Gamma_x\partial_xr+\frac{\Omega^2m}{r}+r^2T_{xx}\Big)=0 $$
The quantities in large brackets above converge to 0 as $x\to 0$, precisely by our choice of initial data satisfying \hyperlink{eqn:2.10}{(2.10)}-\hyperlink{eqn:2.11}{(2.11)}, which was the restriction of \hyperlink{eqn:2.3}{(2.3)}-\hyperlink{eqn:2.4}{(2.4)} to $\{x=0\}$ (see also \hyperlink{sec:2.2}{Section~2.2}). It follows that equations \hyperlink{eqn:2.3}{(2.3)}-\hyperlink{eqn:2.4}{(2.4)} propagate to the whole of $\Sigma_\varepsilon$, as desired.
\\ \\
\ul{Step 5: Gluing to the vacuum region.} It is an algebraic consequence of the vacuum Einstein equations in spherical symmetry (i$.$e$.$ \hyperlink{eqn:2.3}{(2.3)}-\hyperlink{eqn:2.6}{(2.6)} taken with $T_{\mu\nu}\equiv0$) that the Hawking mass $m$ is constant, say $m=m_0$. Indeed, this is a step in the proof of Birkhoff's theorem. Solving the equations in the negative $x$ direction thus reduces to solving the following PDE problem on a suitable domain.
\begin{align*}
(-\partial^2_x+\partial^2_t)\log\Omega^2 =&\frac{4\Omega^2 m_0}{r^3}\tag{\hypertarget{eqn:6.1}{6.1}}\\
(-\partial^2_x+\partial^2_t)r =-&\frac{2\Omega^2 m_0}{r^2}\tag{\hypertarget{eqn:6.2}{6.2}}
\end{align*}
$$\log\Omega^2|_{x=0}=0,\quad \partial_x\log\Omega^2|_{x=0}=2(\Gamma_x)_0, \qquad r|_{x=0}=r_0,\quad \partial_xr|_{x=0}=(\partial_xr)_0\vspace{1mm}$$
If we are to have uniqueness, we are restricted to a region bounded by the null curves $t-x=b$ and $t+x=a$. Since our goal is to solve for a neighbourhood of the caustic curve located at $x=0$, an appropriate domain is the trapezium $\Sigma_{-\varepsilon}$ bounded by $x=0$, $x=-\varepsilon$ and the null curves mentioned above---see the Figure in \hyperlink{thm:1.1}{Theorem~1.1}. Of course, we may need to (for the final time) reduce $\varepsilon$ to achieve this.\\ \\
The existence and uniqueness of a smooth solution $(\Omega^2,r)$ to \hyperlink{eqn:6.1}{(6.1)}-\hyperlink{eqn:6.2}{(6.2)} on this domain (for \textit{some} possibly smaller $\varepsilon>0$), taking the given data, and with $r$ bounded away from 0, follows from a standard, very short, contraction mapping argument which we omit. By Birkhoff's theorem, the solution obtained on this trapezium is isometric to an open subset of the Schwarzschild spacetime with mass $m_0$. We remark that the new $\varepsilon$ depends only on $\delta_2$, $\|(\Gamma_x)_0\|_{C^0}$ and $\|(\partial_xr)_0\|_{C^0}$.
\\ \\
We now have $\Omega^2,r$ defined on $\Sigma_\varepsilon\cup\Sigma_{-\varepsilon}$. Let $U\subset\mathbb{R}^2$ be the union of $\Sigma_\varepsilon$, $\Sigma_{-\varepsilon}$ and the caustic $\{t\in(a,b),x=0\}$, i$.$e$.$ precisely the domain conveyed in \hyperlink{thm:1.1}{Theorem~1.1}. Then $\Omega^2,r$ extend to $C^1$ functions on $U$. The first derivatives $\partial_t\Omega^2$, $\partial_x\Omega^2$, $\partial_tr$, $\partial_xr$ are cusp-regular on $\Sigma_\varepsilon$, as well as smooth on $\Sigma_{-\varepsilon}$ with extension to $x=0$. It follows straightforwardly that they hence belong to the H\"{o}lder class $C^{1/2}$, and so $\Omega^2,r$ themselves belong to $C^{1,1/2}$ as claimed.
\\ \\
It remains to discuss whether the collection $(\Omega^2,r,T_{\mu\nu})$ is a weak solution to Einstein's equation, in the sense of \hyperlink{def:b.1}{Definition~B.1}. $T_{\mu\nu}$ is defined already on $\Sigma_\varepsilon$, and is identically zero on $\Sigma_{-\varepsilon}$, while $\Omega^2,r$, as just discussed, extend to $C^1$ functions on $U$. The conditions of \hyperlink{pro:b.1}{Proposition~B.1} are then clearly satisfied: for the integrability of $T_{\mu\nu}$ across $x=0$, we use the fact that each component is bounded above by a multiple of $x^{-1/2}$, which we can integrate on any compact domain of $\mathbb{R}^2$. We conclude that we have a weak solution of Einstein's equation on the whole domain, and the proof is complete.
\hfill $\square$\\ \\ \\
\newpage
\Large\hypertarget{sec:7}{\textbf{7\hspace{4mm}Static dust caustics}}\label{7}\normalsize\\ \\
We now turn to the static examples which are the subject of \hyperlink{thm:1.2}{Theorem~1.2}. The proof is comparatively simple, as the further assumption of staticity reduces the problem to studying the properties of an ODE system for the dynamical variables, which turns out to be first-order. By choosing a radial parametrization, and identifying several conserved quantities, the final system has only two variables: the lapse function $\Omega^2$ and Hawking mass $m$. See already \hyperlink{eqn:7.8}{(7.8)}-\hyperlink{eqn:7.9}{(7.9)} below.\\ \\
\large\hypertarget{sec:7.1}{\textbf{7.1\hspace{4mm}Reduction to ODE system}}\label{7.1}\normalsize\\ \\
To set up the problem, we return to equations \hyperlink{eqn:2.3}{(2.3)}-\hyperlink{eqn:2.9}{(2.9)}, describing the dynamics of a 2-dust spacetime $(\Omega^2,r,\bm{u}^t_\pm,\bm{u}^x_\pm,\rho_\pm)$ with respect to coordinates $(t,x)$ in which the metric takes the form
$$g=\Omega^2(-dt^2+dx^2)+r^2g_{S^2} $$
We now consider those solutions, if any, which are independent of the $t$-coordinate, representing steady-state flow. This entails that all $\partial_t$-derivatives vanish, implying via \hyperlink{eqn:2.3}{(2.3)} that the energy-momentum component $T^{tx}$ also vanishes. To arrange this, it is sufficient to impose that 
\begin{equation}
\rho_+\equiv \rho_-,\qquad \bm{u}^t_+\equiv \bm{u}^t_-,\qquad \bm{u}^x_+\equiv -\bm{u}^x_-\tag{\hypertarget{eqn:7.1}{7.1}}
\end{equation}
which also reduces the number of variables. That is, the ingoing and outgoing species are exact `mirror images' of each other. We simply write $\rho,\bm{u}^t,\bm{u}^x$ from now on, referring to the outgoing `$+$' species. \\ \\
The system reduces further in view of the following relations. As before, we have that 
\begin{equation}
g(\bm{u},\bm{u}):=\Omega^2\big(-(\bm{u}^t)^2+(\bm{u}^x)^2\big)\equiv -1\tag{\hypertarget{eqn:7.2}{7.2}}
\end{equation}
The $t$-component of the geodesic equation also yields a conserved quantity. Discarding $\partial_t$-derivatives from \hyperlink{eqn:2.7}{(2.7)}, and rearranging, establishes that
\begin{equation}
\partial_x(\Omega^2 \bm{u}^t)=0\quad\implies\quad \Omega^2\bm{u}^t=\text{const.}\tag{\hypertarget{eqn:7.3}{7.3}}
\end{equation}
Meanwhile, the conservation equation (2.9) for the energy density $\rho$ becomes simply
\begin{equation}
\partial_x(\rho\bm{u}^x\Omega^2r^2)=0\quad\implies \quad\rho\bm{u}^x\Omega^2r^2=\text{const.}\tag{\hypertarget{eqn:7.4}{7.4}}
\end{equation}
As in Section~2, the level curve $\{x=0\}$ is to represent a caustic curve $\mathcal{C}$, where in particular $\bm{u}^x=0$, and where we make the gauge choice $\Omega^2|_{x=0}=1$. It follows that the constant in \hyperlink{eqn:7.3}{(7.3)} is simply unity, and inserting this into \hyperlink{eqn:7.2}{(7.2)} yields that
\begin{equation}
\Omega^4(\bm{u}^x)^2=1-\Omega^2\tag{\hypertarget{eqn:7.5}{7.5}}
\end{equation}
This in turn allows us to eliminate the fluid velocity from \hyperlink{eqn:7.4}{(7.4)}. Naming the constant $\varsigma>0$, we have
\begin{equation}
\rho r^2\sqrt{1-\Omega^2}=\varsigma\tag{\hypertarget{eqn:7.6}{7.6}}
\end{equation}
Thus knowledge of the metric components $\Omega^2,r$ allows us to recover the matter variables $(\rho,\bm{u}^t,\bm{u}^x)$ through \hyperlink{eqn:7.3}{(7.3)}, \hyperlink{eqn:7.5}{(7.5)}, \hyperlink{eqn:7.6}{(7.6)}.\\ \\
The variables $\Omega^2(x),r(x)$ obey an ODE system in the independent variable $x$. However, a simpler system of equations is satisfied by $\Omega^2(r),m(r)$, taking $r$ as the independent variable. In particular, the ODE are first-order. To perform this radial reparametrization, note that the definition \hyperlink{eqn:2.2}{(2.2)} of Hawking mass now reads
\begin{equation}
1-\frac{2m}{r}=\Omega^{-2}(\partial_xr)^2\tag{\hypertarget{eqn:7.7}{7.7}}
\end{equation}
Making $\partial_xr$ the subject of this equation allows us to relate $\partial_x$-derivatives to $\partial_r$. Note that we still write $\partial_x$, $\partial_r$ etc$.$ even though these are now all ordinary derivatives.\\ \\
We next observe that, when $\partial_t$-derivatives vanish, equation \hyperlink{eqn:2.4}{(2.4)} already determines $\Omega^2$. Moreover, the equation is first order, making it preferable to \hyperlink{eqn:2.5}{(2.5)}. Rearranging \hyperlink{eqn:2.4}{(2.4)}, and applying equations \hyperlink{eqn:7.5}{(7.5)}-\hyperlink{eqn:7.7}{(7.7)}, one arrives at the evolution equation for $\Omega^2$
\begin{equation}
\partial_r\Omega^2=\bigg(1-\frac{2m}{r}\bigg)^{-1}\bigg(\frac{2\Omega^2m}{r^2}+\frac{4\varsigma}{r}\sqrt{1-\Omega^2}\bigg)\tag{\hypertarget{eqn:7.8}{7.8}}
\end{equation}
Finally, we close the system by deriving an equation for $m$. It is here that we make use of equation \hyperlink{eqn:2.6}{(2.6)}. Differentiating \hyperlink{eqn:7.7}{(7.7)} with respect to $r$, and rearranging with the help of \hyperlink{eqn:2.6}{(2.6)} and \hyperlink{eqn:7.6}{(7.6)}-\hyperlink{eqn:7.8}{(7.8)}, we obtain
\begin{equation}
\partial_rm=\frac{2\varsigma}{\sqrt{1-\Omega^2}}\bigg(1+\frac{1}{\Omega^2}\bigg)\tag{\hypertarget{eqn:7.9}{7.9}}
\end{equation}
We see that the pair \hyperlink{eqn:7.8}{(7.8)}-\hyperlink{eqn:7.9}{(7.9)} is a closed ODE system for $(\Omega^2,m)$, making sense on $r>0$ and whenever the variables satisfy $$0<\Omega^2<1,\qquad 1-\frac{2m}{r}\neq0$$
Given a solution, the calculations above can be reversed to first obtain the parameter $x$, and then uniquely recover a full collection $(\Omega^2,r,\bm{u}^t_\pm,\bm{u}^x_\pm,\rho_\pm)$ of functions of $x$ and (trivially) of $t$, which satisfy \hyperlink{eqn:2.3}{(2.3)}-\hyperlink{eqn:2.9}{(2.9)}.\\ \\
\large\hypertarget{sec:7.2}{\textbf{7.2\hspace{4mm}Obtaining a 3-parameter family of solutions}}\label{7.2}\normalsize\\ \\
We proceed to study the properties of \hyperlink{eqn:7.8}{(7.8)}-\hyperlink{eqn:7.9}{(7.9)} and its solutions in more detail. The solutions we will obtain all take the form indicated in the Figure below, and in fact these comprise all the solutions of \hyperlink{eqn:7.8}{(7.8)}-\hyperlink{eqn:7.9}{(7.9)} for which $1-2m/r>0$ (but we will not quite need to show this).\\ \\
\begin{minipage}[t]{0.55\textwidth}We refer to the endpoints of the domain as the \textit{inner (resp$.$ outer) radius} $r_\text{in}$ ($r_\text{out}$). At these endpoints, $\Omega^2\to 1$, corresponding (see \hyperlink{eqn:7.5}{(7.5)}) to a caustic curve. The Hawking mass $m$ is strictly increasing with $r$, and $\partial_rm$ diverges at the endpoints, in agreement with the results of \hyperlink{thm:1.1}{Theorem~1.1}. However, $m$ reaches a finite limit there, the \textit{inner (resp$.$ outer) mass} $m_\text{in}$ ($m_\text{out}$).\\ \\
The three indicated regions (I)-(III) are always non-empty. In region (I), $\Omega^2$ is strictly decreasing, with $m<0$. In regions (II) and (III), $\Omega^2$ is strictly increasing, with $m<0$ in (II) and $m>0$ in (III).\end{minipage}\hspace{0.03\textwidth}
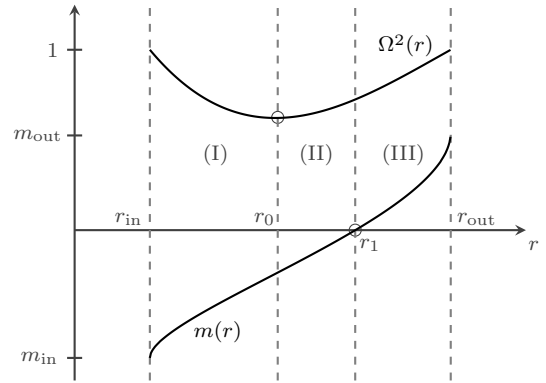
\begin{minipage}[t]{0.42\textwidth}
\vspace{-5mm}
\begin{figure}[H]
\begin{center}
\begin{tikzpicture}[scale=1]
\filldraw[color=darkgray, fill=white](3.7,1.5) circle (0.08);
\filldraw[color=darkgray, fill=white](4.73,0) circle (0.08);
\draw [-stealth, darkgray, thick](1.0,-2) -- (1.0,3);
\draw [-stealth, darkgray, thick] (1.0, 0) -- (7.0, 0);
\draw [darkgray, thick] (0.9, 1.26) -- (1.1, 1.26);
\draw [darkgray, thick] (0.9, 2.4) -- (1.1, 2.4);
\draw [darkgray, thick] (0.9, -1.7) -- (1.1, -1.7);
\draw [gray, thick, dashed] (2, -2) -- (2, 3);
\draw [gray, thick, dashed] (3.7, -2) -- (3.7, 3);
\draw [gray, thick, dashed] (4.73, -2) -- (4.73, 3);
\draw [gray, thick, dashed] (6, -2) -- (6, 3);
\draw [black, thick, domain=2:6, samples=300] plot ({\x},{1.2+0.3*(\x-4)^2+0.04*(\x-2)^2*(6-\x)});
\draw [black, thick, domain=2:6, samples=300] plot ({\x},{0.3+0.5*(\x-2)^(0.5)-(6-\x)^(0.5)});
\node[align=left, darkgray] at (0.05, 1.26) [anchor = west] {$_{m_\text{out}}$};
\node[align=left, darkgray] at (0.15, -1.7) [anchor = west] {$_{m_\text{in}}$};
\node[align=left, darkgray] at (0.5, 2.4) [anchor = west] {$_{1}$};
\node[align=left, black] at (2.45, -1.35) [anchor = west] {$_{m(r)}$};
\node[align=left, black] at (4.9, 2.45) [anchor = west] {$_{\Omega^2(r)}$};
\node[align=left, darkgray] at (2.57, 1) [anchor = west] {$_\text{(I)}$};
\node[align=left, darkgray] at (3.85, 1) [anchor = west] {$_\text{(II)}$};
\node[align=left, darkgray] at (4.95, 1) [anchor = west] {$_\text{(III)}$};
\node[align=left, darkgray] at (1.4, 0.15) [anchor = west] {$_{r_\text{in}}$};
\node[align=left, darkgray] at (3.25, 0.15) [anchor = west] {$_{r_0}$};
\node[align=left, darkgray] at (4.65, -0.2) [anchor = west] {$_{r_1}$};
\node[align=left, darkgray] at (5.95, 0.15) [anchor = west] {$_{r_\text{out}}$};
\node[align=left, darkgray] at (6.9, -0.2) [anchor = west] {$_{r}$};
\end{tikzpicture}
\end{center}
\end{figure}
\end{minipage}\vspace*{6mm}\\
We begin by obtaining a small piece of the above solution, near the lower endpoint, with $r_\text{in}>0$, $m_\text{in}<0$ prescribed. Since the ODE system \hyperlink{eqn:7.8}{(7.8)}-\hyperlink{eqn:7.9}{(7.9)} is only defined for $\Omega^2\in(0,1)$, and $\Omega^2\to 1$ at $r=r_\text{in}$, we cannot simply apply standard results on the local existence of ODEs, starting from $r=r_\text{in}$. However, since local existence near caustic curves is precisely what \hyperlink{thm:1.1}{Theorem~1.1} achieves, we obtain this solution as a corollary of that result. On the other hand, the work needed to establish \hyperlink{thm:1.1}{Theorem~1.1} is far more involved than is needed for this ODE application, and one could employ a simple fixed point argument instead. The present choice is only made for brevity.\\ \\
\hypertarget{cor:7.1}{\textbf{Corollary 7.1.}} \textit{Let $m_\text{in}<0$ and $\varsigma,r_\text{in}>0$ be constants. Then there exists $\varepsilon>0$ and a unique solution $\Omega^2,m:(r_\text{in},r_\text{in}+\varepsilon)\to\mathbb{R}$ to the system \hyperlink{eqn:7.8}{(7.8)}-\hyperlink{eqn:7.9}{(7.9)} with constant $\varsigma$ and satisfying
 $$\lim_{r\to r_\text{in}}\Omega^2=1\qquad \lim_{r\to r_\text{in}}m=m_\text{in}$$
 In particular, the solution is inextendible to the left, since the ODE is only defined for $\Omega^2\in(0,1)$.}\\ \\
\textit{Proof.} Fix $[a,b]=[0,1]$, say, and define the (constant) functions
$$r_0(t)\equiv r_\text{in},\qquad \sigma_0(t)\equiv \frac{\varsigma}{r_\text{in}(-2m_\text{in})^{1/2}}\Big(1-\frac{2m_\text{in}}{r_\text{in}}\Big)^{1/4}>0 $$
with $m_0=m_\text{in}$. This choice of $\sigma_0$ ensures precisely that $\lim_{x\to 0}\big(\rho\bm{u}^x\Omega^2 r^2\big)=\varsigma$. These data satisfy the conditions in \hyperlink{sec:2.3}{Section~2.3}, and we may thus invoke \hyperlink{thm:1.1}{Theorem~1.1}.
\\ \\
In the resulting solution in the given $(t,x)$ coordinates, all variables are $t$-independent. Indeed, this already holds for the iterates $(\Omega^2_n,r_n,(\bm{u}_\pm)_n,(\sigma_\pm)_n)^\infty_{n=1}$ that converge, in the proof of \hyperlink{thm:1.1}{Theorem~1.1}, to the solution $(\Omega^2,r,\bm{u}_\pm,\sigma_\pm)$. This is manifestly true for the initial iterate (see Step~1 in \hyperlink{sec:6.2}{Section~6.2}) and is easily seen to be be preserved by the contraction map of \hyperlink{lem:6.1}{Lemma~6.1}. Hence we have $\Omega^2=\Omega^2(x)$, $r=r(x)$, $m=m(x)$, defined on some $x\in(0,\varepsilon)$. Since $\partial_xr|_{x=0}>0$, we have $\partial_xr>0$ on some possibly smaller domain. Thus we may reparametrize to $\Omega^2=\Omega^2(r)$, $m=m(r)$, and, by reversing the calculations of \hyperlink{sec:7.1}{Section~7.1}, these indeed satisfy equations \hyperlink{eqn:7.8}{(7.8)}-\hyperlink{eqn:7.9}{(7.9)} with the correct limits at $r=r_\text{in}$.\hfill$\square$\\ \\
\hypertarget{pro:7.2}{\textbf{Proposition 7.2.}} \textit{Let $m_\text{in}<0$ and $\varsigma,r_\text{in}>0$ be constants. The solution $\Omega^2,m$ to the system
\begin{equation}
\partial_r\Omega^2=\bigg(1-\frac{2m}{r}\bigg)^{-1}\bigg(\frac{2\Omega^2m}{r^2}+\frac{4\varsigma}{r}\sqrt{1-\Omega^2}\bigg)\tag{\hypertarget{eqn:7.8b}{7.8}}
\end{equation}
\begin{equation}
\partial_rm=\frac{2\varsigma}{\sqrt{1-\Omega^2}}\bigg(1+\frac{1}{\Omega^2}\bigg)\tag{\hypertarget{eqn:7.9b}{7.9}}
\end{equation}
obtained in \hyperlink{cor:7.1}{Corollary 7.1} extends uniquely to a maximal solution on $(r_\text{in},r_\text{out})$ for some $r_\text{out}<\infty$. Moreover, we have
$$\lim_{r\to r_\text{out}}\Omega^2=1\qquad \lim_{r\to r_\text{out}}m\in(0,\tfrac{1}{2}r_\text{out})$$}\\
\textbf{Remark 7.1. }\textit{We emphasize the content of this proposition: not only must the maximal solution always terminate after a finite interval, but this always occurs due to the appearance of a second (interior) caustic curve, and with positive lower bounds on the quantity $1-2m/r$, guaranteeing the absence of trapped surfaces in these examples.}\\ \\
\textit{Proof of Proposition 7.2. }By standard ODE theory, the solution $\Omega^2,m:(r_\text{in},r_\text{in}+\varepsilon)\to\mathbb{R}$ in \hyperlink{cor:7.1}{Corollary 7.1} extends uniquely as a smooth solution to the ODE \hyperlink{eqn:7.8b}{(7.8)}-\hyperlink{eqn:7.9b}{(7.9)} to a maximal domain $(r_\text{in},r_\text{out})$, where at the moment possibly $r_\text{out}=\infty$. Moreover, via the Escape Lemma (see e$.$g$.$ [\hyperlink{Lee13}{Lee13}, Lemma~9.19]), one sees that, if the solution terminates after a finite interval, i$.$e$.$ $r_\text{out}<\infty$, then at least one of the following must hold:
 $$\lim_{r\to r_\text{out}}\Omega^2=1,\quad \lim_{r\to r_\text{out}}\Omega^2=0,\quad \lim_{r\to r_\text{out}}m=-\infty,\quad\lim_{r\to r_\text{out}}\Big(1-\frac{2m}{r}\Big)=0,$$
 and we will use this characterization in the sequel. Note that we consider only those solutions satisfying $m(r)<r/2$ throughout their domain. This already holds near $r=r_\text{in}$ (since $m_\text{in}<0$), and it can be shown that if $\lim_{r\to r_*}(1-2m/r)=0$ during evolution, then necessarily $\lim_{r\to r_*}\partial_r\Omega^2=\infty$, so that we exit the category of smooth solutions. In fact, $\lim_{r\to r_*}(1-2m/r)=0$ is impossible within the maximal solution, as we will presently show.\\ \\
 \underline{Region (I): Evolution until $\partial_r\Omega^2=0$:} We first claim that, for some $r_0\in(r_\text{in},r_\text{out})$, we have $\partial_r\Omega^2(r_0)=0$. It is clear from \hyperlink{eqn:7.8b}{(7.8)} that $\lim_{r\to r_\text{in}}\partial_r\Omega^2<0$, so that, were this claim false, we would have $\partial_r\Omega^2<0$ throughout $(r_\text{in},r_\text{out})$. Suppose for contradiction, then, that $\partial_r\Omega^2<0$ on $(r_\text{in},r_\text{out})$. It follows also that $m<0$ throughout $(r_\text{in},r_\text{out})$, since $m<0$ near $r=r_\text{in}$, and whenever $0\leq m< r/2$, \hyperlink{eqn:7.8b}{(7.8)} implies that $\partial_r\Omega^2>0$. The fact that $m<0$ on $(r_\text{in},r_\text{out})$ already precludes $r_\text{out}=\infty$, because we see from \hyperlink{eqn:7.9b}{(7.9)} that $\partial_rm\geq 4\varsigma>0$, so $m$ must reach 0 within a finite interval. The only way for the solution to terminate at $r=r_\text{out}$, then, is that $\lim_{r\to r_\text{out}}\Omega^2=0$ (this function being decreasing). However, we then have
 $$\lim_{r\to r_\text{out}}\partial_r\Omega^2=\bigg(1-\frac{2\lim_{r\to r_\text{out}}m}{r_\text{out}}\bigg)^{-1}\frac{4\varsigma}{r_\text{out}}>0 $$
 a contradiction, since $\partial_r\Omega^2<0$ on $r\in(r_\text{in},r_\text{out})$. Hence $\partial_r\Omega^2$ indeed vanishes somewhere in $(r_\text{in},r_\text{out})$. Let $r_0$ be the first such value. Note that we still have $m(r_0)<0$, again because $m(r_0)=0$ would would make $\partial_r\Omega^2(r_0)>0$, but we have $\partial_r\Omega^2(r_0)=0$.
 \\ \\
 \underline{Region (II): Evolution until $m=0$:} We now study the maximal forwards solution starting from $r=r_0$, at which $\partial_r\Omega^2(r_0)=0$, $m(r_0)<0$, and claim first that, for all $r>r_0$, we have $\partial_r\Omega^2>0$. Indeed, suppose that, at some arbitrary $r=r_*$, we have $\partial_r\Omega^2(r_*)=0$ and $m(r_*)\leq0$. Then we have, from \hyperlink{eqn:7.8b}{(7.8)},
 \begin{equation}
 \bigg(\frac{2\Omega^2m}{r^2}+\frac{4\varsigma}{r}\sqrt{1-\Omega^2}\bigg)\bigg|_{r=r_*}=0\tag{\hypertarget{eqn:7.10}{7.10}}
 \end{equation}
 Differentiating this bracketed expression, and evaluating at $r=r_*$, we may discard $\partial_r\Omega^2$ terms, yielding
 \begin{align*}
 \partial_r\bigg(\frac{2\Omega^2m}{r^2}+\frac{4\varsigma}{r}\sqrt{1-\Omega^2}\bigg)\bigg|_{r=r_*}&=\bigg(\frac{2\Omega^2\partial_rm}{r^2}-\frac{4\Omega^2m}{r^3}-4\varsigma\frac{\sqrt{1-\Omega^2}}{r^2}\bigg)\bigg|_{r=r_*}\\
 &=\bigg(\frac{2\Omega^2\partial_rm}{r^2}-\frac{2\Omega^2m}{r^3}\bigg)\bigg|_{r=r_*}-\frac{1}{r}\bigg(\frac{2\Omega^2m}{r^2}+\frac{4\varsigma}{r}\sqrt{1-\Omega^2}\bigg)\bigg|_{r=r_*}\\
 &=\bigg(\frac{2\Omega^2\partial_rm}{r^2}-\frac{2\Omega^2m}{r^3}\bigg)\bigg|_{r=r_*}>0
 \end{align*}
 where we applied \hyperlink{eqn:7.10}{(7.10)} in the last equality. It follows, that, for some $\delta>0$, the bracketed expression in \hyperlink{eqn:7.10}{(7.10)}, and thus $\partial_r\Omega^2$ itself, is negative on $(r_*-\delta,r_*)$, positive on $(r_*,r_*+\delta)$. It follows that $\partial_r\Omega^2>0$ holds not only on some $(r_0,r_0+\delta)$, but also thereafter. Indeed, let $r=r_*$ be the first subsequent point at which $\partial_r\Omega^2=0$. Applying the argument above yields that $\partial_r\Omega^2<0$ immediately earlier, a contradiction.\\ \\
We secondly claim that $m=0$ is reached before the solution terminates, leading us to region (III). Note again that, since $\partial_rm\geq 4\varsigma>0$, we reach $m=0$ after a finite interval, unless the solution terminates earlier, which can only be due to $\lim_{r\to r_\text{out}}\Omega^2= 1$, in view of our first claim above. Suppose for contradiction, then, that $\lim_{r\to r_\text{out}}\Omega^2= 1$ and $m<0$ on $r\in(r_0,r_\text{out})$. Since $m<0$, we have from \hyperlink{eqn:7.8b}{(7.8)} that
$$\partial_r\Omega^2\leq\frac{4\varsigma}{r}\sqrt{1-\Omega^2} $$ on $r\in(r_0,r_\text{out})$. This enables us to estimate the non-negative improper integrals
$$\int^{r_\text{out}}_{r_0}\frac{dr}{\sqrt{1-\Omega^2(r)}}=\int^1_{\Omega^2(r_0)}\frac{d\Omega^2}{\sqrt{1-\Omega^2}(\partial_r\Omega^2)}\geq\int^1_{\Omega^2(r_0)}\frac{d\Omega^2}{(1-\Omega^2)}\frac{r}{4\varsigma}\geq \frac{r_0}{4\varsigma}\int^1_{\Omega^2(r_0)}\frac{d\Omega^2}{1-\Omega^2}=\infty $$
where, in the first equality, we changed variable to $\Omega^2$. On the other hand, the LHS of 
$$m(r_\text{out})-m(r_0)=\int^{r_\text{out}}_{r_0}\partial_rm\hspace{1pt}dr\geq 4\varsigma\int^{r_\text{out}}_{r_0}\frac{dr}{\sqrt{1-\Omega^2(r)}} $$ is bounded, by $|m(r_0)|<\infty$, a contradiction. It follows that the maximal solution must reach $m=0$ (at some $r=r_1$, say) strictly before it terminates.
\\ \\
\underline{Region (III): Termination of the solution:} We now study the maximal forwards solution starting from $r=r_1$, at which $m(r_1)=0$, $\Omega^2(r_1)\in(0,1)$. Note that whenever $m\geq 0$, we have both $\partial_r\Omega^2>0$ and $\partial_rm>0$, so both variables are strictly increasing as we evolve. Since $m$ is non-negative, we in fact have
$$\partial_r\sqrt{1-\Omega^2} \equiv\frac{-\tfrac{1}{2}\partial_r\Omega^2}{\sqrt{1-\Omega^2}}\leq -\frac{2\varsigma}{r}$$
Integrating up, we have that for $r>r_1$,
$$\sqrt{1-\Omega^2(r)}\leq \sqrt{1-\Omega^2(r_1)}-2\varsigma\log(r/r_1) $$
Since $\log r$ diverges as $r\to\infty$, the RHS must become negative after a finite interval, say by $r=r_*:=r_1e^{1/2\varsigma}$, whence $\Omega^2\to 1$ must have already occurred. So $r_\text{out}<\infty$.\\ \\
It remains to show that the solution indeed terminates due to $\Omega^2\to 1$, \textit{strictly before} $1-2m/r\to0$. For this, it is useful to have a positive lower bound for $m$. Starting from $r=r_1$, evolve forwards by a small amount to $r=r_2>r_1$, where still $\Omega^2(r_2)\in(0,1)$ and $1-2m(r_2)/r_2>0$, but $m(r_2)>0$. Now, for $a>0$ constant, consider the function
$$h_a(r):=\log\Big(1-\frac{2m}{r}\Big)-a\sqrt{1-\Omega^2} $$
We will deduce the conclusion from the fact that this $h_a$ is increasing, at least for large enough $a$. Using \hyperlink{eqn:7.8b}{(7.8)}-\hyperlink{eqn:7.9b}{(7.9)}, one computes the derivative and discards positive terms as follows:
\begin{align*} 
h'_a(r)&=\frac{1}{r-2m}-\frac{1}{r}+\frac{1}{(r-2m)\sqrt{1-\Omega^2}}\bigg(a\Big(\frac{\Omega^2m}{r}+2\varsigma\sqrt{1-\Omega^2}\Big)-4\varsigma\Big(1+\frac{1}{\Omega^2}\Big)\bigg)\\
&\geq \frac{1}{(r-2m)\sqrt{1-\Omega^2}}\bigg(\frac{a\Omega^2m}{r}-4\varsigma\Big(1+\frac{1}{\Omega^2}\Big)\bigg)
\end{align*}
If we now choose 
$$a =\frac{4\varsigma r_*}{\Omega^2(r_2)m(r_2)}\bigg(1+\frac{1}{\Omega^2(r_2)}\bigg) $$
then, since $\Omega^2,m$ are increasing, we will have $h'_a\geq 0$, and thus $h_a$ increasing, on $r\in(r_2,r_*)$. In particular we have the lower bound 
$$\log\Big(1-\frac{2m}{r}\Big)\geq a\sqrt{1-\Omega^2}+h_a(r_2)\geq h_a(r_2) $$
and so $1-2m/r$ is strictly bounded away from zero on $r\in(r_2,r_*)$. Since $\Omega^2\to1$ already by $r=r_*$, we deduce that the maximal solution indeed terminates at $r=r_\text{out}<\infty$ with $m(r_2):=\lim_{r\to r_2}m(r)<\tfrac{1}{2}r_\text{out}$. Note that the latter limit exists since $m$ is increasing. \hfill$\square$\\ \\
\textbf{Remark 7.2.} \textit{We believe that any choice of three out of the four quantities $r_\text{in},r_\text{out},m_\text{in},m_\text{out}$ may serve to parametrize the family of solutions. For example, a comparison argument should suffice to show that, with $r_\text{in},m_\text{in}$ fixed, the endpoint $r_\text{out}=r_\text{out}(\varsigma)$ of the maximal domain is a continuous strictly decreasing function of $\varsigma$, taking all values in $(r_\text{in},\infty)$, thus exchanging the parameter $\varsigma$ for $r_\text{out}$.}\\ \\ \\
\large\hypertarget{sec:7.3}{\textbf{7.3\hspace{4mm}Proof of Theorem 1.2}}\label{7.3}\normalsize\\ \\
\textit{Proof of Theorem 1.2.} For each $m_\text{in}<0$ and $\varsigma,r_\text{in}>0$, \hyperlink{pro:7.2}{Proposition~7.2} gives us a unique maximal solution $\Omega^2,m:(r_\text{in},r_\text{out})\to\mathbb{R}$ to \hyperlink{eqn:7.8b}{(7.8)}-\hyperlink{eqn:7.9b}{(7.9)}. We can use \hyperlink{eqn:7.7}{(7.7)} to recover the coordinate $x$, which varies from $x=0$ at $r=r_\text{in}$ to some $x=x_\text{out}$ at $r=r_\text{out}$. Since $\Omega^2$ and $1-2m/r$ are bounded above and below on $(r_\text{in},r_\text{out})$, we see that $x_\text{out}<\infty$. Reversing the steps of \hyperlink{sec:7.1}{Section~7.1} yields a solution $(\Omega^2,r,\bm{u}^t_\pm,\bm{u}^x_\pm,\sigma_\pm)$ to \hyperlink{eqn:2.3}{(2.3)}-\hyperlink{eqn:2.9}{(2.9)} on the coordinate domain $(t,x)\in\mathbb{R}\times(0,x_\text{out})$.\\ \\
Next, we glue, as in Step~5 of the proof of \hyperlink{thm:1.1}{Theorem~1.1}, to vacuum regions which are isometric to Schwarzschild spacetimes with masses $m_\text{in},m_\text{out}$ respectively. A global $(t,x)$ coordinate system, taking values in $\mathbb{R}\times(0,\infty)$ (with $x=0$ redefined to coincide with the $r=0$ singularity of the former) covers the resulting spacetime $(M,g_{\mu\nu})$, yielding the Penrose diagram as depicted in the statement of \hyperlink{thm:1.2}{Theorem~1.2}. It remains to prove the other advertised properties of $(M,g_{\mu\nu})$.\\ \\
\ul{No (anti-)trapped surfaces:} We argue that $(M,g_{\mu\nu})$ is free of (anti-)trapped \textit{symmetry spheres}, and refer the reader to [\hyperlink{KU22}{KU22}, Appendix B] for the extension to general spacelike 2-surfaces. There are no trapped or anti-trapped symmetry spheres in a negative mass Schwarzschild spacetime, or in the exterior region of a positive mass Schwarzschild spacetime. Hence in $(M,g_{\mu\nu})$ we need only study the annular region $x\in[x_\text{in},x_\text{out}]$. However, we have already shown that $1-2m/r$ and $\Omega^2$, and by consequence $\partial_xr$ (see \hyperlink{eqn:7.7}{(7.7)}), are bounded below away from zero. Rotating to double null coordinates $(u,v)$, this is precisely what is needed to ensure that $\partial_ur<0$ and $\partial_vr>0$ hold throughout the annular region.
\\ \\
\ul{Not globally hyperbolic:} The argument is essentially that $(M,g_{\mu\nu})$ has the same Penrose diagram as negative mass Schwarzschild, but we include the details for completeness. Suppose, towards a\\
\begin{minipage}[t]{0.735\textwidth}contradiction, that $(M,g_{\mu\nu})$ admits a Cauchy surface $\Sigma$. Let $\gamma_1$ be an inextendible radial timelike curve in $(M,g_{\mu\nu})$, terminating at $r=0$ to the past and future. $\gamma_1$ must intersect $\Sigma$ at a unique point $p\in M$, because $\Sigma$ is a Cauchy surface. Since $\Sigma$ is by definition achronal, it does not intersect the chronological future $I^+(p)$. However, we can choose $\gamma_2$ with the same properties as $\gamma_1$, but lying entirely in $I^+(p)$, for example by translating $\gamma_1$ suitably in the $t$ coordinate. (In the Figure, $I^+(p)$ is a \textit{subset} of the shaded region, but every point $q$ whose sphere lies in the shaded region and which shares the angular coordinates of $p$ must belong to $I^+(p)$.) This is a contradiction, as $\gamma_2$ must also intersect $\Sigma$.
\hfill$\square$\end{minipage}\hspace{0.025\textwidth}
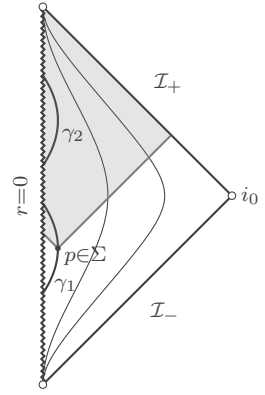
\begin{minipage}[t]{0.24\textwidth}
\vspace{-8mm}
\begin{figure}[H]
\begin{center}
\begin{tikzpicture}[scale=1]
\draw [darkgray, domain=-8*pi:8*pi, samples=500] plot ({0.8*(rad(atan((\x)*abs(\x)+0.6))-rad(atan((\x)*abs(\x)-0.6))},({0.8*(rad(atan((\x)*abs(\x)+0.6))+rad(atan((\x)*abs(\x)-0.6))});
\draw [darkgray, domain=-8*pi:8*pi, samples=500] plot ({0.8*(rad(atan((\x)*abs(\x)+1.6))-rad(atan((\x)*abs(\x)-1.6))},({0.8*(rad(atan((\x)*abs(\x)+1.6))+rad(atan((\x)*abs(\x)-1.6))});
\draw [darkgray, thick, domain=-1:1, samples=500] plot ({0.2-0.2*(\x)^2},({0.6*(\x)-0.7});
\draw [darkgray, thick, domain=-1:1, samples=500] plot ({0.2-0.2*(\x)^2},({0.6*(\x)+1});
\draw [darkgray, thick] (0, -0.8*pi) -- (0.8*pi,0);
\draw [darkgray, thick] (0.8*pi,0) -- (0,0.8*pi);
\fill[opacity=0.2, gray] (0.2,-0.7) -- (1.7,0.8) -- (0,0.8*pi) -- (0.0,-0.5) -- cycle;
\draw [gray, thick] (0.2,-0.7) -- (1.7,0.8);
\draw [gray, thick] (0.2,-0.7) -- (0,-0.5);
\filldraw[color=darkgray, fill=darkgray](0.2,-0.7) circle (0.03);
\node[darkgray] at (1.35, 1.5) [anchor = west] {$_{\mathcal{I}_+}$};
\node[darkgray] at (1.3, -1.55) [anchor = west] {$_{\mathcal{I}_-}$};
\node[darkgray] at (2.5, 0) [anchor = west] {$_{i_0}$};
\node[darkgray] at (0.12,0.8) [anchor = west] {$_{\gamma_2}$};
\node[darkgray] at (0.02,-1.2) [anchor = west] {$_{\gamma_1}$};
\node[darkgray] at (0.15,-0.8) [anchor = west] {$_{p\in\Sigma}$};
\node[darkgray,rotate=90] at (-0.3, -0.4) [anchor = west] {$_{r=0}$};
\draw[darkgray, decorate,decoration={zigzag,segment length=0.8mm, amplitude=.2mm},thick] (0, -0.8*pi) -- (0, 0.8*pi);
\filldraw[color=darkgray, fill=white](0,0.8*pi) circle (0.05);
\filldraw[color=darkgray, fill=white](0,-0.8*pi) circle (0.05);
\filldraw[color=darkgray, fill=white](0.8*pi,0) circle (0.05);
\end{tikzpicture}
\end{center}
\end{figure}\end{minipage}
\\ \\ \\
\Large\hypertarget{sec:A}{\textbf{A\hspace{4mm}Useful inequalities}}\label{A}\normalsize\\ \\
We first record some standard estimates that are used repeatedly in Sections 3 and 4. Each of Lemmas A.1-A.3 follows by a simple induction strategy, similar to \hyperlink{lem:3.3}{Lemma~3.3}, and we omit the proofs.\\ \\
\hypertarget{lem:a.1}{\textbf{Lemma A.1. }}\textit{Let $k\in\mathbb{N}_0$ be a non-negative integer. For both statements below, the constant $C_k>0$ depends only on $k$. (Each $f,g$ is to be defined on a suitable open subset of $\{x>0\}\subset\mathbb{R}^2$.)
\begin{enumerate}
\item For all $f,g\in C^\infty_\text{cusp}$,
$$\|fg\|_{C^k_\text{cusp}}\leq C_k \|f\|_{C^k_\text{cusp}}\|g\|_{C^k_\text{cusp}} $$
Alternatively, the statement holds for regular $\|\cdot\|_{C^k}$ norms: for all $f,g\in C^\infty$,
$$\|fg\|_{C^k}\leq C_k \|f\|_{C^k}\|g\|_{C^k} $$
\item For all $f\in C^\infty_\text{cusp}$ and $F\in C^\infty(\mathbb{R})$,
$$\|F(f)\|_{C^k_\text{cusp}}\leq C_k\|F\|_{C^k(\text{range}(f))}\big(1+\|f\|_{C^k_\text{cusp}}\big)^k $$
Again, the statement holds for regular $\|\cdot\|_{C^k}$ norms: for all $f\in C^\infty$ and $F\in C^\infty(\mathbb{R})$,
$$\|F(f)\|_{C^k}\leq C_k\|F\|_{C^k(\text{range}(f))}\big(1+\|f\|_{C^k}\big)^k $$
\end{enumerate}}
\vspace{-10mm}\hfill$\square$\\ \\
\hypertarget{lem:a.2}{\textbf{Lemma A.2. }}\textit{Let $k\in\mathbb{N}_0$ be a non-negative integer. Then there exists $C_k>0$, depending only on $k$, with the following property. Let $i+j\leq k$. For all $f\in{C^\infty_\text{cusp}}$ and $\bm{u}=\bm{u}^t\partial_t+\bm{u}^\xi\partial_\xi$ with $\bm{u}^t,\bm{u}^\xi\in C^\infty_\text{cusp}$,}
$$\|\partial^i_\xi\partial^j_t(\bm{u}(f))-\bm{u}(\partial^i_\xi\partial^j_tf)\|_{C^0} \leq C_k\|f\|_{C^k_\text{cusp}}\big(\|\bm{u}^t\|_{C^k_\text{cusp}}+\|\bm{u}^\xi\|_{C^k_\text{cusp}}\big)$$
\textit{(Again $f,\bm{u}^t,\bm{u}^\xi$ are to be defined on a suitable open subset of $\{x>0\}\subset\mathbb{R}^2$.)}\hfill$\square$\\ \\
\hypertarget{lem:a.3}{\textbf{Lemma A.3.} }\textit{Let $k\in\mathbb{N}_0$ be a non-negative integer. Then there exists $C_k>0$, depending only on $k$, with the following property. For all $f_0\in C^\infty[a,b]$ and $\varepsilon\leq1$, the functions $f_0(u)$, $f_0(v)$,  which are defined on $\Sigma_\varepsilon$, satisfy the following bound.} 
$$\hspace{4.1cm}\|f_0(u)\|_{C^k_\text{cusp}(\Sigma_\varepsilon)},\|f_0(v)\|_{C^k_\text{cusp}(\Sigma_\varepsilon)}\leq C_k\|f_0\|_{C^k[a,b]}\hspace{4.1cm}\square$$
The next two estimates are essential to the results of Sections 3 and 4.\\ \\
\hypertarget{lem:a.4}{\textbf{Lemma A.4. }}\textit{Let $k\in\mathbb{N}_0$ be a non-negative integer. Let $\Theta\subset\mathbb{R}^2$ be a comoving domain (see \hyperlink{def:4.1}{Definition~4.1}) contained in $\{0<\tau<1\}$ and let $x\in C^\infty(\Theta)$ be smooth with extension to $\{\tau=0\}$, satisfying $x|_{\tau=0}=\partial_\tau x|_{\tau=0}=0$ and the bounds
$$\tfrac{1}{4}\delta_1\leq \partial^2_\tau x\leq 2\|\Gamma\|\qquad\text{ on }\Theta $$
Then $\xi(\tau,\chi):=\sqrt{x(\tau,\chi)}$ obeys the estimate
$$\|\xi\|_{C^k(\Theta)},\hspace{1mm}\|\partial_\tau\xi\|_{C^k(\Theta)}\leq C(k,\delta_1,\|\partial^2_\tau x\|_{C^k(\Theta)}) $$
where the constant $C>0$ depends only on the indicated quantities. (It need not depend on $\|\Gamma\|$ because $\|\partial^2_\tau x\|_{C^k(\Theta)}$ is also an upper bound for $\partial^2_\tau x$.)}\\ \\
\textit{Proof.} To see that $\xi(\tau,\chi)$ is smooth with extension to $\{\tau=0\}$, see comments before \hyperlink{lem:4.2}{Lemma~4.2}. The estimate is essentially Hardy's inequality: because $x(\tau,\chi)$ vanishes to second order at $\tau=0$, we can control $x/\tau^2$ in terms of $\partial^2_\tau x$. We then estimate all the terms that arise from differentiating $\xi(\tau,\chi)=\tau(x/\tau^2)^{1/2}$.\\ \\
For this, we use the following standard identity for smooth functions vanishing to second order at 0. Let $f:[0,\infty)\to\mathbb{R}$ be smooth, and suppose $f(0)=f'(0)=0$. Then we have, for each $n\in\mathbb{N}_0$, the identity
$$\frac{d^n}{d\tau^n}\bigg(\frac{f}{\tau^2}\bigg)=\frac{1}{\tau^{n+2}}\int^\tau_0\tilde{\tau}^n(\tau-\tilde{\tau})\bigg(\frac{d^{n+2}f}{d\tau^{n+2}}\bigg)(\tilde{\tau})d\tilde{\tau} $$
This yields the $C^k$ version of Hardy's inequality:
\begin{align*}
\bigg\|\frac{x}{\tau^2}\bigg\|_{C^{k}(\Theta)}&=\sum_{i+j\leq k}\sup_\Theta\bigg|\partial^i_\tau\partial^j_\chi\bigg(\frac{x}{\tau^2}\bigg)\bigg|
=\sum_{i+j\leq k}\sup_\Theta\bigg|\frac{1}{\tau^{i+2}}\int^\tau_0\tilde{\tau}^i(\tau-\tilde{\tau})(\partial^{i+2}_\tau\partial^j_\chi x)(\tilde{\tau},\chi)d\tilde{\tau}\bigg|\\
&\leq \sum_{i+j\leq k}\sup_\Theta |\partial^{i+2}_\tau\partial^j_\chi x|=\|\partial^2_\tau x\|_{C^k(\Theta)}
\end{align*}
The extra prefactor of $\tau$ in $\xi(\tau,\chi)=\tau(x/\tau^2)^{1/2}$ allows us to control one extra derivative. More specifically, if we integrate the identity by parts, then for any $i,j$,
\begin{align*}
\bigg|\tau\partial^{i+1}_\tau\partial^j_\chi\bigg(\frac{x}{\tau^2}\bigg)\bigg|&=\frac{\tau}{\tau^{i+3}}\int^\tau_0\tilde{\tau}^{i+1}(\tau-\tilde{\tau})\partial^{i+3}_\tau\partial^j_\chi x(\tilde{\tau},\chi)d\tilde{\tau}\\
&=-\frac{1}{\tau^{i+2}}\int^\tau_0\big((i+1)\tilde{\tau}^i\tau-(i+2)\tilde{\tau}^{i+1}\big)\partial^{i+2}_\tau\partial^j_\chi x(\tilde{\tau},\chi)d\tilde{\tau}
\end{align*}
so that we have 
$$\bigg|\tau\partial^{i+1}_\tau\partial^j_\chi\bigg(\frac{x}{\tau^2}\bigg)\bigg|\leq 2\sup_{\Theta}|\partial^{i+2}_\tau\partial^j_\chi x| $$
We now examine 
\begin{align*}
\|\partial_\tau\xi\|_{C^k(\Theta)}&=\sum_{i+j\leq k}\sup_{\Theta}\bigg|\partial^{i+1}_\tau\partial^j_\chi\bigg(\tau\bigg(\frac{x}{\tau^2}\bigg)^{1/2}\bigg)\bigg|\\
&=\sum_{i+j\leq k}(i+1)\sup_{\Theta}\bigg|\partial^{i}_\tau\partial^j_\chi\bigg(\bigg(\frac{x}{\tau^2}\bigg)^{1/2}\bigg)\bigg|+\sum_{i+j\leq k}\sup_{\Theta}\bigg|\tau\partial^{i+1}_\tau\partial^j_\chi\bigg(\frac{x}{\tau^2}\bigg)^{1/2}\bigg|
\end{align*}
with the two terms corresponding to whether the $\tau$ is differentiated or not. We can estimate the first term using \hyperlink{lem:a.1}{Lemma~A.1}(ii):
\begin{align*}
\sum_{i+j\leq k}(i+1)\sup_{\Theta}\bigg|\partial^{i}_\tau\partial^j_\chi\bigg(\frac{x}{\tau^2}\bigg)^{1/2}\bigg|&\leq  (k+1)C_k\hspace{2pt}\|(\cdot)^{1/2}\|_{C^k[\tfrac{1}{4}\delta_1,\|\Gamma\|)}\bigg(1+\bigg\|\frac{x}{\tau^2}\bigg\|_{C^k(\Theta)}\bigg)^k\\
&\leq (k+1)C_k\hspace{2pt}\|(\cdot)^{1/2}\|_{C^k[\tfrac{1}{4}\delta_1,\|\Gamma\|)}(1+\|\partial^2_\tau x\|_{C^k(\Theta)})^k
\end{align*}
For the second term, the extra $\tau$ factor plays a role: for each $i,j$, $\partial^{i+1}_\tau\partial^j_\chi(x/\tau^2)^{1/2}$ expands into a polynomial expression in derivatives of $(\cdot)^{1/2}$, evaluated in $[\tfrac{1}{4}\delta_1,\|\Gamma\|)$, and derivatives $\partial^a_\tau\partial^b_\chi(x/\tau^2)$. Each $\partial^a_\tau\partial^b_\chi(x/\tau^2)$ is bounded above by $\|\partial^{a+2}_\tau\partial^b_\chi x\|_{C^0}$, which is itself bounded above by $\|\partial^2_\tau x\|_{C^k(\Theta)}$, except for the top-order derivative, that is when $a=i+1$, $b=j$ and $i+j=k$. For these terms we have
$$\bigg|\tau\partial^{i+1}_\tau\partial^j_\chi\bigg(\frac{x}{\tau^2}\bigg)\bigg|\leq 2\|\partial^2_\tau x\|_{C^k(\Theta)} $$
by our calculation above. Putting this together, we have indeed
$\|\partial_\tau\xi\|_{C^k(\Theta)}\leq C(k,\delta_1,\|\partial^2_\tau x\|_{C^k(\Theta)})$.\\ \\
This nearly extends directly to $\|\xi\|_{C^k(\Theta)}$, except for terms $\partial^i_\chi\xi$. However, since these satisfy $\partial^i_\chi\xi|_{\tau=0}$, we have $$|\partial^i_\chi\xi(\tau,\chi)|\leq \int^\tau_0|\partial_\tau\partial^i_\chi \xi(\tilde{\tau},\chi)|d\tilde{\tau}\leq\tau\|\partial_\tau\xi\|_{C^k(\Theta)}$$
Altogether, we have 
$$\|\xi\|_{C^k(\Theta)}\leq\|\partial_\tau\xi\|_{C^k(\Theta)}+\sum^k_{i=0}\|\partial^i_\chi\xi\|_{C^0(\Theta)}\leq (1+(k+1)\tau))\|\partial_\tau\xi\|_{C^k(\Theta)}\leq (k+2)\|\partial_\tau\xi\|_{C^k(\Theta)}  $$using the fact that $\Theta\subset\{\tau<1\}$. Thus, after relabelling $C$, we have the desired result. \hfill$\square$
\\ \\
\hypertarget{lem:a.5}{\textbf{Lemma A.5. }}\textit{Let $k\in\mathbb{N}_0$ be a non-negative integer. Let $\Theta\subset\mathbb{R}^2$ be a comoving domain (see \hyperlink{def:4.1}{Definition~4.1}) and let $x\in C^\infty(\Theta)$ be smooth with extension to $\{\tau=0\}$, satisfying $x|_{\tau=0}=\partial_\tau x|_{\tau=0}=0$ and the bounds
$$\tfrac{1}{4}\delta_1\leq \partial^2_\tau x\leq 2\|\Gamma\|\qquad\text{ on }\Theta $$
Let $i,j\in\mathbb{N}_0$ satisfy $i+j=k+1$, and suppose that there exists $C_1>0$ such that, on $\Theta$, 
\begin{equation*}|\partial^2_\tau(\partial^i_\tau\partial^j_\chi x)(\tau,\chi)|\leq C_1e^{C_1\tau} \tag{\hypertarget{eqn:A.1}{A.1}}
\end{equation*}Then $\xi(\tau,\chi):=\sqrt{x(\tau,\chi)}$ obeys the estimate
$$|\partial^i_\tau\partial^j_\chi \xi(\tau,\chi)|\leq 4\delta_1^{-1/2} e^{C_1\tau}+C_2(k,\delta_1,\|\partial^2_\tau x\|_{C^k(\Theta)}) $$
where the constant $C_2>0$ depends only on the indicated quantities. (It need not depend on $\|\Gamma\|$ because $\|\partial^2_\tau x\|_{C^k(\Theta)}$ is also an upper bound for $\partial^2_\tau x$.) Moreover, the estimate still holds \ul{at $\tau=0$} even without the assumption \hyperlink{eqn:A.1}{(A.1)}.}
\\ \\
\textit{Proof.} We  again expand $\partial^i_\tau\partial^j_\chi(\tau(x/\tau^2)^{1/2})$, yielding a polynomial expression in $\tau$, derivatives of $(\cdot)^{1/2}$ (themselves bounded in terms of $k,\delta_1,\|\Gamma\|$) and derivatives of $\partial^a_\tau\partial^b_\chi(x/\tau^2)$ itself. For the top-order derivative $\partial^i_\tau\partial^j_\chi(x/\tau^2)$ which appears in only one term, we have
\begin{align*}
\bigg|\partial^i_\tau\partial^j_\chi\bigg(\frac{x}{\tau^2}\bigg)\bigg|&=\bigg|\frac{1}{\tau^{i+2}}\int^\tau_0\tilde{\tau}^i(\tau-\tilde{\tau})\partial^i_\tau\partial^j_\chi(\partial^2_\tau x)d\tilde{\tau}\bigg|\\
&\leq\frac{1}{\tau^{i+2}}\int^\tau_0\tilde{\tau}^i(\tau-\tilde{\tau})C_1e^{C_1\tilde{\tau}}d\tilde{\tau}\\
&\leq\frac{1}{\tau}\int^\tau_0C_1e^{C_1\tilde{\tau}}d\tilde{\tau}\leq\frac{1}{\tau}(e^{C_1\tau}-1)
\end{align*}
where it is critical that integration removes the factor of $C_1$. Next, expanding $\partial^i_\tau\partial^j_\chi(\tau(x/\tau^2)^{1/2})$ and writing $C_2(k,\delta_1,\|\Gamma\|,\|\partial^2_\tau x\|_{C^k(\Theta)})$ (as in \hyperlink{lem:a.4}{Lemma~A.4}) to bound the collection of terms not involving the top-order $\partial^i_\tau\partial^j_\chi(x/\tau^2)$, we have
\begin{align*}
|\partial^i_\tau\partial^j_\chi\xi|&\leq 2\tau(\tfrac{1}{4}\delta_1)^{-1/2}\frac{1}{\tau}e^{C_1\tau}+C_2(k,\delta_1,\|\Gamma\|,\|\partial^2_\tau x\|_{C^k(\Theta)})\\
&=4\delta_1^{-1/2}e^{C_1\tau}+C_2(k,\delta_1,\|\Gamma\|,\|\partial^2_\tau x\|_{C^k(\Theta)})
\end{align*}
where the $\delta_1^{-1/2}$ comes from the derivative of $(\cdot)^{1/2}$ evaluated at $\tfrac{1}{4}\delta_1$.\\ \\
Finally, if the assumption \hyperlink{eqn:A.1}{(A.1)} is not made, simply choose a large enough $C_1>0$ for it to hold (which exists because $x$ is assumed smooth with extension), and then evaluate at $\tau=0$ to deduce
$$|\partial^i_\tau\partial^j_\chi\xi(0,\chi)|\leq 4\delta^{-1/2}_1 +C_2(k,\delta_1,\|\partial^2_\tau x\|_{C^k(\Theta)})$$
which is what was to be shown.\hfill$\square$
\\ \\ \\
\Large\hypertarget{sec:B}{\textbf{B\hspace{4mm}Weak formulation of Einstein's equation}}\label{B}\normalsize\\ \\
The following definition, specific to the coordinate chart $(t,x)$, i$.$e$.$ for which the metric takes the form $$g=\Omega^2(-dt^2+dx^2)+r^2g_{\mathbb{S}^2} $$ is equivalent to the definition of Geroch-Traschen \hyperlink{GT87}{$[$GT87$]$} and is really a coordinate-independent condition.\vspace*{4mm}\\
\hypertarget{def:b.1}{\textbf{Definition B.1}} (Weak solution of Einstein's equation)\textbf{.} \textit{Let $U\subset\mathbb{R}^2$ be an open subset and let $r,\Omega^2:U\to\mathbb{R}_{>0}$ be locally bounded above and below and weakly differentiable, with locally square-integrable weak derivatives. Let $T_{xx}$, $T_{tx}$, $T_{tt}:U\to\mathbb{R}$ be locally integrable functions. Then we say that $r,\Omega^2, T$ \underline{satisfy Einstein's equation weakly on U} if for each $\phi\in C^\infty_c(U)$, the following integral conditions hold.}
\begin{equation*}
\int_U\Big(\partial_xr\partial_x\phi-\partial_tr\partial_t\phi+\frac{2\Omega^2m}{r^2}\phi-r(T_{tt}-T_{xx})\phi\Big)dx =0 \tag{\hypertarget{eqn:B.1}{B.1}}
\end{equation*}
\begin{equation*}
\int_U\Big(\partial_x\log\Omega^2\partial_x\phi-\partial_t\log\Omega^2\partial_t\phi-\frac{4\Omega^2m}{r^3}\phi+2 (T_{tt}-T_{xx})\phi\Big)dx =0 \tag{\hypertarget{eqn:B.2}{B.2}}
\end{equation*}
\begin{equation*}
\int_U\Big(\partial_xr\partial_t\phi+\big(\tfrac{1}{2}\partial_t\log\Omega^2\partial_xr+\tfrac{1}{2}\partial_x\log\Omega^2\partial_tr-rT_{tx}\big)\phi\Big)dx=0 \tag{\hypertarget{eqn:B.3}{B.3}}
\end{equation*}
\begin{equation*}\int_U\Big(\partial_tr\partial_t\phi+\big(\tfrac{1}{2}\partial_t\log\Omega^2\partial_tr+\tfrac{1}{2}\partial_x\log\Omega^2\partial_xr-\frac{\Omega^2m}{r^2}-rT_{xx}\big)\phi\Big)dx=0 \tag{\hypertarget{eqn:B.4}{B.4}}
\end{equation*}
\textit{The derivatives of $r,\Omega^2$ appearing in the integrals are understood to be the weak derivatives.}\\ \\
If $r,\Omega^2,T$ are smooth, the integral conditions are equivalent to the Einstein equations \hyperlink{eqn:2.3}{(2.3)}-\hyperlink{eqn:2.6}{(2.6)} holding classically (see also below). This may be seen by integrating the first terms in each integrand by parts. \vspace*{3.5mm} \\
We now identify a condition for $r,\Omega^2,T$ to meet the hypothesis of \hyperlink{def:b.1}{Definition~B.1} which is sufficient for our purposes. Namely, if we solve Einstein's equation (classically) in the domains $\{x>0\}$ and $\{x<0\}$ with $T$ integrable \textit{across $x=0$} , and we glue the two solutions along $\{x=0\}$ in a $C^1$ manner, then this is a weak solution on $U$.
\vspace*{3.5mm} \\
\hypertarget{pro:b.1}{\textbf{Proposition B.1.}} \textit{Let $U\subset \mathbb{R}^2$ be an open subset and let $r,\Omega^2,T_{xx},T_{xt},T_{tt}\in C^\infty(U\cap\{x\neq0\})$ be defined on $\{x\neq0\}$. Moreover, suppose that:
\begin{enumerate}
\item $r,\Omega^2$ extend to $C^1$ functions on $U$ which are locally bounded below.
\item $T_{xx},T_{xt},T_{tt}$, if extended arbitrarily to $U$, are locally integrable on $U$.
\item on $U\cap\{x\neq0\}$, the Einstein equations hold classically:
$$(-\partial^2_x+\partial^2_t)r=-\frac{2\Omega^2m}{r^2}+r(T_{tt}-T_{xx})\vspace{-2mm} $$
$$(-\partial^2_x+\partial^2_t)\log\Omega^2=\frac{4\Omega^2m}{r^3}-2(T_{tt}-T_{xx}) \vspace{2mm} $$
$$\partial^2_tr-\tfrac{1}{2}\partial_t\log\Omega^2\partial_xr-\tfrac{1}{2}\partial_x\log\Omega^2\partial_tr=-rT_{tx} $$
$$\partial_x\partial_tr-\tfrac{1}{2}\partial_t\log\Omega^2\partial_tr-\tfrac{1}{2}\partial_x\log\Omega^2\partial_xr=-\frac{\Omega^2m}{r^2}-rT_{xx} $$
\end{enumerate}
Then $r,\Omega^2,T$ satisfy Einstein's equation weakly on $U$.}\\ \\
\textit{Proof. }In view of conditions (i)-(ii), we regard $r,\Omega^2,T$ as being defined on $U$. The conditions on $r,\Omega^2$ in Definition B.1 are clearly satisfied, e$.$g$.$ continuity of $\partial r$ certainly implies local square integrability. Hence each integral \hyperlink{eqn:B.1}{(B.1)}-\hyperlink{eqn:B.4}{(B.4)} under concern is well-defined, and we study the first one \hyperlink{eqn:B.1}{(B.1)} only---the argument is identical for the others.\\ \\
Fix $\phi\in C^\infty_c(U)$. We want to show that 
$$\int_U\Big(\partial_xr\partial_x\phi-\partial_tr\partial_t\phi+\frac{2\Omega^2m}{r^2}\phi- r(T_{tt}-T_{xx})\phi\Big)dx =0 $$
Shortening the integrand to $(\hspace{1pt}\cdots)$, we are permitted to split up the integration domain as follows:
$$\int_U(\hspace{1pt}\cdots)\hspace{2pt}dx=\int_{U\cap\{x>0\}}(\hspace{1pt}\cdots)\hspace{2pt}dx+\int_{U\cap\{x<0\}}(\hspace{1pt}\cdots)\hspace{2pt}dx=\lim_{\varepsilon\downarrow0}\bigg(\int_{U\cap\{x>\varepsilon\}}(\hspace{1pt}\cdots)\hspace{2pt}dx+\int_{U\cap\{x<-\varepsilon\}}(\hspace{1pt}\cdots)\hspace{2pt}dx\bigg) $$
since we are really integrating on the (compact) support of $\phi$ on which the functions are all integrable.\\ \\
Now fix $\varepsilon>0$. On $U\cap\{x>\varepsilon\}$, all the terms appearing in $(\hspace{1pt}\cdots)$ are smooth. Integrating by parts, we obtain
$$\int_{U\cap\{x>\varepsilon\}}(\hspace{1pt}\cdots)\hspace{2pt}dx=-\int_{U\cap\{x=\varepsilon\}}\partial_xr\phi\hspace{2pt} dt+\int_{U\cap\{x>\varepsilon\}}\phi\Big(-\partial^2_xr+\partial^2_tr+\frac{2\Omega^2m}{r^2}-r(T_{tt}-T_{xx})\Big)dx $$
By condition (iii), the latter integral vanishes. So we finally have 
$$\int_U(\hspace{1pt}\cdots)\hspace{2pt}dx=\lim_{\varepsilon\downarrow0}\bigg(-\int_{U\cap\{x=\varepsilon\}}\partial_xr\phi\hspace{2pt} dt+\int_{U\cap\{x=-\varepsilon\}}\partial_xr\phi\hspace{2pt} dt\bigg)$$
We now use the fact that the $C^1$ limits on both sides agree, and argue via Dominated Convergence that this limit therefore vanishes.\hfill$\square$
\newpage
\Large\textbf{References}\normalsize \\ \\
\setlength{\tabcolsep}{0em}
\begin{tabularx}{\textwidth} { 
   >{\raggedright\arraybackslash \hsize=.1\textwidth}X 
   >{\raggedright\arraybackslash}X  }
$[$BG73$]$ & M. S. Baouendi and C. Goulaouic. \hypertarget{BG73}{Cauchy Problems with Characteristic Initial Hypersurface.} \textit{Commun. Pure App. Math.}, 26:455-475, 1973.\vspace*{2mm} \\
$[$Chr84$]$ & D. Christodoulou. \hypertarget{Chr84}{Violation of cosmic censorship in the gravitational collapse of a dust cloud.} \textit{Commun. Math. Phys.}, 93:171-195, 1984. \vspace*{2mm} \\
$[$CN98$]$ & C. M. Claudel and K. P. Newman. \hypertarget{CN98}{The Cauchy problem for quasi-linear hyperbolic evolution problems with a singularity in the time.} \textit{Proc. R. Soc. Lond. A}, 454, 1998.\vspace*{2mm} \\
$[$COD92$]$ & C. Clarke and N. O'Donnell. \hypertarget{COD92}{Dynamical extension through a space-time singularity.} \textit{Rend. Sem. Mat. Univ. Politec. Torino}, 50(1):39-60, 1992. \vspace*{2mm} \\
$[$GT87$]$ & R. Geroch and J. Traschen. \hypertarget{GT87}{Strings and other distributional sources in general relativity.} \textit{Phys. Rev. D}, 36:1017-1031, 1987.\vspace*{2mm} \\
$[$Isr65$]$ & W. Israel. \hypertarget{Isr65}{Singular hypersurfaces and thin shells in general relativity.} \textit{Nuovo Cimento B (1965-1970)}, 44:1-14, 1965.\vspace*{2mm} \\
$[$Lee13$]$ & J. Lee. \hypertarget{Lee13}{\textit{Introduction to Smooth Manifolds (2nd ed.)}}. Springer, 2013.\vspace*{2mm} \\
$[$Kic96$]$ & S. Kichenassamy. \hypertarget{Kic96}{Fuchsian equations in Sobolev spaces and blow-up.} \textit{Journal of Differential Equations}, 125:299-327, 1996.\vspace*{2mm} \\
$[$KU22$]$ & C. Kehle and R. Unger. \hypertarget{KU22}{Gravitational collapse to extremal black holes and the third law of black hole thermodynamics.} \textit{J. Eur. Math. Soc.}, 2022.\vspace*{2mm} \\
$[$Man00$]$ & T. Mandai. \hypertarget{Man00}{The method of Frobenius to Fuchsian partial differential equations.} \textit{J. Math. Soc. Japan}, 52(3), 2000.\vspace*{2mm} \\
$[$Mon81$]$ & V. Moncrief. \hypertarget{Mon81}{Global properties of Gowdy spacetimes with $T^2\times\mathbb{R}$ topology.} \textit{Ann. Phys.} (NY) 132:87-107, 1981. \vspace*{2mm} \\
$[$Nol03$]$ & B. Nolan. \hypertarget{Nol03}{Dynamical extensions for shell-crossing singularities.} \textit{Class. Quant. Grav.}, 20:575-586, 2003.\vspace*{2mm} \\
$[$OS39$]$ & J. R. Oppenheimer and H. Snyder. \hypertarget{OS39}{On continued gravitational contraction.} \textit{Phys. Rev.} 56:455-459, 1939.\vspace*{2mm} \\
$[$PH67$]$ & A. Papapetrou and A. Hamoui. \hypertarget{PH67}{Surfaces caustiques d\'eg\'en\'er\'es dans la solution de Tolman. La singularit\'e physique en relativit\'e g\'en\'erale.} \textit{Annal. Inst. Hen. Poin.}, 6:343-364, 1967.\vspace*{2mm} \\
$[$Pen69$]$ & R. Penrose. \hypertarget{Pen69}{Gravitational collapse: the role of general relativity.} \textit{Rivista del Nuovo Cimento}, 1:252–276, 1969.\vspace*{2mm} \\
$[$Ren00$]$ & A. Rendall. \hypertarget{Ren00}{Fuchsian analysis of singularities in Gowdy spacetimes beyond analyticity.} \textit{Class. Quant. Grav.}, 17:3305-3316, 2000. \vspace*{2mm} \\
$[$Ren04$]$ & A. Rendall. \hypertarget{Ren04}{Fuchsian methods and spacetime singularities.} \textit{Class. Quant. Grav.}, 21:S295-S304, 2004.\vspace*{2mm} \\
$[$Sbi18$]$ & J. Sbierski. \hypertarget{Sbi18}{The $C^0$-inextendibility of the Schwarzschild spacetime and the spacelike diameter in Lorentzian geometry.} \textit{Journal of Differential Geometry}, 108(2):319-378, 2018.\vspace*{2mm} \\
\end{tabularx}
\begin{tabularx}{\textwidth} { 
   >{\raggedright\arraybackslash \hsize=.1\textwidth}X 
   >{\raggedright\arraybackslash}X  }
$[$Smi25$]$ & M. Smith. \hypertarget{Smi25}{Shell crossing and the dynamic formation of thin shells in the collapse of self-gravitating relativistic dust.} In preparation.\vspace*{2mm} \\
$[$Teg11$]$ & S. Tegai. \hypertarget{Teg11}{On a weak solution of Einstein equations for expanding dust.} \textit{J. Sib. Fed. U.}, 4:43-49, 2011.\vspace*{2mm} \\
$[$YSM73$]$ & P. Yodzis, H. Seifert and H. Mueller zum Hagen. \hypertarget{YSM73}{On the occurrence of naked singularities in general relativity.} \textit{Commun. Math. Phys.}, 34:135-148, 1973.
\end{tabularx}
\end{document}